\documentclass[aps,nofootinbib, preprint, tightenlines]{revtex4}

\usepackage[mathscr]{eucal}
\usepackage{amsfonts,amssymb,amsmath,bbm}

\usepackage[all]{xy}  
\usepackage{tikz}  

\usepackage[english]{babel}

\usepackage[latin1]{inputenc}

\usepackage{graphics}

\newcommand{\cA}{{\mathcal A}}

\newcommand{\cM}{{\mathcal M}}

\newcommand{\cO}{{\mathcal O}}
\newcommand{\cP}{{\mathcal P}}

\newcommand{\cS}{{\mathcal S}}

\newcommand{\q}{\quad}

\usepackage{amsfonts}
\def\ka{\kappa}
\newcommand{\su}{\mathfrak{su}}
\newcommand{\an}{\mathfrak{an}}
\newcommand{\so}{\mathfrak{so}}
\newcommand{\SU}{\mathrm{SU}}
\newcommand{\AN}{\mathrm{AN}}

\newcommand{\SO}{\mathrm{SO}}

\def\kk{{\cal K}}
\newcommand{\R}{\mathbb{R}}
\newcommand{\C}{\mathbb{C}}
\newcommand{\N}{\mathbb{N}}

\newcommand{\be}{\begin{equation}}
\newcommand{\ee}{\end{equation}}
\newcommand{\bes}{\begin{eqnarray}}
\newcommand{\ees}{\end{eqnarray}}
\newcommand{\f}{\frac}

\def\tl{\widetilde}

\def\ka{\kappa}

\def\la{\langle}
\def\ra{\rangle}
\def\vphi{\varphi}
\renewcommand{\hat}{\widehat}
\def\kk{{\cal K}}

\def\ss{{\cal S}}
\def\tlF{\tl{F}}

\def\mn{{\mu\nu}}

\newcommand{\tF}{{\tilde F}}

\newcommand\Cyl{{\rm Cyl}}

\def\inv{{\mbox{\tiny -1}}}
\def\minus{{\mbox{\small -}}}
\def\plus{{\mbox{\tiny +}}}
\def\ka{\kappa}
\newcounter{letter} \newcounter{numeral} \newcounter{Numeral}

\def\nn{\nonumber}

\newcommand\Tr{\mathrm{Tr}}
\def\extd{\mathrm {d}}
\def\vphi{\varphi}
\def\vphihat{\widehat{\varphi}}
\newcommand\e{{\mbox{e}}}
\newcommand{\g}{{\mathfrak{g}}}

\newcommand{\id}{\mathbb{I}}

\begin{document}

\title{The microscopic dynamics of quantum space as a group field theory}


\author{\bf Daniele Oriti}
\address{Max Planck Institute for Gravitational Physics (Albert Einstein Institute)
\\ Am M\"{u}hlenberg 1
D-14476 Golm,
Germany, EU \\ daniele.oriti@aei.mpg.de}

\begin{abstract}
We provide a rather extended introduction to the group field theory approach to quantum gravity, and the main ideas behind it. We present in some detail the GFT quantization of 3d Riemannian gravity, and discuss briefly the current status of the 4-dimensional extensions of this construction. We also briefly report on some recent results obtained in this approach, concerning both the mathematical definition of GFT models as bona fide field theories, and  possible avenues towards extracting interesting physics from them. \end{abstract}

\maketitle

\tableofcontents

\section{Introduction and basic ingredients}

The field of non-perturbative and background independent quantum gravity has progressed considerably over the past few decades \cite{libro}. New research directions are being explored, new important developments are taking place in existing approaches, and some of these approaches are converging to one another. As a result, ideas and tools from one become relevant to another, and trigger further progress. The group field theory (GFT) formalism \cite{laurentgft,iogft,iogft2, VincentRenorm} nicely captures this convergence of approaches and ideas. It is a generalization of the much studied matrix models for 2d quantum gravity and string theory \cite{mm}. At the same time, it generalizes them, as we are going to explain, by incorporating the insights coming from canonical loop quantum gravity and its covariant spin foam formulation, and so it became an important part of this approach to the quantization of 4d gravity \cite{carlo,review,alex,thesis}. Furthermore, it is a point of convergence of the same loop quantum gravity approach and of simplicial quantum gravity approaches, like quantum Regge calculus and dynamical triangulations \cite{iogft}, in that the covariant dynamics of the first takes the form, as we are going to see, of simplicial path integrals. More recently, tools and ideas from non-commutative geometry have been introduced as well in the formalism, and this has helped attempts to make tentative contact with effective models and quantum gravity phenomenology.

The goals of this paper are to explain the general idea behind the GFT formalism and its roots, discuss its relation with other current approaches to quantum gravity, detail to some extent the construction and features of GFT models in 3 and 4 dimensions, and finally report briefly on some recent results, concerning both an improved mathematical understanding of it and results with possible bearing on phenomenology. We do not intend to provide an up-to-date review or a status report of the subject, which has progressed enormously in the last few years, and we will refer instead to the literature for most of these recent developments. The models we will discuss in some more detail are in Euclidean signature, but the whole construction can be performed in the Lorentzian setting as well\footnote{Doing so does not require a modification of the general formalism, but only of some ingredients, and of course much additional care to the mathematical issues of dealing with non-compact groups.}.

\subsection{A general definition}

In very general terms, group field theories are an attempt to define quantum gravity in terms of {\it combinatorially non-local quantum field theories on group manifolds or on the corresponding Lie algebras}, related to the Lorentz or rotation group. The formalism itself is mostly characterized by the combinatorial non-locality that we are going to specify in the following, and the choice of group manifolds as domain of definition of the field, and in particular of the Lorentz/rotation groups, is dictated by the wish to model quantum gravity. Other choices could be devised easily, for different purposes (e.g. describing matter or gauge fields, capturing topological structures, etc). From this point of view, group field theories are a special case of tensor models \cite{tensor}, corresponding to a special choice of domain space for the fundamental tensors and, possibly, peculiar symmetries. We stick in fact to quantum gravity models in this contribution. Before introducing the formalism or specific models, let us motivate in some detail these basic ingredients: a quantum field theory framework, the use of group structures, the combinatorial non-locality. 

\subsection{A quantum field theory for quantum gravity?}

It is actually easy to see why we may want to use a quantum field theory formalism (QFT) also for quantum gravity. Quantum field theory is the best formalism we have for describing physics at both microscopic and mesoscopic  scales, both in high energy particle physics and in condensed matter physics, for both elementary systems and many-particle ones. And actually, even at large and very large scales, it is still field theory that we use, only we are most often able to neglect quantum aspects (e.g. General Relativity itself). 

So, probably, a more relevant question is: can we still hope for a formulation of quantum gravity, for  description of the microscopic structure of quantum space, in terms of a quantum field theory? The rationale behind such question is the following. We only know how to define quantum field theories on given background manifolds and we have under full control (including renormalization etc), only quantum field theories on flat spaces. Moreover, we have already tried to apply this formalism to gravity, formulating it as a quantum field theory of massless spin-2 particle propagating on a flat space, the gravitons, thought of as the carriers of the gravitational interaction, coupled universally to other matter and gauge fields. It has been the first approach to quantum gravity ever developed, by the great scientists of the past century \cite{carlohistory}. We know it does not work. The field theory defined by this approximation is not renormalizable. Quantum gravity, beyond the effective field theory level, is not such a quantum field theory. 

This historically well founded objection, however, is not a no-go theorem, of course. In particular, the non-renormalizability result does not rule out the use of a quantum field theory formalism as such. What it rules out is the specific idea of a field theory of gravitons as a fundamental definition of the dynamics of quantum space. To the eyes of many, it rules out also the idea that the requirement of background independence with respect to spacetime is a dispensable one, supporting instead the belief that this should be the defining property of any sound quantum gravity formalism \cite{carlodiffeo,leeback}, and possibly manifestly realized\footnote{In fact, even a theory formulated perturbatively around a given background can be background independent, if the physical observables computed from it turn out to be independent of the background used to compute them. However, this is not easy to realize and to verify explicitly, so the requirement of manifest background independence becomes more than just an aesthetic choice.}.

The more serious objection to the idea of using a quantum field theory formalism for quantum gravity, in fact, is that a good theory of quantum gravity should be background-independent, because it should explain origin and properties {\it of spacetime} itself, of its geometry and, maybe, its topology. But, as said, we know how to formulate quantum field theories only on given backgrounds. 

Again, this does not rule out the use of the QFT formalism. It means however, that, if an (almost) ordinary QFT it should be, quantum gravity can only be a QFT on some auxiliary, internal or \lq\lq higher-level\rq\rq space\footnote{An alternative to this conclusion is represented by the Asymptotic Safety program \cite{asympsafety}. Although clearly based on a very different starting point and a different language, this program is not in contradiction with the GFT approach. We are however not going to discuss their possible relations, which would lead us astray from the focus of this paper.}. 

So we can then look at General Relativity (GR) itself and try to identify some background (non-dynamical) structures that are already present in it and could provide or characterize such space. After all, GR {\it is} a classical field theory, and it is background independent with respect to the geometry of spacetime, which is the minimum of what we may want to have as dynamical quantum degrees of freedom in a quantum gravity theory.

\subsection{Background structures in classical and quantum GR}

The first background structure that comes to mind is the spacetime dimension. We have of course overwhelming experimental evidence for a 4-dimensional spacetime. But it is also true that we do not have a clear enough understanding of {\it why} this dimensionality should hold true at high energy, microscopic length scales or when all quantum effects of space dynamics are taken into account. So, it makes sense to look for alternatives, i.e. for a formalism in which the spacetime dimension is dynamical. Group field theories (as loop quantum gravity or simplicial quantum gravity and tensor models) fix the kinematical dimension of space at the very beginning, at least in the present formulation. However, on the one hand these approaches are not subjected to any obstruction to dimensional generalization; on the other hand this kinematical choice does not ensure that the dynamic dimension of spacetime in some effective continuum and classical description will match the kinematical one. The example of dynamical triangulations \cite{DT}, in fact, show that this matching is far from trivial, and that its achievement can be considered in fact an important result.

\

Another background structure is the internal, local symmetry group of the theory, i.e. the Lorentz group, which provides the local invariance under change of reference (tetrad) frame, and that is at the heart of the equivalence principle. It is a sort of \lq\lq background internal space \rq\rq of the theory. This gives the primary conceptual motivation for using the Lorentz group (and related) in GFT. At the same time, as we are going to discuss in the following, this choice allows to incorporate in the GFT formalism what we have learned from canonical loop quantum gravity, as well as many of its results \cite{carlo,thomas}. In fact, the role of the Lorentz group (and of its $\SU(2)$ rotation subgroup) is brought to the forefront in the LQG formalism (see chapter by H. Sahlmann), which is based on the classical reformulation of GR as a background independent (and diffeomorphism invariant) theory of a Lorenz connection. Upon quantization, it gives a space of states which is an $L^2$ space of generalized $\SU(2)$ connections, obtained as the projective limit of the space of $L^2$ cylindrical functions of finite numbers of $\SU(2)$ group elements, representing parallel transports of the same connection along elementary paths in space. Thus spacetime geometry is encoded, ultimately, in these group-theoretic structures. 

\

A background structure of GR is, in fact, also its configuration space, seen from the Hamiltonian perspective, and regardless of the specific variables used to parametrize it: the space $\mathcal{S}$ of (spatial) geometries on a given (spatial) topology, coined \lq\lq superspace\rq\rq by Wheeler. In the ADM variables, this is a metric space in its own right \cite{giulini} and could be considered indeed a sort of \lq\lq background meta-space: a space of spaces\rq\rq. Let us sketch briefly how this background structure enters the quantization of the classical theory, at least at the formal level, in both canonical and covariant approaches. Loop quantum gravity, spin foam models and simplicial quantum gravity reformulate and make more precise and successful, to different degrees and in different ways, these 'historic' approaches.

The canonical approach starts with a globally hyperbolic spacetime, with topology $\Sigma \times \mathbb{R}$. For simplicity, $\Sigma$ is usually chosen to be compact and simply connected, with the topology of the 3-sphere $S_3$. The wave function of canonical geometrodynamics are functionals on the space of 3-metrics on the 3-sphere, $\Psi(h_{ij})$, and kinematical observables are functionals of the phase space variables, themselves built from the conjugate 3-metric $h_{ij}$ and extrinsic curvature  $K_{ij}$, turned into (differential) operators acting on the wave functions. Gravity being (classically) a totally constrained system, the dynamics is imposed by identifying the space of states (and associated inner product), i.e. the space of functionals on the space of 3-geometries (metrics up to spatial diffeomorphisms), which satisfy also the Hamiltonian  constraint, and thus the Wheeler-deWitt equation $\mathcal{H}\Psi(h_{ij})\,=\,0$. This, together with the identification of physical observables, defines the theory from a canonical perspective. A covariant definition of the dynamics can instead be looked for in sum-over-histories framework. Given the same (trivial) spacetime topology, and the same choice of spatial topology,
consider all the possible geometries (spacetime metrics up to
diffeomorphisms) that are compatible with it. Transition
amplitudes (defining either the physical inner product of the canonical theory or \lq causal\rq ~transition amplitudes, and thus Green functions for the canonical Hamiltonian constraint operator \cite{teitelboim}), for given boundary configurations of the field (i.e. possible 3-geometries on the 3-sphere): $h$ and $h'$, would be given by a sum over spacetime geometries like:
\be
Z_{QG}\left(h,h'\right)=\int_{g(\mathcal{M}\mid h,h')}\mathcal{D}g \,e^{i\,S_{GR}(g,\mathcal{M})}
\ee
i.e. by summing over all 4-geometries inducing the given 3-geometries
on the boundary, with the amplitude modified by boundary terms if needed.
The expression above is obviously purely formal, for a variety of well-known reasons. In any case, it looks like a prototype of a background independent quantization of spacetime geometry, for given spatial and spacetime topology, and given space $\mathcal{S}$ of possible 3-geometries. Also, the physical
interpretation of the above quantities presents several challenges,
given that the formalism seems to be bound to a cosmological setting, where our usual interpretation of quantum
mechanics is rather dubious. A good point about group field theory, and about LQG, spin foams and simplicial gravity, is that they seem to provide a more rigorous definition of the above formula, which is also {\i local} in a sense to be clarified below.

\subsection{Making topology dynamical: the idea of 3rd quantization}

Making sense of a path integral
quantization of gravity on a fixed spacetime is difficult and ambitious enough. However, one may wish to treat also topology as a dynamical
variable in the theory, and try to implement a sort of
\lq\lq sum over topologies\rq\rq alongside a sum over geometries, thus
extending the latter to run over {\it all} possible spacetime
geometries and not only over those which can live on a given topology. This has
consequences on the type of geometries one can consider, in the
Lorentzian case, given that a non-trivial spacetime topology implies
spatial topology change \cite{fay} and this in turn forces the metric
to allow either for closed timelike loops or for geometries which are degenerate (at least) at isolated points. This argument was made stronger by
Horowitz \cite{horo} concluding that if
degenerate metrics are included in the (quantum) theory, then topology
change is not only possible but unavoidable and non-trivial
topologies therefore must be included in the quantum
theory. There are several other results that suggest the need for topology change in quantum
gravity, including work on topological geons \cite{fayrafael}, in string theory \cite{greene}, and on wormholes as a possible explanatory mechanism for the small value of the cosmological constant
\cite{banks}. Moreover, the possibility has been raised that {\it all}
constants of nature can be seen as computable dynamical vacuum expectation values in a theory in which topology change is allowed \cite{coleman}. 

All this, together with the analogy
with string perturbation theory and the aim to solve some problems of
the canonical formulation of quantum gravity, prompted the proposal of
a \lq\lq third quantization\rq\rq ~formalism for quantum gravity
\cite{kuchar, giddingsstrominger, guigan}. The general idea is to define a (scalar) field theory on superspace $\mathcal{S}$ for a given choice of
basic spatial manifold topology, e.g. the 3-sphere. This means turning the wave function of the canonical
theory into an operator (acting on a new Hilbert space) $\phi(h_{ij})$, whose dynamics is defined by
an action of the type:
\be
S(\phi)=\int_{\mathcal{S}}\mathcal{D} h\,\phi(h)\Delta\phi(h) + \lambda \int_{\mathcal{H}}\mathcal{D} h\,\mathcal{V}\left(\phi(h)\right)
\ee
with $\Delta$ being the Wheeler-DeWitt differential operator of canonical
gravity here defining the kinetic term (free propagation) of the
theory, while $\mathcal{V}(\phi)$ is a generic, e.g. cubic,
and generically non-local (in superspace) interaction term for the
field, governing the topology changing processes. Notice that because
of the choice of basic spatial topology needed to define the 3rd
quantized field, the topology changing processes described here are
those turning $X$ copies of the 3-sphere into $Y$ copies of the same.

The quantum theory
is \lq\lq defined\rq\rq  ~by the partition function $Z=\int\mathcal{D}\phi 
e^{-S(\phi)}$, in its perturbative expansion in \lq\lq Feynman diagrams\rq\rq:

\begin{figure}[here]
\includegraphics[width=11cm]{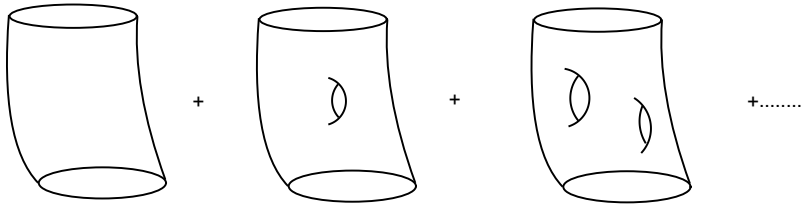}
\caption{The perturbative expansion of the 3rd quantized field theory in interaction processes for universes}
\end{figure}

The Feynman amplitudes will be given by the quantum gravity path integral (sum over geometries) for each spacetime topology (identified with a particular interaction process of universes). The one for trivial topology will represent a sort
of one particle propagator, thus a Green function for the
Wheeler-DeWitt equation\footnote{Note that this is in analogy with what happens in ordinary quantum field theory of point particles. Here the Feynman diagrams represent possible interaction processes of a certain number of (virtual) particles, and the Feynman amplitudes can be written in the form of sum over histories for the particles involved in these processes.}. 

Other features of this (very) formal setting are: 1) the
full classical equations of motions for the scalar field on superspace will be a
non-linear extension of the Wheeler-DeWitt equation of canonical
gravity, due to the interaction term in the action, i.e. of the
inclusion of topology change; 2) the perturbative 3rd quantized vacuum
of the theory will be the \lq\lq no spacetime\rq\rq ~state, and not any
state with a semiclassical geometric interpretation in terms of a
smooth geometry, say a Minkowski state. 

We will see that the group field theory approach shares these general features, on top of proposing a new context to realize them \cite{3rd}.
      
Notice that in this formalism for spacetime topology change, the spatial topology of each single-universe sector remains fixed, and the superspace $\mathcal{S}$ itself remains a background structure for the \lq\lq 3rd quantized\rq\rq field theory.

Notice also that, if one was to attempt to reproduce this type of setting in terms of the variables used in LQG, two background structures of classical GR, i.e. the Lorentz group and the superspace $\mathcal{S}$ would be somehow unified, as superspace would have to be identified with the space of (generalized) Lorentz (or $\SU(2)$) connections on some given spatial topology. 

Something of this sort happens in group field theory, which can be seen as a sort of discrete non-local field theory on the space of geometries for {\it building blocks} of space. in turn given by group or Lie algebra variables. Before getting to the details of the GFT formalism, however, we want to motivate further the use of discrete structures and the associated non-locality.

\subsection{A finitary substitute for continuum spacetime?}
However good the idea of a path integral for gravity and its extension
to a third quantized formalism may be, there has been no definite
success in realizing them rigorously (beyond the minisuperspace-reduced contexts or semi-classical approximation).
One is tempted to identify the main reason for the difficulties encountered to be the use of
a {\it continuum} for describing spacetime, both at the topological
and at the geometrical level. One can advocate the use of {\it
  discrete structures} as a way to regularize and make computable the
above expressions, or to provide a more fundamental definition of the
theory, with the continuum description emerging only in a
continuum {\it approximation} of the corresponding discrete quantities, like hydrodynamics for large ensembles of many particles.  This
was in fact among the motivations for discrete approaches to quantum
gravity as matrix models, or dynamical triangulations or quantum Regge
calculus. And various arguments have been put
forward to support the point of view that discrete structures provide
a more {\it fundamental} description of spacetime. One possibility,
suggested by various approaches to quantum gravity such as string
theory or loop quantum gravity, is the existence of a fundamental length scale that sets a
lower bound for distances and thus makes the notion of a
continuum loose its physical meaning. Also, one can argue on both philosophical and mathematical
grounds \cite{isham} that the very notion of \lq\lq point\rq\rq can
correspond at most to an idealization of the nature of spacetime (see. e.g. non-commutative models of quantum
gravity \cite{majid, QGPhen}). Spacetime
points are indeed to be replaced, from this point of view, by small
but finite regions and a more fundamental model of spacetime should take these local regions as
basic building blocks. Also, the results of black hole thermodynamics
seem to suggest that there should be a finite number of fundamental
spacetime degrees of freedom associated to any region of spacetime,
the apparent continuum being the result of the microscopic (Planckian)
nature of them \cite{sorkinBH}. In other
words, a finitary topological space \cite{sorkin} would constitute a
better model of spacetime than a smooth manifold. These arguments also favor a simplicial description of
spacetime, with the simplices being a finitary
substitute of points. And these same arguments are reinforced by the results of LQG whose kinematical states are labelled by graphs \cite{thomas}.  

Here is where GFTs provide a discrete or finitary implementation of the 3rd quantization idea, which is also a more \lq local\rq one, in the following sense. In GFT the spatial manifold is to be thought of as a collection of (glued) building blocks, akin to a many-particle state, and the field theory should be defined on the space of possible geometries of each such building block. Spatial topology is also allowed to change arbitrarily, if the building blocks are allowed to combine arbitrarily by the theory. These building blocks can be depicted as either fundamental simplices or (pieces of) finite graphs, as we will see, and the space of geometries of such discrete structures is then necessarily finite dimensional. So the basic dynamical objects whose interactions produce all possible geometries and topologies of space are fundamental constituents of a universe, as opposed to the \lq global\rq framework of formal 3rd quantization, where the system interacting is the whole universe itself.

So the GFT formalism incorporates insights from other approaches (loop quantum gravity and simplicial gravity) also in answering a second natural question that comes to mind when suggesting a quantum field theory formalism for the microstructure of space: a QFT of which fundamental quanta? Again, we know that these cannot be gravitons. They have to be quanta {\it of space} itself, excitations around a vacuum that corresponds to the {\it absence of space} . 

To introduce how their dynamics is identified, and in the process motivate the peculiar {\it combinatorial non-locality} of GFT interactions (which is the price to pay for trying to maintain both \lq physical locality\rq and topology change),  we take a short detour and discuss first briefly their lower-dimensional predecessors: matrix models.

\subsection{A combinatorial non-locality: from point particles to extended combinatorial structures}

Consider a point particle in 0+1 dimensions, with action $S(X) = \frac{1}{2} X^2\,+\,\frac{\lambda}{3} X^3$. This action defines a trivial dynamics (for a trivial system), of course. What interests us here, however, is the combinatorial structure of its \lq\lq Feynman diagrams\rq\rq, i.e. the graphs that can be used as a convenient book-keeping tool in computing the corresponding partition function $Z = \int dX\, e^{- S_{\lambda}(X)}$ perturbatively in $\lambda$. These are simple 3-valent (because of the order of the \lq\lq interaction\rq\rq) graphs.

\begin{figure}[here]
  \includegraphics[width=6cm]{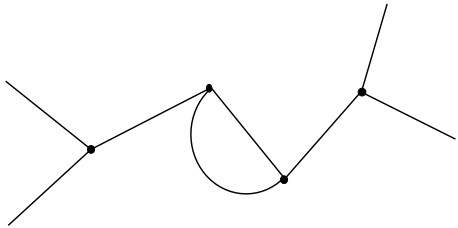}
  \caption{A Feynman graph for a point particle and the corresponding field theory}
\end{figure}

The fact that the Feynman diagrams of the theory are simple graphs like the above follows from 1) the {\it point-like} nature of the particle, and 2) the {\it locality} of the corresponding interaction, encoded in the identification of $X$ variables in the interaction part of the action.
A less trivial system would be given by a relativistic particle, for example, or, better, a system of such interacting particles\footnote{Notice also that the relativistic particle is often taken as a sort of General Relativity in $0+1$ dimensions, and while this analogy has strong limitations (like all analogies), it is indeed very useful to grasp an intuitive understanding of several issues that show up in the (quantum) gravity context \cite{teitelboim}.}. The combinatorial structure of the Feynman diagrams, now weighted by non-trivial amplitudes (convolutions of Feynman propagators for each interacting particle with identified initial/end points), will be the same as long as we do not allow for non-local interactions.

The same structure of diagrams is maintained, because the local nature of the interaction and the point-like nature of the corresponding quanta are maintained, also when moving to a field theory setting. Going from the above particle dynamics to the corresponding field theory (still dynamically rather trivial) means allowing for the creation and annihilation of particles, and then infinite number of degrees of freedom, because of the arbitrary number of  particles that can be involved in any interaction process. Assume for simplicity that the dynamics is governed by the (trivial) action: $S(\phi) = \frac{1}{2}\,\int dx\,\phi(x)^2\,+\,\frac{\lambda}{3}\,\int dx\,\phi(x)^3$. Now we have integrations over the position variables labeling the vertices, or, in momentum space, lines are {\it labeled} by momenta that sum to zero at vertices, and that are integrated over, reflecting a (potential) infinity of degrees of freedom in the theory. Still, the combinatorics of the diagrams is the same as in the corresponding particle case.

\subsubsection{Matrix models and discrete and continuum 2d gravity}

Now we move up in combinatorial dimension. Instead of point particles, let us consider 1-dimensional objects, that could be represented graphically by a line, with two end points. We label these two end points with two indices $i,j$, and we represent these fundamental objects of our theory by $N\times N$ {\it matrices} $M_{ij}$ (with $i,j=1,...,N$) \cite{mm}, i.e. arrays of real or complex numbers replacing the \lq\lq point\rq\rq variables $X$ \cite{mm}. For simplicity, assume these matrices to be hermitian.

We want to move up in dimension also in the corresponding Feynman diagrams, i.e. we want to have diagrams that correspond to 2-dimensional structures, instead of 1-dimensional graphs. In order to do so, we have to drop the assumption of {\it locality} (that  is, from a formal point of view, the simultaneous identification of all the arguments of the basic variables/fields appearing in the interaction). We define a simple action for $M$, given by \bes S(M) &=& \frac{1}{2} tr M^2 \, -\, \frac{g}{\sqrt{N}}\, tr M^3 \, = \frac{1}{2}  M^i {}_{j}  M^j {}_{i} \, - \frac{g}{\sqrt{N}}  M^i {}_{j}  M^j {}_{k} M^k {}_i \, = \nonumber \\ &=&\, \frac{1}{2}  M^i {}_{j}  K^{j l }{}_{k i} M^k {}_{l} \, - \frac{g}{\sqrt{N}}  M^i {}_{j}  M^m {}_{n} M^k {}_l \, V^{j n l}{}_{m k i} \nonumber \\  && \text{ with} \hspace{1cm} K^{j l }{}_{k i}  \,=\, \delta^j {}_k\, \delta^l {}_i \;\;\;\;\;\;\;\;   V^{j n l}{}_{m k i}\,=\,  \delta^j {}_m \,\delta^n {}_k\, \delta^l {}_i \;\;\;\;\;\;\;\left( K^{-1} \right)^{j l }{}_{k i}  \, =\, K^{j l }{}_{k i}  .\ees  

We have also identified in the above formula the propagator and vertex term that will give the building blocks of the corresponding Feynman amplitudes. These building block can be represented graphically as follows:

\vspace{1cm}
\hspace{3cm}
\begin{minipage}[c]{4cm}
$ \left( K^{-1}\right)^{j l }{}_{k i}$ \\   {\vspace{0.5cm}} \\  $ V^{j n l}{}_{m k i} $
\end{minipage}
\hspace{0.2cm}
\begin{minipage}[l]{5cm}
\includegraphics[width=3.5cm, height=2.5cm]{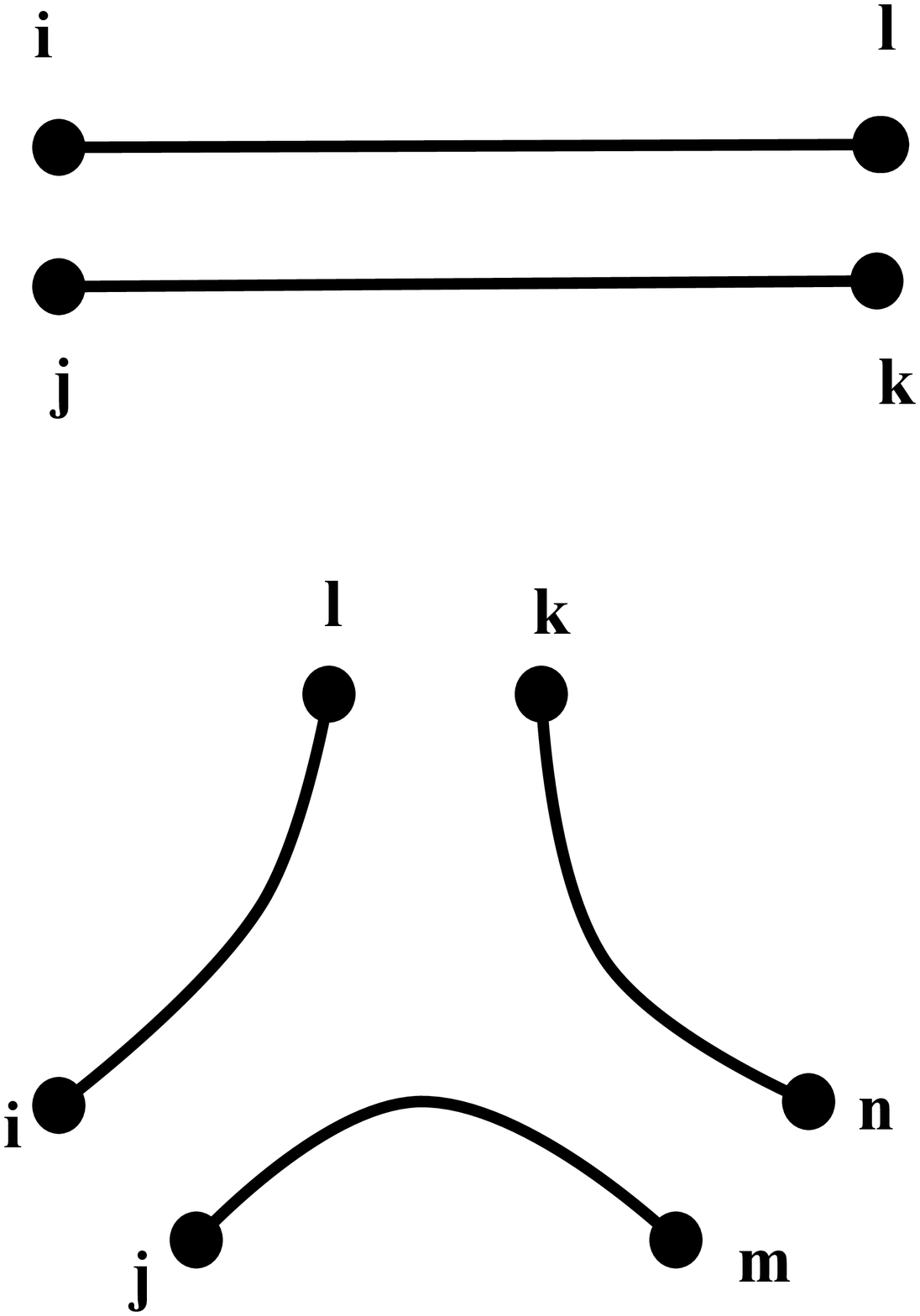}
\end{minipage}

\vspace{1cm}

The composition of these building blocks is performed by taking the trace over the indices $i,j,k$ in the kinetic and vertex term, and it represents the identification of the points labeled by the same indices, in the corresponding graphical representation. This identification of arguments is the higher-dimensional analog of the \lq locality\rq of usual field theory. Diagrams are then made of: (double) lines of propagation (made of two strands), non-local \lq\lq vertices\rq\rq of interaction (providing a re-routing of strands), faces (closed loops of strands) obtained after index contractions. The same combinatorics of indices (and thus of matrices)  can be given a dual simplicial representation as well:

\begin{figure}[here]
 \includegraphics[width=9cm, height=3cm]{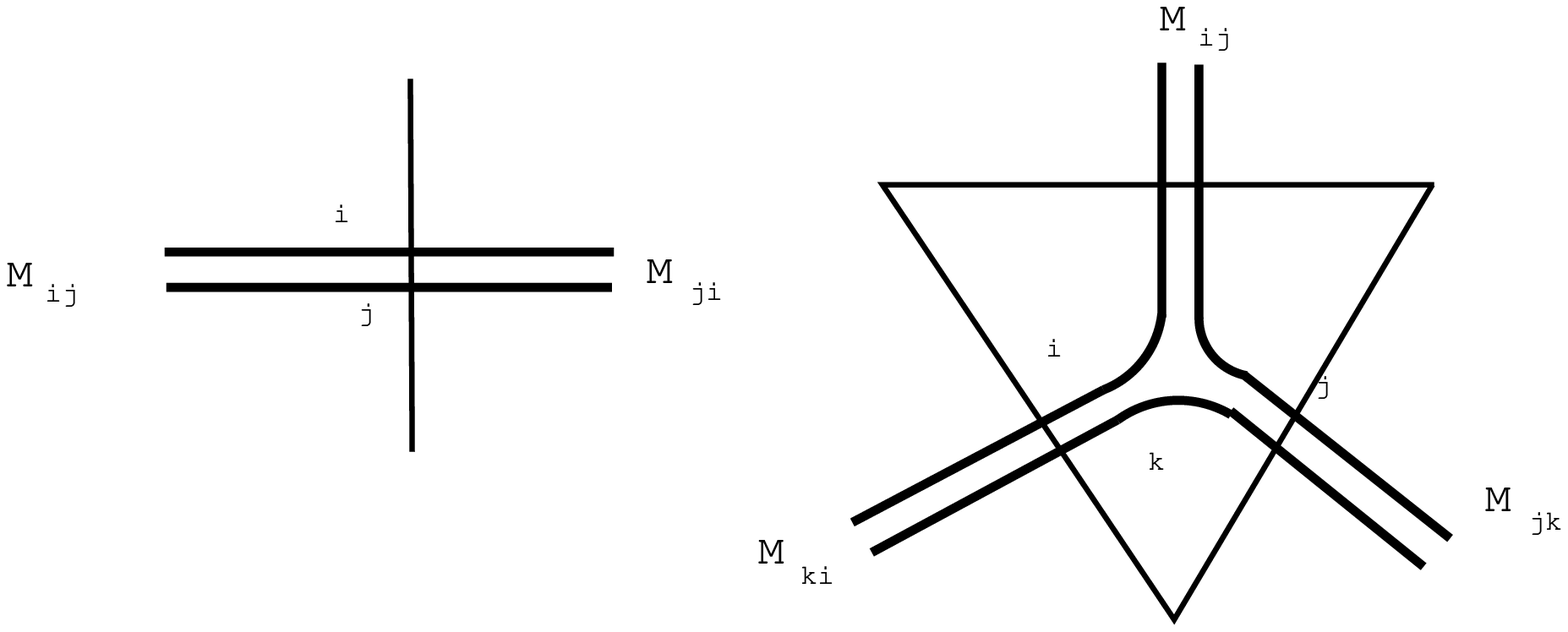}
\end{figure}

Therefore the Feynman diagrams used in evaluating the partition function $Z = \int \mathcal{D}M_{ij} \, e^{- S(M)}$ correspond to 2-dimensional simplicial complexes of arbitrary topology, since they are obtained by arbitrary gluing of triangles (the interaction vertices) along common edges (as dictated by the propagator). Thus a discrete 2d spacetime emerges as a virtual construction, encoding the possible interaction process of fundamentally discrete quanta of space (the edges/matrices).

\begin{figure}[here]
  \includegraphics[width=6cm]{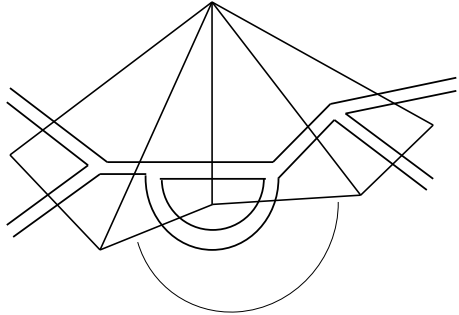}
  \caption{A (piece of) Feynman diagram for a matrix model, of which we give both direct and dual (simplicial) representation; the two parallel lines of propagation correspond to the two indices of the matrix; the extra line on the bottom indicates identification of the two edges of the triangles.}
\end{figure}    

We can easily compute the Feynman amplitudes for the model:

$$ Z\, =\, \sum_\Gamma\, \left( \frac{g}{\sqrt{N}}\right)^\frac{1}{2}\, Z_\Gamma \,=\,  \sum_\Gamma\, g^{V_\Gamma} N^{F_\Gamma - \frac{1}{2} V_\Gamma}$$
where $V_\Gamma$ is the number of vertices of the Feynman diagram, $F_\Gamma$ the number of faces of the latter, and $N$, again, the dimension of the matrices.
We can then use the identity: $ F_\Gamma - \frac{1}{2} V_\Gamma = v - \frac{1}{2} t = v - e +t = \chi = 2 - 2 h $ , where $v,e,t$ are the numbers of vertices, edges and triangles of the simplicial complex dual to the Feynman diagram, $\chi$ is its Euler characteristics and $h$ its genus. Thus,   
$$
Z \,=\, \sum_\Gamma\, g^{V_\Gamma}\, N^{\chi_\Gamma}\; .
$$ 
The same result can be obtained easily using the rescaling: $ M \rightarrow \frac{M}{\sqrt{N}}$, giving $ S(M) \Rightarrow S(M)\,=\, N \frac{1}{2} tr M^2\,-\, N \, g\, tr M^3$.
Now the Feynman expansion gives a factor $N$ for each closed loop of strands (vertex of the simplicial complex), $N^{-1}$ for each propagator (edge), $N g$ for each vertex (triangle), to give the result above using $\chi_\Gamma = V_\Gamma - L_\Gamma + F_\Gamma$.

We ask ourselves now what is the relation with gravity of this expansion and of the Feynman amplitudes of this theory. As long as we remain at the discrete level, we can of course only expect a relation with simplicial gravity. The general idea is that each Feynman amplitude will be associated to simplicial path integrals for gravity, discretized on the associated simplicial complex $\Delta$: $$ Z_\Gamma \simeq \int \mathcal{D} g_\Delta\; e^{-\, S_\Delta(g)} . $$

Now, consider continuum (Riemannian) 2d GR with cosmological constant $\Lambda$, on a 2d manifold $S$. 

Its action is $ \int_{S} d^2 x \,\, \sqrt{g}\, \left( - R (g) \, + \, \Lambda \right)\, =\, -\, 4\pi \, \chi\, +\, \Lambda\, A_S$, where $A_S$ is the area of the surface.
Consider then a simple discretization of the same.
Chop the surface $S$ into equilateral triangles of area $a$. The action will then be given by  $ \frac{1}{G}\int_{S} d^2 x \,\, \sqrt{g}\, \left( - R (g) \, + \, \Lambda \right)\, \rightarrow\, - \, \frac{4\pi}{G} \, \chi\, +\, \frac{\Lambda a}{G} \, t$
Using this discretization, and defining $g=e^{-\frac{\Lambda a}{G}}$ and $ N = e^{+\, \frac{4 \pi}{G}}$, from our matrix model we get:

$$
Z\,=\, \sum_\Gamma\, g^{V_\Gamma}\, N^{\chi}\, =\, \sum_{\Delta}\, e^{+\frac{4\pi}{G}\chi(\Delta) \, -\, \frac{a\Lambda}{G} \, t_\Delta}
$$
In other words, we obtain a (trivial) sum over histories of discrete GR on the given 2d complex, whose triviality is due to the fact that the only geometric variable associated to each surface is  its area, the rest being only a function of topology. In addition to this sum over geometries, from our matrix model we obtain a sum over all possible 2d complexes of all topologies. In other words, the matrix model defines a discrete 3rd quantization of GR in 2d!

Having been assured that this theory has at least some relation with gravity at the discrete level, we can take it seriously and try to see if we can control it and find out if it has nice continuum properties, i.e. if it defines a nice continuum theory from its sum over discrete surfaces. This can be articulated in three basic questions: 1) can we control in any way the sum over triangulations defined by the perturbative expansion of the matrix model? 2) does the model have a critical behaviour corresponding to a continuum limit for trivial topology? 3) can we go further and define a continuum sum over topologies?

The first question is whether we can control the sum over triangulations and over topologies at all. 
The answer is in the affirmative \cite{mm}: the reason is that the sum is manifestly governed by topological parameters and can be organized accordingly:
\bes
Z\,=\, \sum_\Delta\, g^{t_\Delta}\, N^{\chi(\Delta)}\,=\, \sum_\Delta\, g^{t_\Delta}\, N^{2 - 2 h}\, =\, \nonumber \\ =\,\sum_h N^{2- 2 h }\, Z_h(g)\,=\,N^2 \, Z_0(g)\,+\, Z_1(g)\,+ N^{-2}\, Z_2(g) +.....\nonumber
\ees

It is then apparent that, in the limit $N\rightarrow\infty$, only spherical (trivial topology, i.e. planar, i.e. of genus 0) contribute significantly to the sum.

The second question is whether, in this limit of trivial topology, one can also define a continuum limit and match the results of the continuum 2d gravity path integral. In order to study this continuum limit we expand $Z_0(g)$ in powers of $g$, and it can be shown that:

$$ Z_0(g) = \sum_V V^{\gamma -3} \left(\frac{g}{g_c}\right)^V \simeq_{V\rightarrow\infty} \left( g - g_c\right)^{2 - \gamma}$$

so that, in the limit of large number of triangles, and for the coupling constant approaching the critical value $g \rightarrow g_c$ (with critical exponent $\gamma=-\frac{1}{2}$), the free energy (logarithm of the partition function) can be shown to diverge. 
This is a signal of a phase transition. In order to identify this phase transition as a continuum limit we compute the expectation value for the area of a surface, assuming as above that each triangle contributes a constant area $a$:
 $ \la A \ra =\, a\, \la t_\Delta \ra = \,a\, \la V_\Gamma \ra\, =\, a \frac{\partial}{\partial g} \ln{ Z_0(g)} \, \simeq\, \frac{a}{g- g_c}$, for large $V$. We see that it also diverges when $g\rightarrow g_c$.
 
 Thus we can send the area of each triangle to zero by sending the edge lengths $ a$ to zero, $a \rightarrow 0$, and simultaneously the number of triangles to infinity: $t = V \rightarrow \infty$ (continuum limit), while tuning at the same time the coupling constant to its critical value $g \rightarrow g_c$, to get finite continuum macroscopic areas.

This defines a continuum limit of the matrix model. One can then show \cite{mm} that the results obtained in this limit match those obtained with a continuum 2d gravity path integral (when this can be computed explicitly), in the context of Liouville gravity \cite{mm}.

Let us now ask whether the 3rd quantized framework we have (in this 2d case) can also allow to understand and compute the contribution from non-trivial topologies in a continuum limit.
The key technique is the so-called  double-scaling limit \cite{mm}. One can first of all show that the contribution of each given topology of genus $h$ to the partition function is 
$$ Z_h(g) \simeq \sum_V V^{\frac{(\beta - 2) \chi}{2} - 1}  \left( \frac{g}{g_c}\right)^V \simeq \, f_h \, \left( g - g_c \right)^{\frac{(2 -\beta) \chi}{2}}\; ,$$ where the last approximation holds in the limit of many triangles (necessary for any continuum limit, and corresponding to a thermodynamic limit), and where $\beta$ is a constant that can be computed.

Define now $\kappa^{-1} = N \left( g - g_c \right)^{\frac{(2 -\beta)}{2}}$, so that we get:
$$
Z \, \simeq\, \sum_h\, \kappa^{2 h - 2} f_h\, =\, \kappa^{-2} f_0\, +\, f_1\, +\, \kappa^2 f_2\,+\,.......
$$

We can then take the combined limits $N\rightarrow\infty$ and $g \rightarrow g_c$, while holding $\kappa$ fixed. The results of this double limit is a continuum theory to which all topologies contribute, and again match results with continuum Liouville gravity \cite{mm}.

The area of matrix models is vast and rich of results, not only in the 2d quantum gravity context, but ranging from condensed matter physics to hot topics in mathematical physics, from string theory to mathematical biology. For all of this, we can only refer to the literature \cite{mm}.

\subsubsection{Tensor models}

Now we generalize further in combinatorial dimension: from 2d to 3d. This is achieved by going from 1d objects, represented graphically as edges, and mathematically by matrices, to 2d objects, represented graphically as triangles, and mathematically by tensors \cite{tensor}. Obviously, every 2d discrete structure (squares, polygons, etc) could be a possible choice, but any other 2d structure can be built out of triangles, so we stick to what looks like the most fundamental choice. This also means that, in the Feynman expansion of the corresponding theory, we expect to generalize from 2d simplicial complexes to 3d ones. We then define $N\times N\times N$ tensors $T_{ijk}$, with $ i,j,k = 1,.., N$  and an action for them given by

$$ S(T) = \frac{1}{2} tr T^2 \,-\, \lambda\, tr T^4 \,=\, \frac{1}{2} \sum_{i,j,k} T_{ijk} T_{kji} \, -\, \lambda\, \sum_{ijklmn} T_{ijk} T_{klm} T_{mjn} T_{nli}\; , $$
where the choice of combinatorics of tensors in the action, and of indices to be traced out is made so to represent, in the interaction term, four triangles (tensors) glued pairwise along common edges (common indices) to form a closed tetrahedron (3-simplex). Once more, the kinetic term dictates the gluing of two such tetrahedra along common triangles, by identification of the edges. From the action above we read out the kinetic and vertex term: 

$$ K_{ijk i' j'k'} = \delta_{ii'}\delta_{jj'}\delta_{kk'} = (K^{-1})_{ijk i'j'k'}$$

$$ V_{ii' jj' kk' ll' mm' nn'} = \delta_{i i'}\delta_{j j'}\delta_{k k'}\delta_{l l'}\delta_{m m'}\delta_{n n'}$$ 
which can be represented graphically as

\begin{figure}[here]
\includegraphics[width=9.5cm, height=8cm]{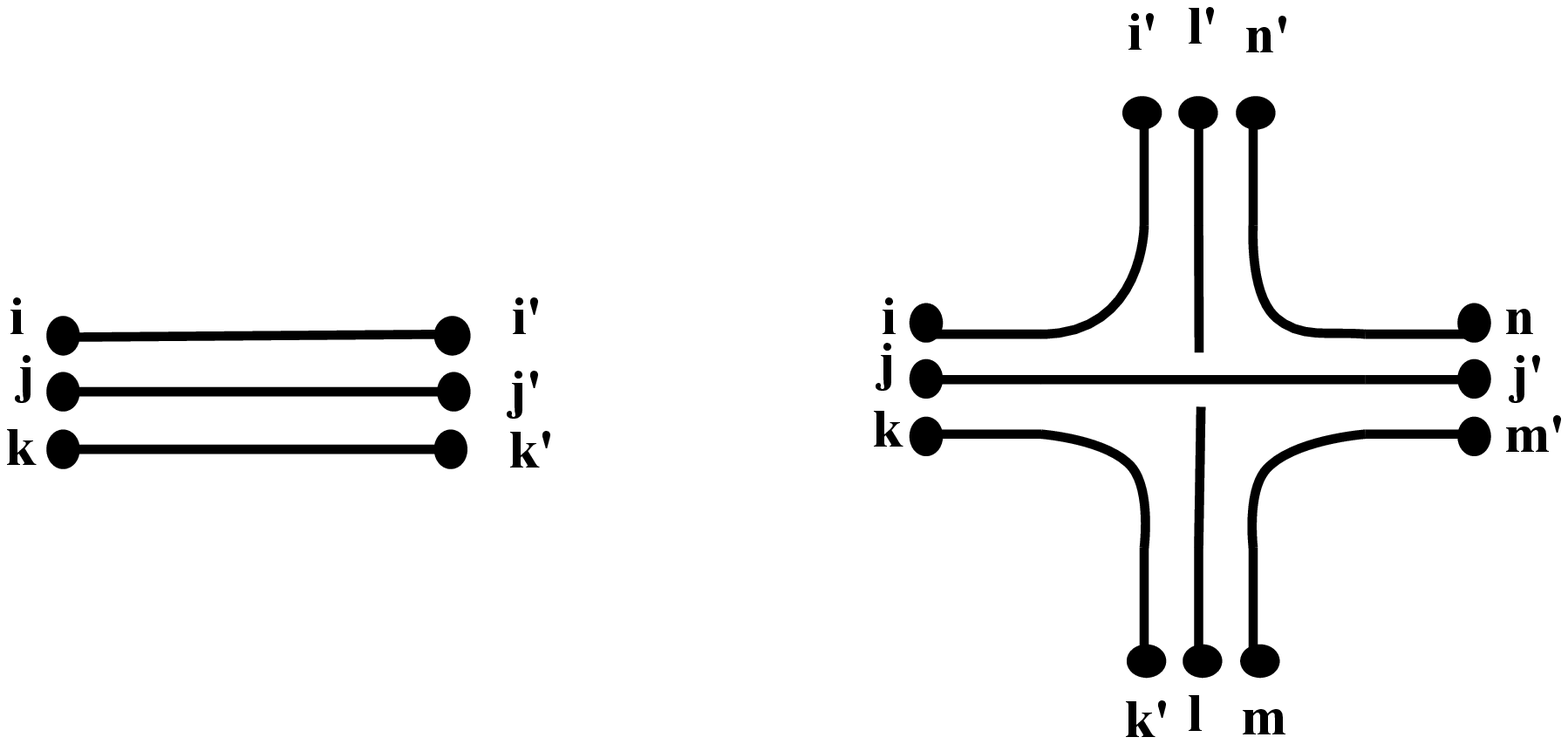}
\label{fig:TensorPropVertex}
\end{figure}

We can use them to expand perturbatively the  partition function:

$$Z\,=\, \int \mathcal{D}T \, e^{- S(T)}\, =\, \sum_{\Gamma} \lambda^{V_\Gamma}\, Z_\Gamma  \;\;.$$

Feynman diagrams are again obtained by contraction of vertices with propagators over internal indices

\begin{figure}[here]
\includegraphics[width=8cm, height=3.5cm]{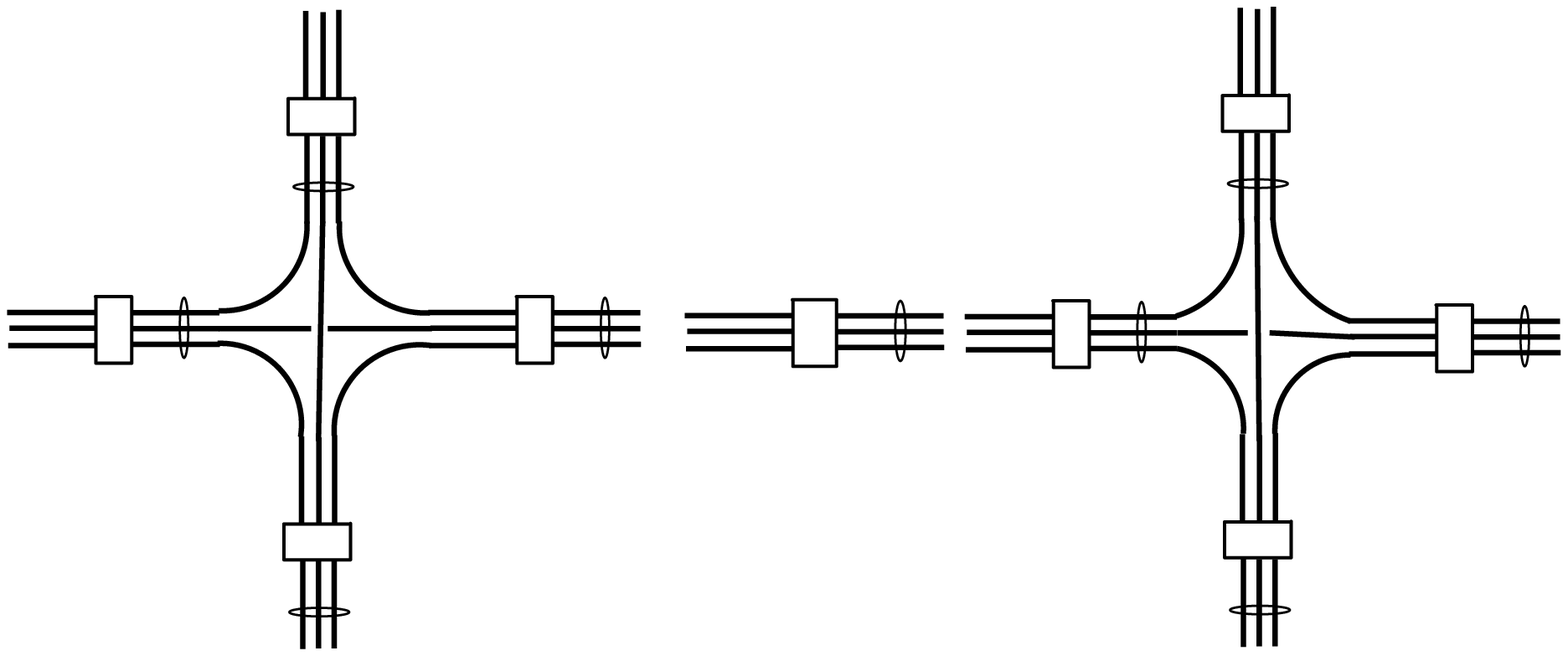}
\end{figure}

By construction, Feynman diagrams are again formed by vertices, lines and faces, but now they can form also \lq\lq bubbles\rq\rq (3-cells), and are dual to 3d simplicial complexes.

\begin{figure}[here]
\includegraphics[width=6cm, height=4.2cm]{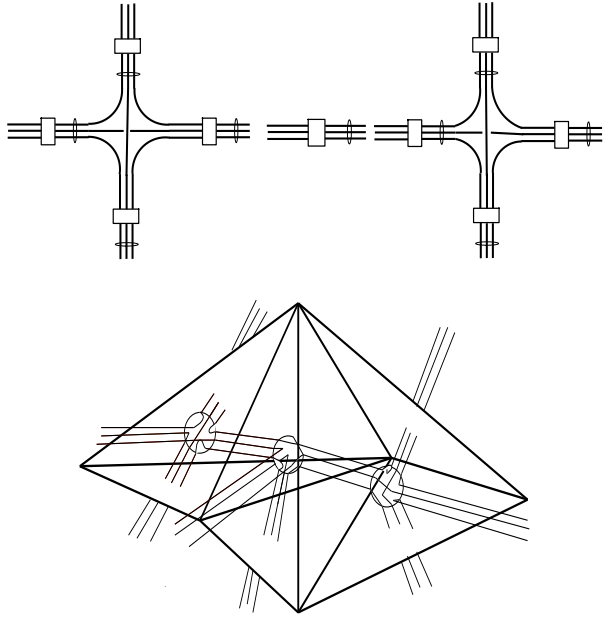}
 \caption{A (piece of) Feynman diagram for a tensor model, of which we give both direct and dual (simplicial) representation; the three parallel lines of propagation (dual to the three edges in the triangles of the simplicial complex) correspond to the three indices of the tensor.}
 \end{figure}
 
As a result $Z$ is defined as a sum over all 3d simplicial complexes including manifolds as well as more singular complexes (i.e. singular complexes such that the neighbourood of some points is not homeomorphic to a 3-Ball), because we impose a priori no restriction on the gluing procedure of vertices by means of propagators.

\

Do these models provide a good definition of 3d quantum gravity? The answer, unfortunately, is no, at least in this simple formulation of them. We will discuss how more refined versions of the same models improve the situation (see also the detailed and up-to-date review \cite{tensorReview}). None of the nice features of matrix models export to these tensor models. First of all there is no strong relation between the Feynman amplitudes of the above tensor model with 3d simplicial (classical or quantum) gravity. Even though 3d gravity is a topological theory, with no local propagating degrees of freedom and only locally flat solutions (in absence of a cosmological constant), it is still a highly non-trivial theory. The amplitudes of tensors models are too simple to capture either the flatness of geometry or the topological character of the quantum gravity partition function (as do Chern-Simons theory or discrete formulations like the Ponzano-Regge spin foam model). They do not have enough data in the amplitudes associated to each simplicial complex, or in boundary states. Second, there is no way to separate the contribution of manifolds from that of pseudo-manifolds, i.e. to suppress singular configurations or even to identify them clearly. Third, the expansion in sum over simplicial complexes cannot be organized in terms of topological invariants, and so there is no control over the topology of the diagrams summed over. Also the last two issues can be thought to be due to the lack of data and structure in the Feynman amplitudes of the theory. For example, one can consider (before going to full GFTs) slightly revised version of the same tensor models with fundamental variables $T^{abc}_{ijk}$, where the indices $(ijk)$ refer to the edges of a triangle and the indices $(abc)$ refer to its vertices \cite{tensor}.
 
 \
 
Clearly, the process of combinatorial generalization can be continued to higher tensor models whose Feynman diagrams will be higher simplicial complexes. It is clear, however, that the difficulties encountered with 3d tensor models are not going to be solved magically if we do not render the structure of the corresponding quantum amplitudes richer and more interesting.
In particular, just as we do in the 1d case, i.e. in the case of particles, we could generalize matrix and tensor models in the direction of adding degrees of freedom, i.e. defining corresponding field theories. 
In the process, the indices of the tensor models will be replaced by variables living in appropriate domain spaces, and sum over indices by appropriate sums or integrals over these domain spaces, while maintaining their combinatorial pairing in the action. This pairing will make the resulting field theories {\it combinatorially non-local}, as we anticipated group field theories are. In fact, once more, this is in many ways the defining properties of group field theories.

The prototype for a field theory of this non-local type and for a choice of domain space $D$ would be, for the 2d case: 

$$
S(\phi)\,=\,\frac{1}{2}\int_D [dg] \phi(g_1,g_2)\phi(g_2,g_1)\, -\,\frac{\lambda}{3!}\,\int [dg] \,\phi(g_1,g_2)\phi(g_2,g_3)\phi(g_3,g_1) , $$ with appropriate integrations over the domain space $D$, and same identification of field arguments as in the indices of the matrix model, while for the field-theoretic generalization of the tensor model for $T_{ijk}$ we get: $$S(\phi)\,=\,\frac{1}{2}\int_D [dg] \phi(g_1,g_2,g_3)\phi(g_3,g_2,g_1)\,-\,\frac{\lambda}{4!}\,\int [dg] \,\phi(g_1,g_2,g_3)\phi(g_3,g_4,g_5) \phi(g_5,g_6,g_1)\phi(g_6,g_4,g_2) . $$

The definition of good group field theory models for quantum gravity, of course, would require a careful choice of domain space and of classical action (kinetic and vertex functions). Here is where the input from other approaches is crucial, and where their characteristic structures are incorporated into the GFT formalism. In particular, the domain space of GFTs for quantum gravity in 3 and higher dimensions is chosen to be either a group manifold (from which the name of the formalism and the choice of notation in the examples of actions given above) or, more recently, the corresponding Lie algebra. This allows also a re-writing of the GFT action (and amplitudes) in terms of group representations, as we will see. We now motivate this choice for the domain space.

\subsection{Ingredients from Loop Quantum Gravity and simplicial topological theories}

Let us first look at Loop Quantum Gravity.  This can be understood as an example of a theory initially defined in the continuum, but that ends up identifying, after quantization, discrete pre-geometric structures as a more fundamental set of building block for such continuum. These discrete, pre-geometric building blocks are then also used as the basic ingredients of GFTs.
As we mentioned already, after passing to connection variables valued in the Lorentz algebra $\so(3,1)$ and suitable frame fixing of the tetrad variables to $\su(2)$ tetrads, GR becomes similar to a gauge theory for the gauge group $\SU(2)$, with classical phase space given by a connection 1-form and a conjugate electric field (triad 1-form). The difference in the Loop Quantum Gravity quantization is that one takes particular care of the diffeomorphism invariance that characterizes the theory \cite{thomas}. Due to the gauge fixing, therefore, what would be initially an $\SO(3,1)$ gauge theory is reduced to an $\SU(2)$ one, with a similar reduction of the conjugate variables to the connection. Let us see briefly (and simplifying considerably the construction) how the kinematical phase space of the theory is defined.

If we assume the $\SU(2)$--bundle to be trivial then every $\SU(2)$--connection can be seen as an $\su(2)$--valued one--form on the three dimensional base manifold $\Sigma$. Being a one--form $A$ can naturally (i.e. without referring to a background metric) be integrated along one--dimensional submanifolds of $\Sigma$, namely along embedded edges $e$:
\be
\int\limits_{e} A := \int\limits_e A_a^j \tau_j dx^a\; .
\ee
The conjugate variable to the connection, the triad
$E$, being an $\su(2)$--valued vector density it has a natural associated 2--form $(*E)^j_{ab}(\sigma) = \epsilon_{abc}E^{cj}(\sigma)$. This 2--form can be integrated along submanifolds of codimension one, namely analytic 2--surfaces $S$ (by means of appropriate parallel transports):
$
E(S,f) := \int\limits_{S}\,(*E)^j \,f_j \;\; .
$
Here $f$ is a smearing functions with values in $\su(2)^*$, the topological dual of $\su(2)$ and $f_j$ are its components in a local basis.\\
To get quantities with a nicer behaviour under $\SU(2)$--transformations one introduces the holonomy
\be
h_e(A) := \cP \exp\left[-\int\limits_{e} A\right]\; ,
\ee
where $\cP$ denotes the path--ordering, which are of course $\SU(2)$ group elements.  For a graph $\gamma$ with $|\gamma|$ edges the holonomy assigns an element $h_e(A) \in \SU(2)$ to every edge. 

The rational for the above is that one is free to choose any parametrization of the classical phase space, provided any point in it can be identified by the coordinates chosen. In particular, one can specify the connection field at every point in the spatial manifold by providing its holonomies along all the paths embedded in the same manifold.

One then defines the space 
\be
\Cyl^\gamma = \{ C^\gamma : \cA \rightarrow \C; A \mapsto C^\gamma(A) \, | \, C^\gamma(A) := c(h_{e_1}(A), h_{e_2}(A), \dots h_{e_{|\gamma|}}(A)) \}   
\ee
of functions called {\it cylindrical with respect to $\gamma$}, i.e. that depend on $A$ only through the holonomies $h_{e_i}(A)$ and $c: \SU(2)^{|\gamma|} \rightarrow \C$ is a continuous complex valued function.\\
The configuration space of the theory is defined to be the space $\Cyl$ (without reference to a specific graph $\gamma$) as the space of functions that are cylindrical with respect to {\it some} graph.
One can then show \cite{thomas} that the fluxes $E(S,f)$ are vector fields on $\Cyl$.

The classical Poisson algebra between cylindrical functions (including single holonomies) can be computed in full generality, i.e. for arbitrary graphs $\gamma$ and surfaces $S$. The basic feature is that holonomies Poisson commute, fluxes and holonomies have non-zero commutators depending on the intersection points between graphs $\gamma$ to which holonomies are associated and surfaces $S$ on which the fluxes are smeared, while fluxes associated to any two surfaces (including coincident ones) do not commute. The non-commutativity of the fluxes even at the classical level is crucial for what follows. The algebra is in general rather complicated, depending on the specific surfaces chosen and their topological relations, to the point that the general commutator between fluxes is not known \cite{acz, ioaristidebiancajohannes}. It simplifies considerably if one considers only \lq\lq elementary\rq\rq surfaces $S_e$ with single intersection points with the edges $e$ of the graph $\gamma$ (this way, one has to label each state by both a graph and a set of dual surfaces to its edges). The fluxes are defined as:

\be
E^e_i= \text{tr} [\tau_i   \int_{S_e} \text{Ad}(h_{e,x})  E(x) ]
\ee
where $h_{e,x}$ is the holonomy along the path from the starting point of the edge $e$ to the point $x$ on the surface $S_e$.

In fact, considering a single link $e$ and a single elementary dual surface, also labelled by $e$, the phase space of the theory reduces to the cotangent bundle of $\SU(2)$, $T^*\SU(2)$, with fundamental variables being the holonomy $h_e$ along the link $e$ and the dual triad $E_e^i$ and fundamental Poisson brackets being 

\bes
\{h_e, h_e'\}_\gamma &=& 0 \nn\\
\{E^e_i, h_e\} &=& \delta^e_{e'} \frac{\tau_i}{2} h_e \nn\\
 \{E^e_i,E^{e'}_j\}&=&-\delta^{ee'} \epsilon_{ijk} E^e_k  \q ,
\ees  

with $\tau_i$ the generators of the $\su(2)$ Lie algebra. The last bracket among fluxes is clearly the $\su(2)$ bracket.

 \ 

Now, interestingly, this is also the kinematical phase space of discrete topological BF theory \cite{biancajimmy, zapata}, with $\SU(2)$ as gauge group, and thus it sets the basic kinematical stage for the simplicial path integral quantization of that theory. This fact is going to be crucial in the following, when we will discuss the GFT  model for 1st order 3d gravity (which in 3d coincides with BF theory). Also simplicial BF theory, then, is based on a configuration space given by cylindrical functions for the gauge group; the difference with LQG (or with the covariant version of it) is therefore only in the constraints that one imposes on such functions to implement the dynamics.

\subsection{Some mathematical tools}

Given this phase space, and considering for now a single edge of any graph labelling the states, there is a natural Fourier transform that can be introduced and used to map between configuration and \lq\lq momentum\rq\rq space. This is the so-called non-commutative \lq\lq group Fourier transform\rq\rq, introduced first in \cite{PR3}, and whose properties have been analysed in \cite{NCFourier}. Introduced first in the context of spin foam models, it has been recently used extensively in the GFT context \cite{ioaristide, ioaristideflorian}, and then applied also in LQG \cite{ioaristidebiancajohannes}, to give a quantization of the theory in terms of metric (flux) variables. We introduce it briefly here, and then we will show its role in a GFT model for 3d gravity.

The group Fourier transform is based on the definition of plane waves
\be
\e: \SU(2)\times \su(2) \rightarrow \C; (g,x) \mapsto \e_g(x):= e^{i\Tr(xg)}
\ee 
where $x := \vec{x}\cdot\vec{\tau}$ in a basis $\tau^i$ of $\su(2)$ and the trace is taken in the fundamental representation. One can always identify $\su(2)$ with $\R^3$ as a vector space and thus the $e_g$ can be interpreted as elements of $C(\R^3)$. Denote the closure of the linear span of these elements as $C_{\kappa}(\R^3)$. We can then introduce a non--commutative product on the algebra of functions $C_{\kappa}(\R^3)$, starting from plane waves
\be \label{star_product}
\star \,: C_{\kappa}(\R^3) \times C_{\kappa}(\R^3) \rightarrow C_{\kappa}(\R^3); (\e_{g_1} ,\e_{g_2}) \mapsto \e_{g_1}\star \e_{g_2} := \e_{g_1 g_2} \quad , 
\ee
and extending it to all of $C_{\kappa}(\R^3)$ by linearity. $C_{\kappa}(\R^3)$ endowed with this non--commutative product is refered to as $C_{\star,\kappa}(\R^3)$.\\
Using these plane waves and the Haar measure $dg$ on $\SU(2)$ one defines the group Fourier transform as
\be
F: C(\SU(2)) \rightarrow C_{\star, \kappa}(\R^3); f(g) \mapsto \tilde{f}(x) := \int dg \e_g(x) f(g) \quad .
\ee
As defined above, the Fourier transform is not invertible, but can be made so by suitable modification of the plane waves (which basically amounts to the multiplication by a polarization vector keeping track of the hemisphere of $\SU(2)$ on which the group element $g$ belongs), as shown in \cite{NCFourier}. An alternative is to limit oneself to work with functions on $\SO(3)$ and define plane waves with $|g| = \mbox{sign}(\Tr g) g$ in place of $g$ in the above definition \cite{PR3,NCFourier}; we choose this second modification in the following. However, at times we do not indicate it explicitly in order to keep the notation simple and keep focusing on the main ideas. Therefore, using the star product  (\ref{star_product}),  and these (modified) plane waves, we get a bijection
\bes
F: & &  C(\SU(2)) \rightarrow C_{\star, \kappa}(\R^3); f(g) \mapsto f(x) := \int dg \e_g(x) f(g)\\
F^{-1}: & & C_{\star, \kappa}(\R^3) \rightarrow C(\SU(2)); f(x) \mapsto f(g) := \int dx (\e_g \star f)(x) \quad . 
\ees
which maps functions on $\SU(2)$ onto functions living on $\R^3$ (equivalently, on the $\su(2)$ Lie algebra). The non--commutativity of $\SU(2)$ (and of its Lie algebra) is taken into account via the star product (\ref{star_product}). The extension of this group Fourier transform to functions of arbitrary numbers of group or Lie algebra  elements \cite{ioaristide,ioaristidebiancajohannes}, and to arbitrary groups, will be a crucial element of the GFT formalism, as it provides a duality of representations for the GFT field as a function of group elements or of Lie algebra elements. Clearly, it also plays an important role in dealing with simplicial BF theory, as we will see. In fact, the Feynman amplitudes of the GFT we will present, in the Lie algebra basis, will be given by the simplicial path integral for 3d $\SU(2)$ BF theory (or 3d gravity). For an application of these tools to a simpler example, where the correctness of the result can however be explicitly checked, see \cite{iomatti}.

\

For arbitrary (square integrable) functions on groups, another type of generalization of the usual Fourier transform on $\mathbb{R}^d$ is available. For compact groups (like the rotation group in any dimension), this is given by the Peter-Weyl decomposition of the function itself into irreducible representations. For $\SU(2)$ it gives:

$$
f(g)\,=\,\sum_j \,(2j+1) f_{mn}^j \, D^j_{mn}(g)\;\;,
$$
where (with repeated indices summed over) $j\in \mathbb{N}/2$ labels the irreducible representations of $\SU(2)$, $m,n$ are indices labelling a basis in the vector space of the representation $j$ (and its dual space), and $D^j_{mn}$ are the Wigner representation matrices, here playing the role of plane waves. This maps functions of group elements to functions of representation labels, and can be inverted to give:

$$
f^j_{mn}\,=\, \int dg\, f(g) \left(D^j_{mn}(g)\right)^*\;\;.
$$

The extension of this decomposition to cylindrical functions is the basis of the spin network representation of Loop Quantum Gravity, in which basis states correspond to graphs $\gamma$, whose links are labelled by irreducible representations of $\SU(2)$, and vertices by intertwiners of the group, after imposition of gauge invariance. Similarly, the spin network basis provides an alternative representation for the GFT field and for the corresponding states. In this representation, as we will see, the field is represented not as a fundamental simplex but as a single spin network vertex, the building block for the construction of arbitrary spin networks.

\subsection{The spin foam idea}
 A covariant
path integral quantization of a theory based on spin networks will
have as histories a higher-dimensional analogue of them: a spin foam
\cite{review,alex,thesis}, i.e. a 2-complex (collection of faces
bounded by links joining at vertices) with representations of the
Lorentz (or $\SU(2)$) group attached to its faces, in such a way that any slice or
any boundary of it, corresponding to a spatial hypersurface, will be
given by a spin network. Spin foam models \cite{review, alex,
  thesis} are intended to give a path integral quantization of gravity based on
these purely algebraic and combinatorial structures.  

In most of the current models the combinatorial structure of the spin
foam is restricted to be topologically dual to a simplicial complex of
appropriate dimension, so that to each spin foam 2-complex it
corresponds a simplicial spacetime, with the representations attached
to the 2-complex providing quantum geometric information to the simplicial
complex. The models are then defined
by an assignment of a quantum probability amplitude (here factorised
in terms of face, edge, and vertex contributions) to each spin foam $\sigma$
summed over, depending on the representations $\rho$ labeling it, also
being summed over:
$$
Z=\sum_{\sigma}\,w(\sigma)\sum_{\{\rho\}}\prod_{f}A_{f}(\rho_f)\prod_{e}A_{e}(\rho_{f\mid
  e})\prod_{v}A_{v}(\rho_{f\mid v})\;\; .$$

\begin{figure}[t] 
\setlength{\unitlength}{1cm}
\begin{minipage}[b]{5.278cm}
\includegraphics[width=4.5cm, height=3.5cm]{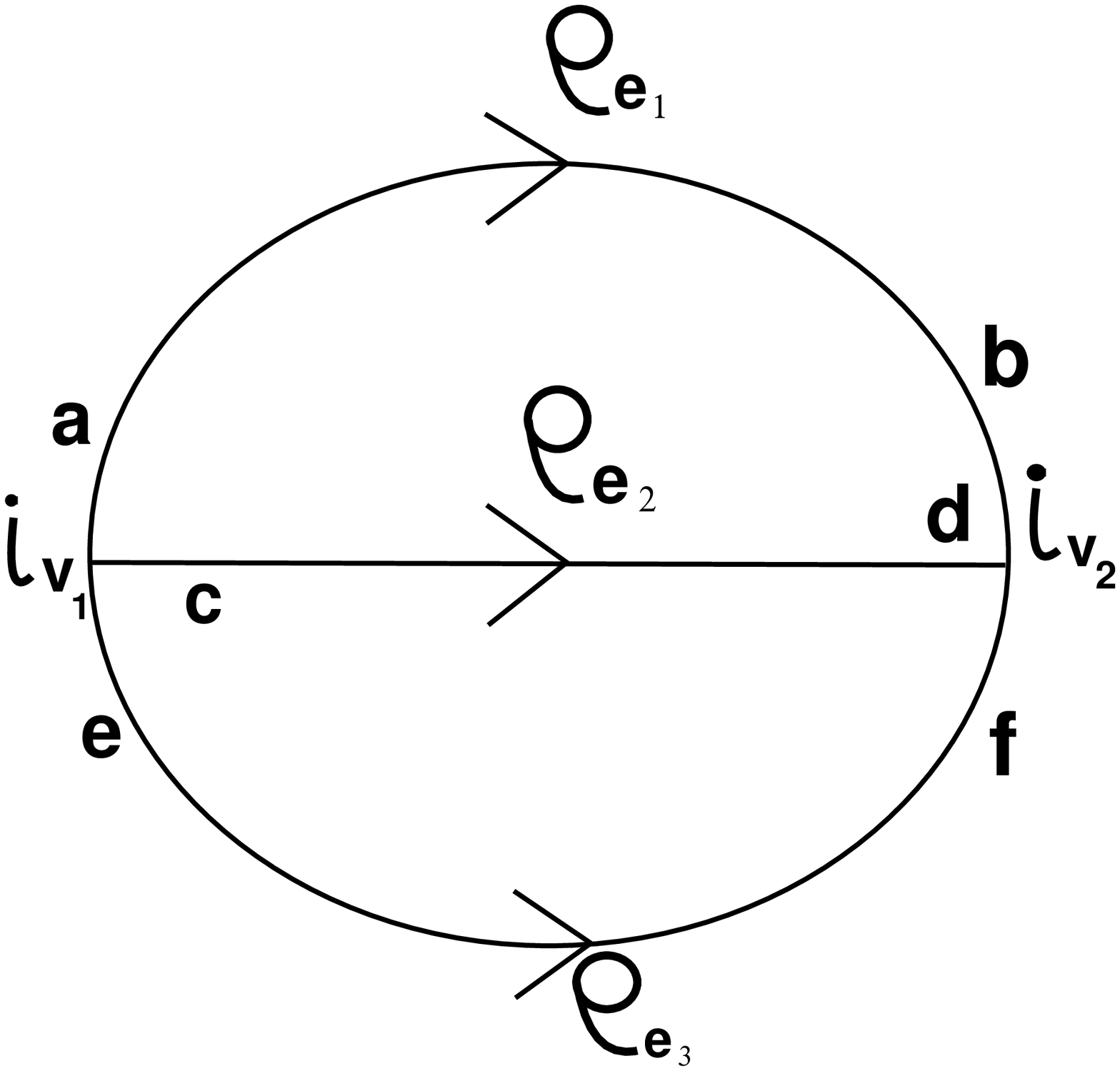}
\caption{A spin network. Representations are here labelled by $\rho$'s and intertwiners by $\iota$'s.}
\end{minipage}
\hspace{3cm}
\begin{minipage}[b]{5.357cm}
\includegraphics[width=5.5cm, height=6cm]{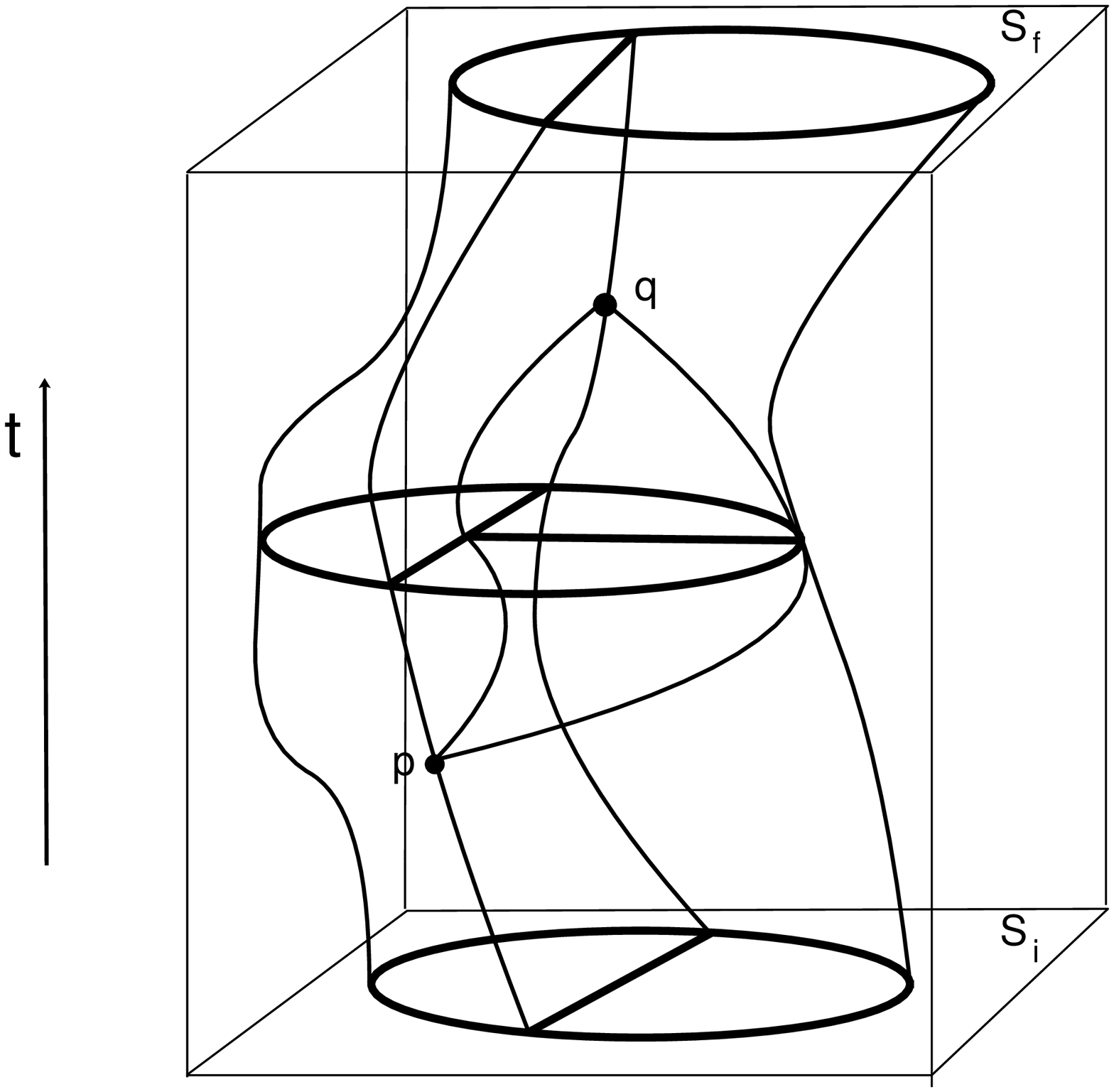}
\caption{\label{fig:spinfoam} A spin foam with two vertices representing a sequence (in a fictitious embedding spacetime) of two elementary transitions between spin networks.}
\end{minipage}
\end{figure}

One then has an implementation of a sum-over-histories for gravity in a
purely combinatorial-algebraic context. We will show that this spin foam representation is characteristic of the GFT Feynman amplitudes, and that it is dual to the representation of the same in the form of a simplicial path integral, a duality stemming from the above duality of representations for functions on group manifolds.
A multitude of results have been already obtained in the spin foam approach, for which we refer to \cite{review,alex,thesis}.

\

Let us summarize ingredients and aims of the GFT formalism, before entering the details of one specific GFT model. We want to define a quantum field theory of fundamental building blocks of quantum space, whose combinations can build up arbitrary spatial topological manifolds, and whose dynamics and interaction processes generate arbitrary spacetime topologies, thanks to a peculiar non-locality of field pairing in the interaction term of the GFT classical action. They are thus a sort of discrete or finitary, and local, 3rd quantization of gravity. In this, they represent a generalization of matrix models to arbitrary dimension. The arguments of the GFT field are given either as group elements or as Lie algebra elements or as group representations. In this, GFTs incorporate the kinematical description of geometry of Loop Quantum gravity and discrete topological BF theories. The Feynman amplitudes of the theory are, as we will see, given by simplicial path integrals or, equivalently, by spin foam models.

\section{Dynamics of 2d quantum space as a group field theory}
We present here in some detail the construction and perturbative analysis of the GFT model for 3d Riemannian gravity, first introduced, in the group picture, by Boulatov \cite{boulatov}. The expansion in group representations of its Feynman amplitudes gives the Ponzano-Regge spin foam model \cite{PR1}. It was recently reformulated in terms of non-commutative Lie algebra variables in \cite{ioaristide}. 

\subsection{The kinematics of quantum 2d space in GFT: quantum simplices and spin networks}

Consider a triangle in $\mathbb{R}^3$. We consider its (2nd quantized) kinematics to be encoded in the GFT field $\varphi$. We work here with real fields for simplicity only. The GFT field can be understood as living on the space of possible geometries for the triangle itself, or on the corresponding conjugate space. We parametrize the possible geometries for the triangle in terms of three $\su(2)$ Lie algebra elements attached to its three edges, and to be thought of as fundamental triad variables obtained by discretization of continuum triad fields along the edges of the same triangle, in line with both the LWG and discrete BF constructions. 

The field is then a function 

$$
\varphi \,:\, (x_1,x_2,x_3) \in \su(2)^3 \longrightarrow \varphi(x_1,x_2,x_3) \in \mathbb{R} 
$$ 

We do not assume any symmetry of the field under permutation of the arguments. Different choices are possible, as the field can be taken to be in any representation of the permutation group acting on its arguments. The choice of the representation made will influence the type of combinatorial complexes generated as Feynman diagrams of the theory \cite{iogft}. This will not concern us here. We will also see that a simple modification of the construction, defining \lq colored GFTs\rq \cite{tensorReview} can be used to make this choice somewhat irrelevant from the point of view of the same combinatorics.
 
Using the non-commutative group Fourier transform \cite{PR3,NCFourier} introduced earlier, the same GFT field can be recast as a function of $\SU(2)$ group elements. 
 
Recapitulating and detailing a bit more its definition, this transform stems from the definition of plane waves $\e_{g}(x) \!=\! e^{i \vec{p}_g\cdot \vec{x}}$ as functions on $\g \sim \R^3$, depending on a choice of coordinates  $\vec{p}_g = \Tr (g \vec{\tau})$ on the group manifold, where $g \! := \! \mbox{sign} (\Tr (g))g$, $\vec{\tau}$ are $i$ times the Pauli matrices and `$\Tr$' is the trace in the fundamental representation \footnote{For more details see \cite{NCFourier, ioaristide, ioaristidebiancajohannes}}.
For $x \!=\! \vec{x} \cdot \vec{\tau}$ and 
$g \!=\! e^{\theta \vec{n} \cdot \vec{\tau}}$, we thus have 
\[
\e_g(x)  =  e^{i \Tr x g} = e^{-2 i\sin \theta \vec{n} \cdot \vec{x}}
\]
and the $\star$-product is the one defined in the previous section. 

As said, Fourier transform and $\star$-product extend straightforwardly to functions of several variables like the GFT field (and generic cylindrical functions) so that
\[
\varphi(x_1, x_2, x_3) =\int [\extd g]^3\, \vphi(g_1, g_2, g_3) \, \e_{g_1}(x_1) \e_{g_2}(x_2) \e_{g_3}(x_3) 
\]
so that the GFT field can also be seen as a function of three group elements, thought of as parallel transports of the gravity connection along fundamental links dual to the edges of the triangle represented by $\varphi$, and intersecting them only at a single point.

In order to define a geometric triangle, the vectors (Lie algebra elements) associated to its edges cannot be independent. Indeed, they have to \lq close\rq ~to form a triangle, i.e. they have to sum to zero. We thus impose a constraint on the field

\[
\vphi = C \star \vphi, \quad C(x_1, x_2, x_3) = \delta_0(x_1 \!+\!x_2\!+\!x_3) 
\] 
by means of the projector $C$, where $\delta_0$ is the element $x=0$ of the family of functions:
\[ 
\delta_x(y)  := \int \extd g \, \e_{g^\inv}(x)  \e_g(y)
\]
These play the role of Dirac distributions in the non-commutative setting, in the sense that\footnote{Although behaving like a proper delta distribution with respect to the star product, under integration, this is a regular function when seen as a function on $\mathbb{R}^3$. In fact,  seen as a function of $\R^3$, $\delta_0$ is the regular function peaked on $x \!=\! 0$ given by $\delta_0(x) \propto J_1(|x|)/|x|$, where $J_1$ is the 1st Bessel function \cite{NCFourier}.}
\[
\int \extd^3 y \, (\delta_x \star f)(y) = \int \extd^3 y\, (f \star \delta_x)(y) = f(x) 
\]
One can also show that $(\delta_x \star f)(y) = (\delta_y \star f)(x)$. 


In terms of the dual field $\varphi(g_1,g_2,g_3)$, the same closure constraint implies invariance under the diagonal (left) action of the group $\SU(2)$ on the three group arguments, imposed by projection $P$:

\be
\varphi(g_1,g_2,g_3) = P \varphi(g_1,g_2,g_3) = \int_{\SU(2)} dh\, \phi( h g_1, h g_2, h g_3) \label{invariance}
\ee

Because of this gauge invariance, which is in fact imposed in the same way as the Gauss constraint is imposed on cylindrical functions in LQG \cite{ioaristidebiancajohannes}, the field can be best depicted graphically as a 3-valent vertex with three links, dual to the three edges of the closed triangle.

\begin{figure}[here]
\includegraphics[width=13cm, height=3cm]{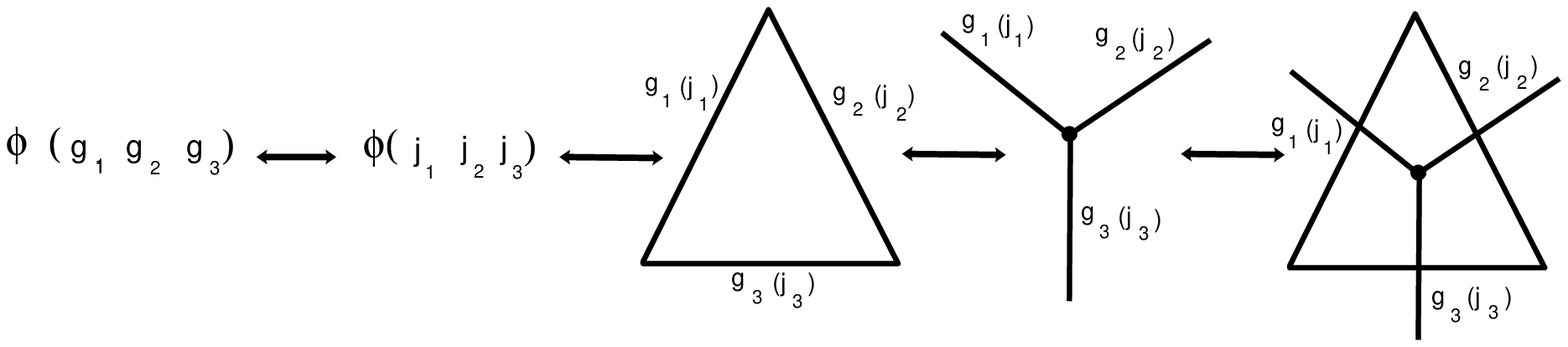}

\end{figure}

This object, both mathematically and graphically, will be the GFT building block of our quantum space.

One obtains another representation of the GFT field by means of Peter-Weyl decomposition into irreducible representations, in the same way as one obtains the spin network expansion of generic cylindrical functions in LQG.

The invariant field decomposes in $\SU(2)$ representations as:

\begin{equation}
\varphi(g_1,g_2,g_3) = \sum_{j_1,j_2,j_3} \, \varphi^{j_1j_2j_3}_{m_1m_2m_3}\, D^{j_1}_{m_1n_1}(g_1)D^{j_2}_{m_2n_2}(g_2)D^{j_3}_{m_3n_3}(g_3)\,C^{j_1 j_2 j_3}_{n_1 n_2 n_3}
\label{peterweyl}
\end{equation}

where 
$ C^{j_1 j_2 j_3}_{n_1 n_2 n_3}$ is the Wigner invariant 3-tensor, the 3j-symbol.

\

This expansion in irreducible group representations can be understood as a representation of the GFT field in terms of the quantum numbers associated to the quantized geometry of the triangle it represents. The $j_i$'s label eigenvalues of the length operators corresponding to its edges, while the angular momentum indices encode directional degrees of freedom. This is confirmed by  canonical analysis of the quantum geometry of the triangle, as well as by geometric quantization methods \cite{barbieri, baezbarrett, eteracarlo}.  

\

Multiple fields can be convoluted (in the group or Lie algebra picture) or traced (in the representation picture) with respect to some common argument. This represents the gluing of multiple triangles along common edges, and thus the formation of more complex simplicial structures, or, dually, of more complicated graphs

\begin{figure}[here]
\includegraphics[width=13cm, height=3cm]{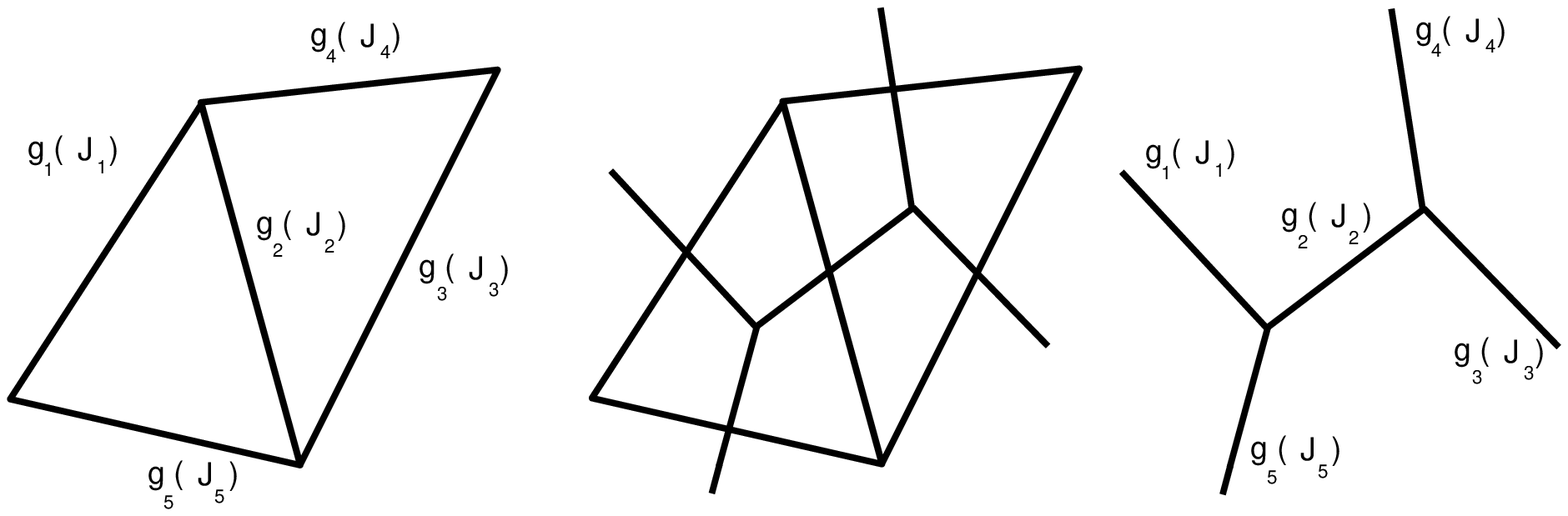}
\end{figure}

The corresponding field configurations represent thus extended chunks of quantum space, or many-GFT-particle states. A generic polynomial GFT observables would be given by this type of construction, and thus be associated with a particular quantum space.
This includes, of course, any open configuration, in which the arguments of the involved GFT fields are not all convoluted or contracted, representing a quantum space (not necessarily connected) with boundary. 

We would like to point out that, as in the case of tensor models, combinatorial generalizations can be considered, since there is no a priori restriction on how many arguments a GFT field can have. Once closure constraint or gauge invariance has been imposed on such generalized field with $n$ arguments, it can be taken to represent a general n-polygon (dual to an n-valent vertex), and glued to others in order give a polygonized quantum space in the same way, as we have outlined for triangles.

\subsection{Classical (3rd quantized) dynamics of 2d space in GFT}
We now define a classical dynamics for the GFT field we have introduced. The prescription for the interaction term, as in tensor models, is simple: four geometric triangles should be glued to one another, along common edges, to form a 3-dimensional geometric tetrahedron.  The kinetic term should encode the gluing of two tetrahedra along common triangles, by identification of their edge variables. There is no other dynamical requirement at this stage.

Thus we take four fields and identify pairwise their edge Lie algebra elements (triad edge vectors), with the combinatorial pattern of the edges of a tetrahedron; this gives the potential term in the action, weighted by an arbitrary coupling constant. And we take two more fields and identify their arguments; this defines the kinetic term in the action. Naming $\varphi_{123}=\varphi(x_1,x_2,x_3)$, the combinatorial structure of the action is then
\[
S =\! \frac{1}{2}\int [\extd x]^3 \,\varphi_{123} \star \varphi_{123} -  
\frac{\lambda}{4!} \! \int [\extd x]^6 \,\varphi_{123} \star \varphi_{345} \star \varphi_{526} \star \varphi_{641}
\]
where it is understood that $\star$-products relate repeated indices as $\varphi_i \star \varphi_i \! :=\! (\varphi \star \varphi_{\minus})(x_i)$, with $\varphi_{\minus}(x) \! = \! \varphi(\minus x)$.

Notice once more that there is no difficulty (as there is none in matrix models) in defining or dealing with a combinatorial generalization of the action; given a building block $\varphi(x_i)$, one can add other interaction terms corresponding to the gluing of triangles (or polygons) to form general polyhedra, or even more pathological configurations (e.g. with multiple identifications among triangles). The only restriction may come from the symmetries of the action, and for the wish to keep things manageable.

The structure of this action is best visualized in terms of diagrams, similar to those used in the discussion of tensor models. Kinetic and interaction terms identify a propagator and a vertex with combinatorial structure as
\be \label{Feynmanrules}
\begin{array}{c}
\begin{tikzpicture}[scale=2.4]
\draw (-0.3,-0.2) rectangle (0.3, 0.2);
\path (0, 0) node {\small $t$};
\draw (-0.2, -1) -- (-0.2,-0.2);
\draw (0,-1) -- (0, -0.2);
\draw (0.2, -1) -- (0.2, -0.2);
\path (-0.2, -1.2) node {\tiny $x_1$};
\path (0, -1.2) node  {\tiny $x_2$};
\path (0.2, -1.2) node {\tiny $x_3$};
\path (0, -1.5) node  {\tiny $ $};

\draw (-0.2, 0.2) -- (-0.2, 1);
\draw (0, 0.2) -- (0, 1);
\draw (0.2, 0.2) -- (0.2, 1);
\path (0.2, 1.2) node {\tiny $y_3$};
\path (0, 1.2) node  {\tiny $y_2$};
\path (-0.2, 1.2) node {\tiny $y_1$};
\path (0, 1.5) node  {\tiny $ $};

\end{tikzpicture}
\hspace{1cm} 
\begin{tikzpicture}[scale=2.4]
\draw (-1, 0) -- (1,0);
\draw (0,-1) -- (0, -0.1);
\draw (0, 0.2) -- (0, 1);
\draw (-1, 0.2) -- (-0.2, 0.2);
\draw (0.2, 0.2) -- (1, 0.2);
\draw (-1, -0.2) -- (-0.2, -0.2);
\draw (0.2,- 0.2) -- (1, -0.2);
\draw (-0.2, -1) -- (-0.2,-0.2);
\draw (-0.2, 0.2) -- (-0.2, 1);
\draw (0.2, -1) -- (0.2, -0.2);
\draw (0.2, 0.2) -- (0.2, 1); 
\path (0, 0.1) node {\small $\tau$};
\path (-1.2, 0.2) node {\tiny $x_1$};
\path (-1.2, 0) node {\tiny $x_2$};
\path (-1.2, -0.2) node {\tiny $x_3$};
\path (-0.2, -1.2) node {\tiny $y_3$};
\path (0, -1.2) node  {\tiny $x_4$};
\path (0.2, -1.2) node {\tiny $x_5$};
\path (1.2, -0.2) node {\tiny $y_5$};
\path (1.2, 0) node {\tiny $y_2$};
\path (1.2, 0.2) node {\tiny $x_6$};
\path (0.2, 1.2) node {\tiny $y_6$};
\path (0, 1.2) node  {\tiny $y_4$};
\path (-0.2, 1.2) node {\tiny $y_1$};
\path (-1.5, 0) node {\tiny $t_a$};
\path (0, -1.4) node  {\tiny $t_b$};
\path (1.5, 0) node {\tiny $t_c$};
\path (0, 1.4) node  {\tiny $t_d$};
\end{tikzpicture}
\end{array}
\ee
and an expression (obtained as usual in QFT by writing explicitly the kinetic and interaction parts of the action in the form of a convolution of unconstrained fields with a kinetic and an interaction kernel, respectively) given by: 
\be \label{prop&vertex}
\int \extd h_t \, \prod_{i=1}^3 (\delta_{\minus x_i} \star \, \e_{h_t} )(y_i), 
\quad 
\int \prod_t \extd h_t \, \prod_{i=1}^6 ( \delta_{\minus x_i} \star\, \e_{h_{tt'}})(y_i)
\ee
with $h_{tt'} := h_{t\tau} h_{\tau t'}$, where we have used `$t$' for triangle and `$\tau$' for tetrahedron. The group variables $h_{t}$ and $h_{t\tau}$ arise from (\ref{invariance}), and should be interpreted as parallel transports through the triangle $t$ for the former, and from the  center of the tetrahedron $\tau$ to triangle $t$ for the latter. 

The integrands in (\ref{prop&vertex}) factorize into a product of functions associated to strands (one for each field argument), with a clear geometrical meaning. The pair of variables $(x_i,y_i)$ associated to the same edge $i$ corresponds to the edge vectors as seen from the frames associated to the two triangles $t, t'$ sharing it. The vertex functions state that the two variables are identified, up to parallel transport $h_{tt'}$, and up to a sign labeling the two opposite edge orientations inherited from the  triangles $t, t'$. The propagator encodes a similar gluing condition, allowing for a further mismatch between the reference frames associated to the same triangle in two different tetrahedra.
In this non-commutative Lie algebra representation of the field theory, the geometric content of the action is indeed particularly transparent.

\

Using the group Fourier transform we can obtain a pure group representation of the theory (like in the original definition by Boulatov \cite{boulatov}). 

\bes \hspace{-1cm}S_{3d}[\phi]\,=\, \frac{1}{2}\int[dg]^3
\varphi(g_1,g_2,g_3)\varphi(g_3,g_2,g_1) \;-\;\frac{\lambda}{4!}\int [dg]^6
\varphi(g_1,g_2,g_3)\varphi(g_3,g_4,g_5)\varphi(g_5,g_2,g_6)\varphi(g_6,g_4,g_1) \quad  \label{boulatov} . \ees

The combinatorics of arguments in the action has been already discussed above. 
It is also shown in the picture below, for a single tetrahedron (interaction term), where one can see both the tetrahedron, to whose edges are associated Lie algebra elements, and its topological dual\footnote{The topological dual complex is obtained associating a vertex to each tetrahedron, a link to each triangle and a dual face to each edge; the elements of the boundary dual complex have of course a similar relation to those of the boundary simplicial complex.}, to whose boundary links are associated conjugate group elements.

\begin{figure}[here]
\includegraphics[width=6.5cm, height=5.5cm]{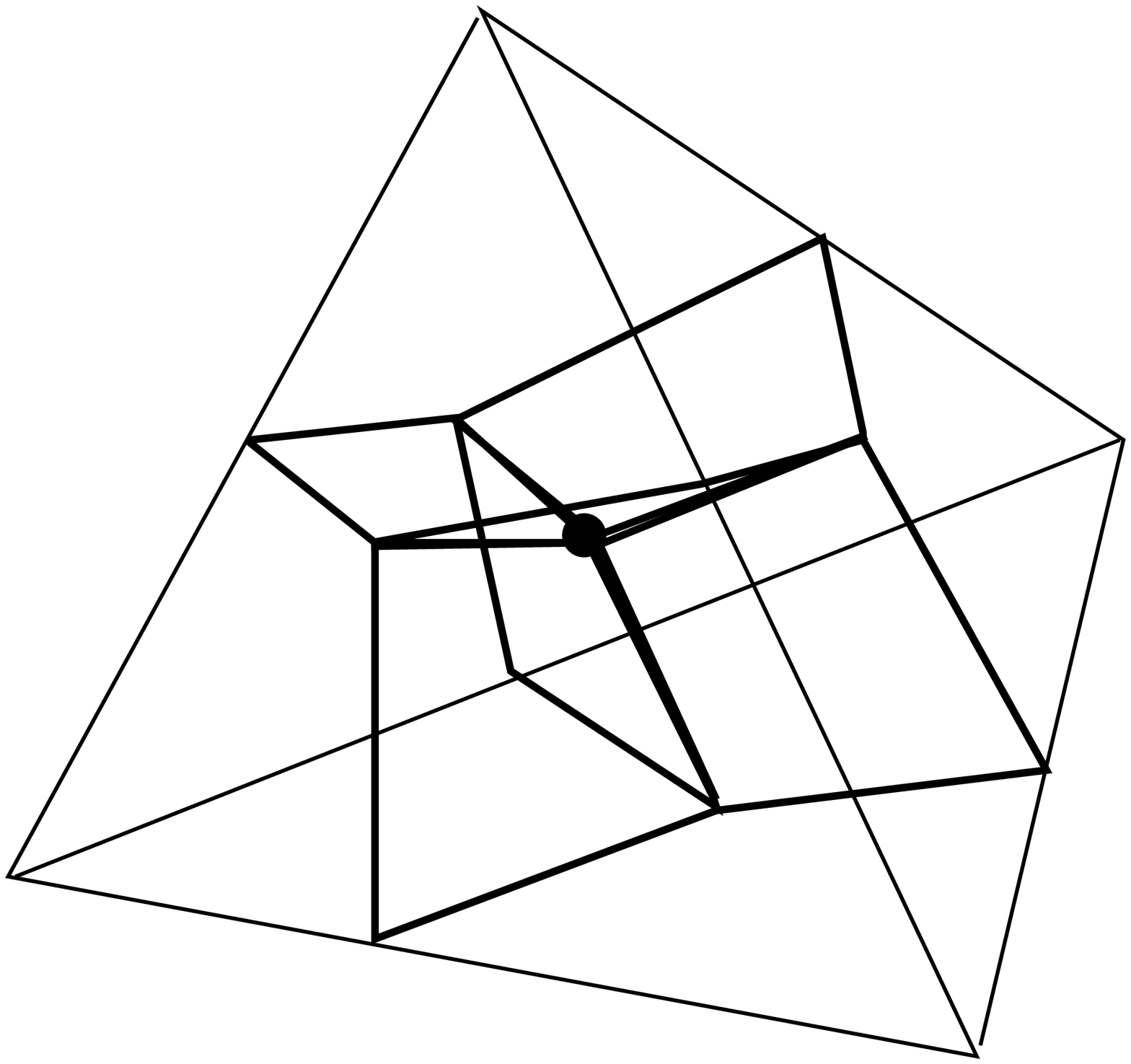}
\end{figure}

Labelling with indices $i,j = 1,\dots , 4$ the triangles in each tetrahedron, and thus the group elements $h$ imposing left diagonal invariance, and by a pair $(ij)$ the edges shared by the triangles $i$ and $j$, and thus the arguments of the field $g_{ij}$ and $g_{ji}$ associated to the same edge in the triangles $i$ and $j$, the kinetic and vertex functions (once more, obtained as the kernels of the convolutions of unconstrained fields in the kinetic and interaction parts of the action, respectively) are:

\begin{equation}
\mathcal{K}(g_i, \tilde{g}_i) = \int dh  \,\prod_{i=1}^{3}\,\delta(g_i h
\tilde{g}_i^{-1}) \hspace{1cm} \mathcal{V}(g_{ij}) =
\prod_{i,j=1}^4\int dh_i \,\prod_{i\neq j}^{} \delta(g_{ij} h_i
h_j^{-1} g_{ji}^{-1}) 
\label{kinpropgroup}
\end{equation}

Also in this set of variables, the geometric content of the model can be read out rather easily: the six delta functions in the vertex term encode the flatness of each \lq\lq wedge\rq\rq, i.e. of the portion of each dual face inside a single tetrahedron \cite{alex,thesis}. This flatness is characteristic of the piecewise-flat context  in which the GFT models are best understood.

The result of the evaluation of the Feynman amplitudes in this representation will be a pure $\SU(2)$ gauge  theory for a discrete connection associated to the various elements of the simplicial complex dual to each Feynman diagram, and  to the Feynman diagram itself. We will show the form of the Feynman amplitudes explicitly below.

Before doing that, however, we also present the form of the action in representation space, obtained after Peter-Weyl decomposition of the GFT field.

Using the expansion of the field (\ref{peterweyl}), performing the appropriate contractions and using standard properties of the $3j$-symbols from $\SU(2)$ recoupling theory, the action becomes:

$$
S(\varphi) \, =\, \frac{1}{2} \sum_{\{ j\}, \{ m\} }\, \varphi^{j_1 j_2 j_3}_{m_1 m_2 m_3}\,\varphi^{j_3 j_2 j_1}_{m_3 m_2 m_1}\;-\;\frac{\lambda}{4!}\,\sum_{\{ j\}, \{ m\} }\, \varphi^{j_1 j_2 j_3}_{m_1 m_2 m_3}\,\varphi^{j_3 j_4 j_5}_{m_3 m_4 m_5}\,\varphi^{j_5 j_2 j_6}_{m_5 m_2 m_6}\,\varphi^{j_6 j_4 j_1}_{m_6 m_4 m_1}\; \left\{
\begin{array}{ccc} 
j_1 &j_2 &j_3
\\ j_4 &j_5 &j_6 
\end{array}\right\}
$$

from which one reads the kinetic and vertex terms:

\begin{eqnarray*}
\mathcal{K} &=& = \mathcal{K}^{-1}\,=\,\delta_{j_1\tilde{j}_1}\delta_{m_1\tilde{m}_1}
\delta_{j_2\tilde{j}_2} \delta_{m_2\tilde{m}_2}
\delta_{j_3\tilde{j}_3}\delta_{m_3\tilde{m}_3} \\ 
\mathcal{V} &=& \delta_{j_1\tilde{j}_1}\delta_{m_1\tilde{m}_1}
\delta_{j_2\tilde{j}_2} \delta_{m_2\tilde{m}_2}
\delta_{j_3\tilde{j}_3}\delta_{m_3\tilde{m}_3}\delta_{j_4\tilde{j}_4}\delta_{m_4\tilde{m}_4}\delta_{j_5\tilde{j}_5} \delta_{m_5\tilde{m}_5}
\delta_{j_6\tilde{j}_6}\delta_{m_6\tilde{m}_6} \left\{
\begin{array}{ccc} 
j_1 &j_2 &j_3
\\ j_4 &j_5 &j_6 
\end{array}\right\} \label{propvertJ}
\end{eqnarray*}  
where $\Delta_j = 2 j + 1$ is the dimension of the representation $j$ and for
each vertex of the 2-complex we have a so-called $6j$-symbol , a
real invariant (under group action on the representation spaces) function of the 6 representations meeting at that vertex. 

The geometry behind the construction, in this formulation, is maybe less transparent. On the other hand, one works directly in terms of quantum numbers labelling the states of the theory, thus can deal more easily with the quantum properties of the model. 

Notice also that the vertex amplitude could have simply been obtained by taking four $\SU(2)$ intertwiners (one per triangle in the boundary of a tetrahedron), associated to spin network vertices with links labelled by representations of $\SU(2)$, which are the known quantum states of BF theory, and gluing them (tracing over corresponding magnetic indices) pairwise along links, with the combinatorial pattern of the edges of the tetrahedron. This type of construction is what is also performed in defining several current spin foam models for 4d gravity.  

\

The classical dynamics of GFT models is basically unknown territory (with a few exceptions, e.g. \cite{eterawinston, noi}). The classical equations of motion for this model are, in group space:

$$
\hspace{-1cm} \int d{h}\,\phi(g_1h,g_2 h,g_3 h) -
\frac{\lambda}{3!}\prod_{i=1}^{3} \int dh_{i}\prod_{j=4}^{6}\int
dg_j\, \phi(g_3 h_1,g_4 h_1,g_5 h_1)
\phi(g_5 h_2,g_6 h_2,g_2 h_2)\phi(g_6 h_3,
g_4 h_3, g_1h_3)=0
$$
but can of course be written both in Lie algebra space, where they also look like complicated integral equations, as well as in the form of purely algebraic equations in representation space.

The role and importance of these
equation from the point of view of the GFT per se are
obvious. They define the classical dynamics of the field theory (classical limit of the quantum effective action), they
would allow the identification of classical background configurations
and non-trivial GFT phases,
etc. From the point of view of quantum gravity, considering the interpretation of the GFT as a discrete \lq 3rd quantisation\rq  of gravity,  these classical GFT equations encode
fully the quantum dynamics of the underlying (simplicial) canonical quantum gravity
theory, or equivalently the quantum dynamics of 1st quantized spin networks, thus implementing both Hamiltonian and diffeomorphism
constraints. This is in analogy with, for example, the Klein-Gordon equation, which represents at the same
time the classical equation of motion of a (free) scalar field theory
and the full quantum dynamics for the corresponding 1st quantised
(free) theory.

\subsection{Quantum dynamics of 2d space in GFT:  duality of simplicial path integrals and spin foam models}
\noindent Let us now turn to the quantum dynamics of group field theories. In absence of a non-perturbative description, this is tentatively defined by the perturbative expansion of the partition function in Feynman diagrams. As we have already discussed, these Feynman diagrams are by construction cellular complexes topologically dual to simplicial complexes. Here we obtain 3-dimensional simplicial complexes. This expansion is given by:

$$ Z\,=\,\int
\mathcal{D}\varphi\,e^{-S[\varphi]}\,=\,\sum_{\Gamma}\,\frac{\lambda^N}{sym[\Gamma]}\,Z(\Gamma),
$$
where $N$ is the number of interaction vertices in the Feynman graph
$\Gamma$, $sym[\Gamma]$ is a symmetry factor for the graph and
$Z(\Gamma)$ the corresponding Feynman amplitude.

\noindent We now sketch the calculation of the Feynman amplitudes of the model. 

\

We first do it in the Lie algebra picture (following \cite{ioaristide}), as this will immediately allow to relate the amplitudes to simplicial gravity. We then use the propagator and vertex terms in (\ref{prop&vertex}). 
In building up the diagram, propagator and vertex strands are joined to one another using the $\star$-product. Each loop of strands bounds a face of the 2-complex $\Gamma$ dual to a 3d simplicial complex $\Delta$, bounded by several links dual to triangles sharing the same edge of $\Delta$, in different tetrahedra. Because of the non-commutativity of the $\star$-product, it is important to choose and then keep track of the ordering between these triangles.

\begin{figure}[here]
\includegraphics[width=10cm, height=4.5cm]{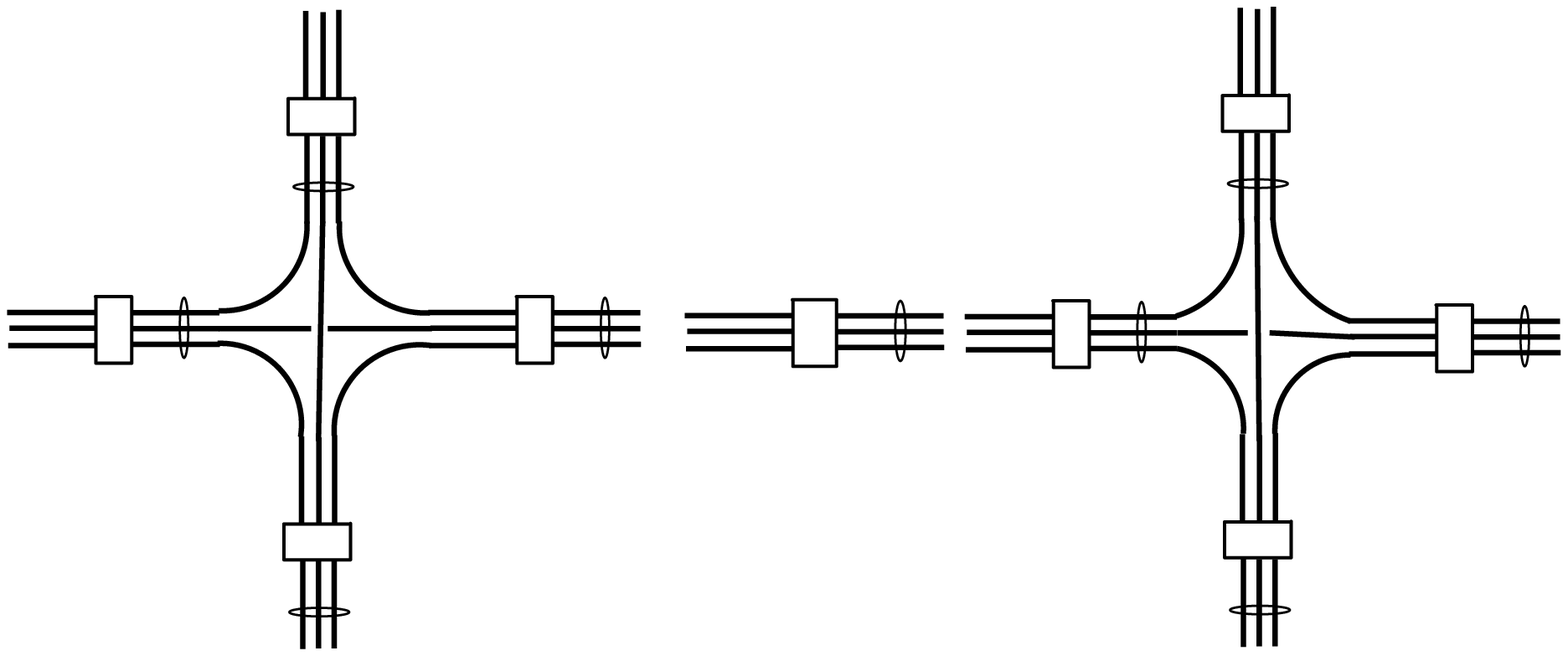}
\end{figure}

A rather involved, but still low order, example of Feynman diagram is represented in the figure below.

\begin{figure}[here]
\includegraphics[width=11.5cm, height=7.5cm]{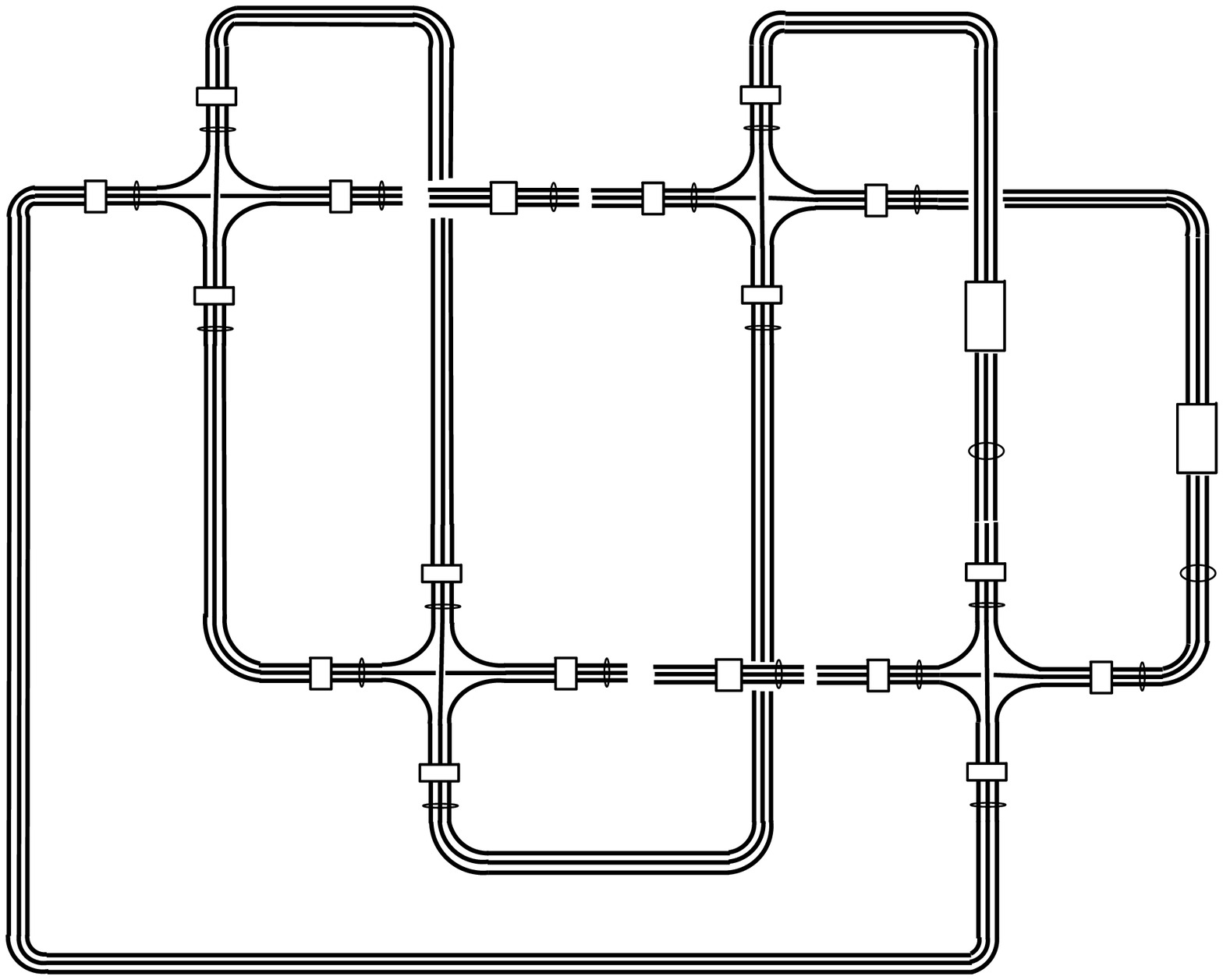}
\end{figure}

We first notice that, under the integration over group elements $h_t$,  the amplitude factorizes into a product of face amplitudes. 
Let then $f$ be a face of the 2-complex, and $A_f[h]$ its associated amplitude.

Choose an ordered sequence $\{\tau_j\}, \{t_j\}_{0\leq N}$,  of triangles (to which propagators are associated) and tetrahedra (to which vertex functions are associated) surrounding the face (dual to an edge of the dual simplicial complex). 
 Each propagator and each vertex function contribute a single \lq 2-point function\rq $ ( \delta_{\minus x_i} \star\, \e_{h})(y_i)$, to give: 
\[ 
A_f[h] = \int \prod_{j=0}^N \extd x_j \vec{\bigstar}_{j=0}^{N + 1} \, (\delta_{x_j} \star e_{h_{jj+1}})(x_{j+1}) 
\]

Next, we integrate over the $N$ variables $x_1, \cdots x_N$. Introducing the total holonomy  $H_{\! 0}:= h_{01} \cdots h_{N0}$ around the boundary of the face, the formula reduces to: 
\[
(e_{H_{\! 0}} \star \delta_{H_{\! 0}^\inv x_0 H_{\! 0}}) (x_{N+1}) = (\delta_{x_0} \star e_{H_{\! 0}})(x_{N+1}) 
\]

We now have to  'close the loop' by setting $x_{N+1} = x_0$; in order to do this, we use $\delta_{x_0} = \int \extd g \, e_{g^\inv}(x_0) e_g$, and get: 
\[
(\delta_{x_0} \star e_{H_{\! 0}})(x_{N+1}) = \int \extd g \, e_{g^\inv}(x_0) e_{gH_{\! 0}}(x_{N+1})
\]
We see that we are left with a single integration over the Lie algebra, and we obtain: 
\[
A_f[h_t] = \int \extd^3 x_0 \int \extd g \, e_{gH_{\! 0}g^\inv}(x_0)
\]
Using $e_{gH_{\! 0}g^\inv}(x_0) = e_{H_{\! 0}}(g^{\mbox{\small -} 1} x_0 g)$ a simple change of variable in $x_0$ leads to the Feynman amplitude: 
\be \label{bf} Z(\Gamma) = \int \prod_t \extd h_t \prod_f A_f [h_t]= \int \prod_t \extd h_t \prod_f \extd x_f  \, e^{i \sum_f \Tr \, x_f H_f} \ee
where $H_f$ is the holonomy along the boundary of the face $f$, calculated for a given choice of a reference tetrahedron. 

This is the usual expression for the simplicial path integral of 3d gravity in 1st order form (or 3d $BF$ theory). The Lie algebra variables $x_f$, one per edge of the simplicial complex, play the role of discrete triad, while the group elements $h_t$, one per triangle or link of the dual 2-complex, play the role of discrete connection, defining the discrete curvature $H_f$ through holonomy around the faces dual to the edges of the simplicial complex \cite{alex, thesis}.

More precisely, consider the 3d gravity action in the continuum:
$$ S(e,\omega)\,=\,\int_{\mathcal{M}} tr \left( e\wedge F(\omega)\right) , $$  
with variables the triad 1-form $e^i(x)\in su(2)$ and the 1-form connection  $\omega^{j}(x)\in su(2)$, with curvature $F(\omega)$. Its discretization proceeds analogously to the definition of smeared phase space variables in LQG.
Introducing the simplicial complex $\Delta$ and its topological dual cellular complex, we can discretize the triad in terms of Lie algebra elements associated to the edges of the simplicial complex as $x_e = x_{f} = \int_e e(x) = x^i \tau_i \in \su(2)$, and the connection in terms of elementary parallel transports along links of $\Gamma$, dual to triangles of $\Delta$, as $h_{L} = h_t = e^{\int_{L}\omega}\in \SU(2)$. The discrete curvature will then be given by the holonomy around the dual face $f$, obtained as ordered product of group elements $h_L$ associated to its boundary links: $H_{f} = H_e\,=\, \prod_{L\in \partial f} h_{L} = e^{F_{f}}\in \SU(2)$, and a discrete counterpart of the continuum action will be given by the action appearing in (\ref{bf}).

The application of the non-commutative group Fourier transform to the simpler quantum mechanics of a point particle living on the group manifold gives similar results, i.e. a correct definition of the amplitude as a path integral \cite{iomatti}, and gives further insights on the technical details. 

\

We have thus shown that the Feynman amplitudes of the GFT model are simplicial path integrals for 3d Riemannian gravity in 1st order form. Moreover, in the non-commutative Lie algebra variables, the underlying simplicial geometry of the model is made transparent. The Lie algebra variables represent discrete triad variables associated to the edges of a triangle, in the corresponding frame, the GFT field being its 2nd quantized wave function. The closure of the triangle is obtained by constraining appropriately these edge vectors. By expressing this constraint in integral form, as it appears in the Feynman amplitudes, one sees that this closure is encoded in the $h$-integration and thus in one of the equations of motion of the simplicial action (the one corresponding to the variations of $h$-variables). The GFT action encodes the correct gluing of triangles in a tetrahedron, and across tetrahedra, by identification of triad vectors up to parallel transport, parametrized by a gauge connection. The quantum dynamics of each interaction process of triangles, corresponding to a given simplicial complex/Feynman diagram, that emerges therefore as a virtual and quantum construction, is defined by the associated amplitude.    

It is instructive to compute the GFT amplitude for given boundary simplicial data, i.e. for open GFT Feynman diagrams.  
The one-vertex contribution to the 4-point functions is the function of twelve metric variables $x_i, x'_i$ obtained by connecting a closure operator $\widehat{C}$ (propagator) on each external leg of the vertex diagram in (\ref{Feynmanrules}), thus building up four triangles $t_a, \cdots t_c$. The resulting amplitude is the exponential of the BF action for a single flat simplex, and thus made out of pure boundary terms, with fixed metric ($x$ variables) on the boundary (and integrated bulk connection). The vertex also relates the metric data of pairs of triangles $t, t'$ via $(\delta_{x_i} \star \e_{h_{tt'}})(x'_i)$. This may be viewed as a constraint on the gauge connection. In a semi-classical limit the constraints become those characterizing a discrete Levi-Civita connection. 

This also implies that: first, for generic simplicial complex with boundary the GFT Feynman amplitudes in the $x$ representation are given by a path integral fort he BF action augmented by the appropriate boundary terms; second, the (exponential of the) BF action for a single simplex is already explicitly present in the interaction term of the GFT action, in the $x$ representation. This can be useful to study the link with semi-classical/continuum gravity directly at the GFT level.

\

We have thus shown that the GFT model we have introduced succeeds in at least one of the points where the simpler tensor models failed, i.e. in defining amplitudes for its Feynman diagrams (identified with discrete spacetimes), arising in perturbative expansion around the \lq no-space state\rq, that correctly encode the classical and quantum simplicial geometry and that can be nicely related to a simplicial gravity action. 

As for the other issues that tensor models face, and that are also shared by GFTs, concerning the combinatorial structures obtained in the same perturbative expansion, and the control over this perturbative sum, we can hope that the additional structure and data in the GFT amplitudes, the same that allows the link with classical discrete gravity, would also help in progressing on this front. We will discuss in the following to what extent recent results either already fulfill or at least give more ground for this hope.

\

The Feynman amplitudes can also be computed in the other representations we have at our disposal. We first do so in the group picture, using the kinetic and vertex terms \ref{kinpropgroup}.
The Feynman amplitudes are obtained by convolution of such vertex functions and propagators. 
As it can be easily checked, the result of these convolutions is a single delta function $\delta (\prod_{L\in\partial f} h_{L} )$ for each 2-cell $f$ in the Feynman diagram, dual to a single edge of the simplicial complex, with argument given by the product of group elements $h_L$ each associated to a link in the boundary  of the 2-cell. In other words, the model imposes flatness of the discrete curvature located on each dual 2-face, i.e. on each edge of the simplicial complex $\Delta$. Once more, this is in accordance with our understanding of 3d simplicial gravity (and of continuum 3d gravity as well). The overall amplitude is then:
\begin{equation}
Z(\Gamma)\,=\, \prod_{L\in \Gamma} \int \,dh_L
\,\,\prod_{f}\,\delta (\prod_{L\in\partial f} h_L )\, \label{amplgroup}.
\end{equation}

Of course, the same result could have been obtained by taking the group Fourier transform of the expression \ref{bf} for the same amplitude, which in this case would simply amount to performing the integral over $x_f$ in \ref{bf}.

\

Similarly, we can compute the spin foam expression of the Feynman amplitudes $Z(\Gamma)$, i.e. their expression in terms of group representations (quantum numbers of geometry). We can obtain it starting from the re-writing of the GFT action in representation space, i.e. from the expressions (\ref{propvertJ}), or by direct Peter-Weyl decomposition of amplitudes (\ref{amplgroup}), and successive group integrations. The result is an assignment of an irreducible $\SU(2)$ representation $j_f$ to each face of $\Gamma$, and of a group intertwiner to each link of the same complex, i.e. a spin foam \cite{review,alex,thesis}. 

The corresponding amplitude reads:
$$ Z(\Gamma)=\left(\prod_{f}\,\sum_{j_{f}}\right)\,\prod_{f}(2j_{f} +1)\,\prod_{v}\, \left\{ \begin{array}{ccc}
j_1 &j_2 &j_3
\\ j_4 &j_5 &j_6
\end{array}\right\}. $$

This is the well-known Ponzano-Regge spin foam model for 3d Riemannian quantum gravity, actually the first spin foam model ever proposed.

Notice that the above expression is formal, in that, depending on the simplicial complex on which it is evaluated, the sum over representations may diverge. While this may be unpleasant, it is to be expected when the same amplitude is understood as a Feynman amplitude of a field theory, as in this GFT context.  
Many results have been obtained about this model, which, upon regularization, provides a topological invariant of 3d manifolds. We refer to the literature for details on such results, and a more detailed analysis of the model \cite{PR1,PR2, barrettPR, review, alex, thesis}, from a spin foam point of view. Its divergences have also be carefully studied both in a spin foam context and from the GFT point of view, in the recent literature \cite{matteovalentin, ren, borel}. 
For example, one can define a quantum group deformation of the Ponzano-Regge model, and of the corresponding GFT, going from the group $\SU(2)$ to the quantum group $\SU_q(2)$ and to the corresponding category of representations, and obtain in this way the so-called Turaev-Viro model, another invariant of 3-manifolds \cite{TV}, related to 3d gravity with positive cosmological constant (also in Chern-Simons formulation of the same).
Also, one can show, for a single spin foam vertex (tetrahedron), that, for large, fixed $j$'s (which can be understood as a semi-classical approximation of the amplitude)
$$
\left\{ \begin{array}{ccc} 
j_1 &j_2 &j_3
\\ j_4 &j_5 &j_6 
\end{array}\right\}_{v*} \simeq \cos{S_R(l_e)} \simeq e^{iS_R} + e^{-i S_R}
$$
where $S_R({l_e=2j+1})$ is the Regge action for simplicial gravity, with edge lengths given by $2 j_e +1$.
Thus the Feynman amplitudes in spin foam representation match expected form of  semi-classical simplicial gravity path integral. 

This is not surprising, given the dual simplicial path integral expression for the same amplitudes in terms of Lie algebra variables $x_e$, at least from a heuristic point of view.  The $x_e$ in fact are interpreted as classical edge vectors (discrete triads), whose absolute value is the length $L_e$ of the edge to which they are associated, and that has instead quantum spectrum (according to the canonical theory) $L_e= \sqrt{j_e (j_e +1)} \approx 2j_e + 1$, where the approximation holds true for large $j_e$ (semiclassical limit). In this approximation, in fact, the quantum numbers (eigenvalues) are expected to match the corresponding classical variables, and thus also the amplitude should have a functional dependence on them that matches the dependence on the triad variables, in the same approximation (large scales). Now, look at the simplicial path integral we obtained in the Lie algebra representation, for  single tetrahedron and for fixed $x$ variables on the boundary. Its semiclassical limit implies a saddle-point approximation of the action (as in any path integral), which, together with the (commutative limit of) the delta functions on the algebra entering the amplitude, impose that the discrete connection is a Levi-Civita one an a function of the triad variables $x$, thus resulting in a 2nd order action in terms of them only, i.e. in the Regge action for simplicial gravity.   
Seen in this light, the asymptotic result for the $6j$-symbol only confirms the geometric meaning of the quantum variables $j_e$ and would suggests in itself (if we didn't have it already) a simplicial gravity path integral formulation of same spin foam amplitudes.

\

We have thus shown that the Feynman amplitudes of the GFT model can be equivalently written in the form of a spin foam model and as simplicial gravity path integral. This is an exact duality that stems from the possibility of using two equivalent representations of the GFT field, represented as a function on group manifolds: as a function of representation labels, following Peter-Weyl decomposition, and as a non-commutative function on Lie algebra variables, using the non-commutative group Fourier transform. 

\

Notice also that a field $\varphi\in\C(\SU(2)^3)$  can be seen as the tensor product of three representations of the quantum (Drinfeld) double $D\SU(2)$, which is a deformation of the Poincar\'e group \cite{NCFourier}. This is the starting point of the analysis of the transformation properties of the Boulatov field, and of the symmetries of the GFT action we have presented above, and which have just recently been discovered \cite{florianetera, ioaristideflorian}.  The symmetries that one can identify at the level of the GFT action translates, at the level of the corresponding Feynman amplitudes, into the known symmetries of simplicial BF theory \cite{laurentdiffeo}. In particular, the so-called translation symmetry of BF theory (and 3d gravity), strictly related to diffeomorphism symmetry has been identified in the GFT action, and related to (generalized) Bianchi identities in the Feynman amplitudes. This opens the door to the study of diffeomorphism symmetries at the GFT level, by an extensive application of field-theoretic techniques to the analysis of their consequences on GFT transition amplitudes.

\ 

Finally, before moving on to the 4-dimensional case, let us comment on the calculation of transition amplitudes. Transition amplitudes or correlation functions between quantum gravity states are defined as insertions of appropriate GFT observables in the partition function of the theory. These observables are, as in any field theory, arbitrary functionals of the fundamental GFT field, compatible with the symmetries of the theory. Polynomial functionals, in particular, can be used as a basis for the space of observables, and are obtained by convolution in the Lie algebra or in the group variables of products of GFT fields, or by the equivalent traces in representation space. As we hinted at above, such observables describe extended simplicial spaces or, equivalently, extended spin networks, endowed with quantum geometric data, and define thus possible quantum spaces. When inserted in the GFT path integrals, and upon expansion of the same in Feynman diagrams, they provide the boundary structures for the open diagrams, and the perturbative sum provides a tentative definition of their mutual correlations (transition amplitudes) \cite{iogft, laurentgft}.

For example, one can consider the spin network observables 
$$ O_{\Psi=(\gamma, j_e,i_v)}(\varphi)=\left(\prod_{(ij)\int}dg_{ij}dg_{ji}\right) \Psi_{(\gamma, j_e,i_v)}(g_{ij}g_{ji}^{-1})\prod_i \varphi(g_{ij}),$$ where  $\Psi_{(\gamma, j_e,i_v)}(g)$ identifies a spin network functional \cite{thomas, carlo} for the spin network labelled by a graph $\gamma$ with representations $j_e$ associated to its edges and intertwiners $i_v$ associated to its vertices, and $g_{ij}$ are group elements associated to the edges $(ij)$ of $\gamma$ that meet at the vertex $i$.
The transition amplitude between boundary data represented by two
spin networks, of arbitrary combinatorial complexity, can be expressed as the expectation value of the field
operators having the same combinatorial structure of the two spin
networks \cite{laurentgft}.

$$
\langle \Psi_1\mid\Psi_2\rangle = \int \mathcal{D}\varphi\,O_{\Psi_1}\,O_{\Psi_2}\,e^{-S(\varphi)} = \sum_{\Gamma/\partial\Gamma=\gamma_{\Psi_1}\cup\gamma_{\Psi_2}}\,\frac{\lambda^N}{sym[\Gamma]}\,Z(\Gamma)
$$
where the sum involves only 2-complexes (spin foams) with boundary given by the two spin networks chosen.

\section{Towards a group field theory formulation of 4d quantum gravity}
Important results have been obtained recently in the attempt to construct interesting GFT and spin foam models for 4-dimensional quantum gravity. Here, we summarize some of them briefly, phrasing them in the context of a GFT formalism.
The general strategy adopted for these constructions is based on the fact that a formulation of gravity that is classically equivalent to General Relativity is obtained starting from BF theory is 4-dimensions and and imposing constraints on the B field, that force it to be a derived quantity from a tetrad field. When re-written in terms of such tetrad field, the action for BF theory coincides in fact with the Palatini formulation of gravity. We will describe this formulation in the following. 
The idea is then to quantize BF theory in a spin foam or GFT setting, and then impose on the resulting theory the (quantum version of the) constraints that reduce it to gravity.

\

Now we start by outlining the extension of the Boulatov model describing quantum 4d BF theory.

\

The Boulatov model can be generalized easily to any dimension (and to any compact group), and one can similarly generalize both the group Fourier transform and the Peter-Weyl decomposition, to write GFT field, action and amplitudes in Lie algebra, group or representation variables. In all cases, this procedure defines a GFT quantization of BF theory in the given dimension and for the given gauge group. In particular, the 4-dimensional extension of the Boulatov model, proposed by Ooguri \cite{Ooguri} is based on a real field: $\varphi(g_1,..,g_4): \SO(4)^{\times 4} \rightarrow\mathbb{R}$, symmetric under diagonal left action: $\varphi(h g_1, h g_2
,h g_3, h g_4 ) = \varphi(g_1, g_2, g_3,g_4)$. Equivalently, it can be expressed as a field over four copies of the Lie algebra $\so(4)$ with the four Lie algebra arguments satisfying the closure condition $x_1 + x_2 + x_3 + x_4 = 0$. 

The GFT action is then:

\bes S[\varphi] &=& \frac{1}{2}\int [\varphi(g_1,g_2,g_3, g_4)]^2\; +\,\nonumber \\   &+&  \frac{\lambda}{5!}\int \varphi(g_1,g_2,g_3, g_4)
\varphi(g_4,g_5,g_6, g_7) \varphi(g_7,g_3,g_8, g_9)\varphi(g_9,g_6,g_2,g_{10})\varphi(g_{10},g_8,g_5,g_1)\,, \hspace{1.4cm} \ees

or the equivalent in Lie algebra variables (with appropriate $\star$-multiplications).

The combinatorics now represents the gluing of 5 tetrahedra (each corresponding to a field $\varphi$) pairwise along boundary triangles (each corresponding to one of the 4 arguments of $\varphi$) to form a 4-simplex. The Feynman diagrams are 2-complexes dual to 4d simplicial complexes. 

Kinetic and vertex terms again encode the identification up to parallel transport of the bivectors associated to the same triangle in different tetrahedral frames, and the computation of Feynman amplitudes proceeds analogously to the 3d case. The result is again a simplicial path integral for BF theory like (\ref{bf}) with integrals now over $\SO(4)$ group and Lie algebra elements \cite{ioaristide}.

The Lie algebra variables are interpreted as the four bivector variables $x_f = (x_f^+, x_f^-)$ associated to the four triangles of each tetrahedron in a simplicial discretization of 4d BF theory, coming from the discretization of the continuum $B$ field \cite{mike, DP-F}. The continuum 1-form connection is instead discretized as in the 3d case, in terms of group elements associated to links of the 2-complex dual to the simplicial complex, to give holonomies (discrete curvature) associated to each 2-cell to a triangle of the same. Notice however that these data do not allow (even at the classical level) to reconstruct a unique geometry for the simplicial complex, as cannot be expected in a non-geometric theory like BF.

Similarly, one can obtain the spin foam expression for the same Feynman amplitude. It is given by:

\bes
Z(\Gamma)&=&\prod_{f}\int_{\so(4)} dx_{f}\prod_{L}\int_{\SO(4)}d h_{L} \,e^{i\,\sum_{f}\,tr\left(x_{f}\,H_{f}\right)} \, =\, \left(\prod_{L\in \Gamma} \int d h_{L}\right)
\,\prod_{f}\,\delta (\prod_{L\in\partial f} h_{L} )\, = \nonumber \\ &=&\, \sum_{\{j_+, j_- \}} \prod_f (2j_+ + 1) (2 j_- +1)\prod_v \, \big\{ 15j \big\}^v_+ \big\{ 15j \big\}^v_- , \ees
where we used the selfdual/anti-selfdual splitting of $\SO(4)$ representations, which gives basis vectors $| j_+, m_+\rangle | j_-, m_-\rangle$, and the symbol for the vertex amplitude is the well-known $15j$-symbol from the recoupling theory of angular momentum. 

We also know how the same spin foam vertex amplitudes can be reconstructed directly from the structure of the spin network boundary states, with $\SO(4)$ labels. As in 3d, one considers one $\SO(4)$ intertwiner for each tetrahedron in the boundary of a 4-simplex, with one representation associated to each triangle of the tetrahedron, and then glues these intertwiners along common magnetic indices, following the combinatorial pattern of triangles in the boundary of a 4-simplex.
What we said about divergences of the Ponzano-Regge amplitudes applies to this model as well, both from a spin foam and a GFT perspective.

\subsection{Gravity as a constrained topological BF theory}

As anticipated, this easy generalization of the GFT and spin foam formalism from 3d to 4d BF theory is very important because classical 4d gravity can be expressed as a constrained BF theory (Plebanski formulation) \cite{mike, DP-F}. Let us illustrate this. For an $\so(4)$ Lie algebra 1-form connection $\omega$ and a 2-form $B$ also with values in the Lie algebra, one can write the classical continuum action:
$$
S(\omega,B,\phi)=\int_{\mathcal{M}}\left[ B^{IJ}\wedge
F_{IJ}(\omega)-\frac{1}{2}\mu_{IJKL}B^{KL}\wedge B^{IJ}\right],
$$
with the symmetries for the Lagrange multipliers $\mu_{[IJ][KL]} = \mu_{[KL][IJ]}$. Variation with respect to these Lagrange multiplier $\mu$ gives constraints on the $B$ variables. The solution of them (in the appropriate sector \cite{alex,thesis}) force the $B$ to be a function of a tetrad field $e$: $B^{IJ}\,=\,\pm\,\frac{1}{2}\,\epsilon^{IJ} {}_{KL}\,e^{K}\,\wedge\,e^{L} $, so that, for these solutions, the action becomes the Palatini action for gravity: 

$$
S(\omega,e)=\int_{\mathcal{M}}\left[ \frac{1}{2}\,\epsilon^{IJ} {}_{KL}\,e^{K}\,\wedge\,e^{L} \wedge
F_{IJ}(\omega)\right] \,\,\,\;.
$$

As we have seen, we know how to discretize and quantize BF theories in any dimension, and in particular we know how to construct the corresponding spin foam models and GFT action.

Furthermore, we also know how to discretize the Plebanski constraint on an arbitrary simplicial complex \cite{BC, mike, DP-F, sergei, new}. 
In fact, a discrete tetrad can be reconstructed for the whole simplicial complex, and it determines the discrete bivectors $B_f$ associated to triangles of the same, if one requires that 

\be \forall \, \text{tetrahedra}\, t\in\Delta \;\;\exists n_t\in S^3 \;/\;
(*B_f)^{IJ} n_{tJ}\,=\,0\;\;\; \forall B_f\, \;f\subset t  \label{simplicity} \ee

In words, the condition is that, for each tetrahedron, its four triangle bivectors all lie in the hypersurface orthogonal to the same normal vector. This is interpreted as a normal vector to the tetrahedron. This should be supplemented by the condition of \lq closure\rq of all the tetrahedra in $\Delta$, plus an orientation and a non-degeneracy condition \cite{BC,new}. The same constraint reads, in the selfdual/anti-selfdual splitting, 
 
\be \forall \, \text{tetrahedra}\, t\in\Delta \;\;\exists n_t\in S^3\simeq \SU(2) \;/\;
b_+^i \,=\, -\, (n \cdot b_-\cdot n^{-1})^i\;\;\; \forall B_f = (b_{f+}, b_{f-})\, \;f\subset t  \label{simplicity2} \ee
where we use the natural action of $\SU(2)$ in the fundamental representation.

All this can be generalized to include an Immirzi parameter $\gamma$ (which plays a crucial role in LQG), adding a term to the Plebanski action:
$$
S(\omega,B,\phi)=\int_{\mathcal{M}}\left[ B^{IJ}\wedge
F_{IJ}(\omega)\,+\frac{1}{\gamma} (*B)^{IJ}\wedge
F_{IJ}(\omega)-\frac{1}{2}\mu_{IJKL}B^{KL}\wedge B^{IJ}\right].
$$ 

After inserting the solution to the constraints, this becomes the Holst action for gravity:

$$
S(\omega,e)=\int_{\mathcal{M}}\left[ \frac{1}{2}\,\epsilon^{IJ} {}_{KL}\,e^{K}\,\wedge\,e^{L} \wedge
F_{IJ}(\omega)\,+\,\frac{1}{2\gamma}e^{I}\,\wedge\,e^{J} \wedge
F_{IJ}(\omega)\right] \,\,\,\;.
$$
which is classically equivalent to Palatini gravity, as the additional term vanishes on-shell (when imposing metricity and torsion free-ness of the connection) \cite{Holst}. This action is the classical starting point for the loop quantization leading to LQG.

Also the discrete theory can be easily generalized to include the Immirzi parameter. 

One has only to impose on the BF action the new constraints:

$$ \forall \, \text{tetrahedra}\, t\in\Delta \;\;\exists n_t\in S^3 \;/\;
(B_f - \gamma*B_f)^{IJ} n_{tJ}\,=\,0\;\;\; \forall B_f\, \;f\subset t $$

\subsection{A basic idea and two strategies for 4d gravity model building in GFT}

A natural procedure\footnote{But certainly not the only one \cite{cubulations, artem}.} for defining a GFT or spin foam model for 4d gravity, given the knowledge we have on BF models, is then to start from the GFT action for 4d BF theory, or from the corresponding simplicial path integral or spin foam expression, and impose suitable restriction on the dynamical variables representing the discrete (classical or quantum) $B$ field variables, to impose the discrete version of the Plebanski constraints. The result, if things are done properly, should be a good encoding of 4d simplicial (and possibly continuum) geometry.

On the basis of the duality between simplicial gravity path integrals and spin foam models, naturally realized in a GFT context, as we have seen, we can follow two possible strategies:
\begin{enumerate}

\item find a quantum version of Plebanski constraints as operator equations, impose them on BF spin network states to get gravity spin network states; then, construct spin foam/GFT amplitudes (thus, in the $j$-representation) from these states; finally, check that they encode correctly simplicial geometry, for example by relating them to a simplicial gravity path integral.

\item start from the BF GFT model in terms of Lie algebra variables (bivectors), or from the resulting (non-commutative) simplicial path integral, and impose the geometric constraints on them; in this way, one obtains a model whose Feynman amplitudes are manifestly simplicial path integrals for BF theory with constraints; then, one can also re-write the same amplitudes as spin foam models and identify the corresponding space of boundary states in a spin network basis.

\end{enumerate}

As in any quantization procedure, one is forced to make choices and resolve quantization ambiguities. This applies to both strategies outlined above. Their respective advantages may lead to prefer one or the other, but the resulting amplitudes should in any case be tested against a detailed study of the way they encode simplicial geometry and simplicial gravity dynamics, and explicit calculations of the semi-classical approximation and of quantities like lattice correlators or simplicial gravity observables. And it goes without saying that all this would be in any case only a first test, the real issue being a direct and compelling link with effective continuum physics, gravitational physics in particular.

Let us discuss very briefly these two strategie. We refer to the vast literature for details on the results they led to.

\subsubsection{The state sum strategy and spin foam models}

The first strategy can be based on geometric quantization of simplicial structures, on the use of Master constraints techniques, or on a re-writing of spin network states in terms of group coherent states \cite{BC, new, AlexLiving}. The main idea is the following. The classical $B$ variables, at the quantum level, are identified with generators $T^{IJ}=(T_+^i,T_-^i)$ of the $\so(4)$ Lie algebra (or equivalently the corresponding selfdual/anti-selfdual generators $T_\pm^{IJ}$), which then act as operators on the $\SO(4)$ spin network states labelled by representation of the same group. The classical constraints, functions of the $B$'s, are then also turned into operators and imposed at the quantum level on such spin networks, to give restrictions on both the representations attached to their links and on the intertwiners associated to their vertices \cite{new}. Clearly, choices of operator orderings and details of the quantization scheme adopted influences strongly the result one obtains in the end.

We limit ourselves to one simple example of an operator equation encoding some of the needed constraints, the so-called diagonal simplicity constraint. Classically, for all triangles $f$ in the tetrahedron $t$, one requires that the modulus of the selfdual vector component of $B^{IJ}$ is equal to the modulus of the anti-selfdual vector component. Quantum mechanically, in the chosen quantization scheme, this becomes the operator equation:

\be
\hat{T}_+^{IJ}\hat{T}_{+\,IJ}\,-\, \hat{T}_+^{IJ}\hat{T}_{+\,IJ}\,=\,0
\ee 
In turn, and depending again on the quantization scheme chosen, this equation can be imposed strongly on quantum states, giving the resulting constraint on representation labels $j_+\,=\,j_-$, or weakly as a constraint on the mean value of the above operator, restricting the model to include only states where this mean value vanishes.

The resulting spin networks satisfying these restrictions are identified as {\it gravity spin networks}, built out of {\it gravity intertwiners} (constrained BF intertwiners) and used to construct new spin foam vertex amplitudes. This construction follows the analogous one for the BF vertex amplitude: take five gravity spin network vertices (corresponding to the five tetrahedra in the boundary of a 4-simplex); \lq\lq glue them together\rq\rq ~by tracing out internal (vector) variables in each common representation space (corresponding to the triangle common to two tetrahedra). This results in the spin foam amplitude for a single 4-simplex (see \cite{new,AlexLiving} for more details). The rest of the spin foam amplitudes (weights for tetrahedra or triangles) has to be determined by other means \cite{new,AlexLiving}.

An alternative, but closely related way of imposing the same constraints \cite{new} makes use of the Perelomov coherent states for Lie groups \cite{perelomov}. It amounts roughly to the following. One starts by re-writing the BF spin foam amplitudes in terms of $\SO(4)$ coherent states $| j^+, j^- , (n^+,n^-)\rangle$, where $n^{\pm}\in S^2$. Next,  one assumes the coherent states parameters $(n^{+},n^-)$ to be the semi-classical counterpart of classical $B$ bivectors, and imposes the gravity constraints strongly on them. This is sensible because the number of components matches,  and because of their behavior under $\SO(4)$ rotations. Also their semi-classical behaviour matches. In fact, $$ \langle j^+, j^- , (n^+,n^-) | ( T^i_+,  T^i_-)   | j^+, j^- , (n^+,n^-)\rangle = (j^+ n^i_+, j^- n_-^i)\, .$$ These constraints on coherent state parameters become then constraints on the quantum states allowed in the model.

At the GFT level, the construction proceeds analogously, and it amounts, with respect to the BF model, to restricting the GFT fields to expand into a subset of the $\SO(4)$ modes only, those singled out by the imposition of constraints on the spin network states according to the above procedure; the contraction of constrained intertwiners follows indeed the same combinatorial pattern of the identification of field arguments in the GFT action, so that the GFT interaction kernel is exactly the spin foam vertex amplitude obtained by the above procedure. The GFT kinetic term dictates then the rule for matching boundary data across 4-simplices and determines the remaining parts of the spin foam measure.  

In order to appreciate the differences between the various models that have been constructed using this strategy, it is convenient to use the decomposition of each representation $(j^+,j^-)$ of $\SO(4)$ into irreducible representations of the diagonal $\SU(2)$ subgroup $k$. There are ten such representations, one per triangle in the 4-simplex, each triangle being shared by two tetrahedra. The amplitudes all depend on such ten representations of $\SO(4)$, on the ten representations of  $\SU(2)$, and on the corresponding intertwiners, one for each of the five tetrahedra in the 4-simplex, ensuring $\SO(4)$ gauge invariance. The general form of the vertex (4-simplex) amplitude is:

$$ A_v(k_f, i_e)=\sum_{j^+_{ab}, j^-_{ab}}\, \left\{15j^+\right\} \left\{ 15j^-\right\}\, \prod_{(ab)=1,..,10} f (j_{ab}^-j_{ab}^+, k_{ab}) \prod_{a = 1,..,5} i_a(k_{ab})  , $$
where the indices $a,b$ label the five tetrahedra and the pairs $(ab)$ label triangles. The functions $f$ (usually called \lq\lq fusion coefficients\rq\rq) are appropriate mapping coefficients from $\SO(4)$ to $\SU(2)$ representations (they correspond to constrained Clebsch-Gordon coefficients). Because the decomposition of $\SO(4)$ representations into $\SU(2)$ ones is always available and the combinatorial pattern of 4-simplex boundary data, combined with gauge invariance, automatically results in the appearance of $15j$-symbols, the constraints can always be pushed to be completely captured by the specification of these mapping coefficients. Thus they capture all the differences between models, at least a far as the vertex amplitude is concerned.

The resulting models, for various choices of $\gamma$ are the following \cite{new}:

\begin{itemize}
\item in the geometric sector ($\gamma \rightarrow\infty$), one obtains the Barrett-Crane vertex, characterized by: $j_{ab}^+ = j_{ab}^-$, $k_{ab}=0$; using the coherent state method, one instead obtains the Freidel-Krasnov vertex, with a more involved dependence of $k_{ab}$ on the $j_{ab}^{\pm}$ (but with a dominant contribution from the $k=0$ configuration, thus from the Barrett-Crane vertex);

\item in the so-called topological sector ($\gamma \rightarrow 0$), one obtains the Engle-Pereira-Rovelli vertex, with $k_{ab}= 2 j_{ab}^+ = 2 j_{ab}^-$;

\item in the case of finite Immirzi parameter, one obtains the Engle-Pereira-Rovelli-Livine vertex, with $j^+_{ab} = \frac{\gamma + 1}{ |\gamma - 1|} j^-_{ab}$ and $k_{ab} = j^+ \pm j^{-}_{ab}$ in the cases $\gamma < 1$ and $\gamma > 1$ respectively; using coherent state method one obtains the more complicated Freidel-Krasnov model with Immirzi parameter (again characterized by precise but more involved fusion coefficients functions of $\gamma$), which, however, for $\gamma<1$ coincides with the EPRL one.
\end{itemize}

To conclude, we point out the good and less good points of this first strategy. This methods provides nicely the structure of boundary states, and a reasonably simple vertex amplitude compatible with the symmetries (gauge invariance) and with the gravity constraints (at least in a semi-classical approximation). Thanks to the use of Master constraints techniques \cite{thomas}, it takes nicely care of constraint classes, and so imposes different classes of constraints appropriately. In the coherent state method, it allows to keep under control the geometric interpretation of the amplitudes which can also be reformulated as a (rather non-standard) simplicial path integral of BF-type. On the negative side, it does not specify uniquely the other contributions to the spin foam amplitudes, i.e. triangle and tetrahedral amplitudes, which have to be specified by other arguments. Also, the simplicial geometry behind the construction is not so manifest. The situation in this respect improves when using coherent states, whose geometric interpretation can really be justified only in a semi-classical sense, and in terms of which the quantum amplitudes do not really match a standard simplicial path integral even in the BF case. 
For completeness, we mention that a series of criticisms towards these new models has been put forward in \cite{sergeiReview}, from the perspective of Dirac canonical quantization of constrained systems. Although far from being conclusive, these criticisms should be carefully investigated.

These models, currently under active investigations, have been already shown to have (at least) two very nice properties: 

1) their boundary spin networks are of the same type (in the limited sense of having the same representation labels, in a given (time) gauge), as those obtained in canonical loop quantum gravity, and also the kinematical geometric operators such as areas and volumes have the same spectrum as that obtained in loop quantum gravity \cite{YouDing}; this is a nice matching between the canonical and the covariant construction; this obviously holds true only in models with finite $\gamma$ parameter, the only case in which a correspondence can be expected (otherwise, the loop quantization is simply not defined);

2) the vertex amplitudes reduce to the (cosine of) the Regge action in a semi-classical (large-$j$) limit for all models, confirming that the models correctly capture simplicial geometry at least for a single 4-simplex and at least in such approximation \cite{asymptotics}; this results has been also extended (under some assumptions) for the spin foam models with finite Immirzi parameter \cite{asymptotics}.

Moreover, all the above models have a nice Lorentzian counterpart, which is of course a very important asset.
Despite these successes, and leaving aside technical issues in their derivation, the construction of the new models is still part of the story, from the GFT point of view. They can be given a GFT formulation \cite{newGFT}, in representation space, but we have no clear understanding of the geometric translation of the constraints on representations in terms of group (connection) variables or in terms of Lie algebra variables (the variables most directly identified with classical bivectors). As a consequence we do not have under full control the geometric meaning of the new models, nor their simplicial path integral representation, in terms of a constrained BF theory with Lie algebra and group variables, with a definite choice of measure (beside the effective simplicial path integral in coherent state variables of \cite{laurentflorian}).

\



\subsubsection{The (non-commutative) geometric strategy and simplicial path integrals}
Let us now pass to the second strategy, which starts from the Lie algebra representation of the Ooguri GFT model. This provides in fact a convenient starting point for imposing in a geometrically transparent manner the discrete gravity constraints \cite{new,constrainedBF}. Notice that simplicial BF in 4d (and euclidean signature) has a classical phase space, which decomposes into $T^*SO(4)$ for each triangle of the simplicial complex (similarly for the continuum theory, quantized a la LQG). This justifies the use of the $\SO(4)$ non-commutative group Fourier transform.  This line of research is less developed, having started only recently \cite{ioaristide,ioaristide2, ioaristide3}. We report briefly on its current status.

\

One can easily impose that the four bivectors in each tetrahedron are orthogonal to the same normal vector $n$ to the tetrahedron, by means of the constraint (encoded in a non-commutative delta function):
\[
\widehat{S}_n(x^\minus_j, x^\plus_j) =  \prod_{j=1}^4 \delta_{\minus nx^\minus_j n^\inv}(x^\plus_j)
\]
where we have used the decomposition of the $\so(4)$ algebra into selfdual and anti-selfdual components, and we indicate with $\widehat{.}$ functions on the Lie algebra. One recognizes it immediately as the quantization (whose details are hidden in the chosen space of functions on $\so(3)$ and $\star$-product) of the simplicity constraints in the form (\ref{simplicity2}).  
One obtains a constrained field: 
\begin{align*}
(\widehat{S}_n \star \vphihat)(x) &= \int_{\SO(4)} \extd g \, \vphi(g) \int_{\SO(3)} \extd u \, \e_{n^\inv u n g^\minus}(x^\minus) \e_{u g^\plus}(x^\plus) \\
&= \int _{\SO(4)} \extd g \, S_n\vphi(g) \, \e_{g^\minus}(x^\minus) \e_{g^+}(x^\plus) 
\end{align*}
where
\[
S_n\vphi(g) := \int_{\SO(3)} \vphi(n^\inv un g^\minus, u g^\plus) 
\]
having decomposed the $\SO(4)$ element $g$ into a pair of $\SO(3)$ elements $g^{\pm}$ following the splitting of the covering group $Spin(4)\sim SU(2) \times \SU(2)$.
Hence $\widehat{S}_n$ is the dual of the projector onto the fields on a homogeneous space $\cS^3 \sim \SO(4) \backslash \SO(3)_n$ obtained as a quotient of $\SO(4)$ with  the subgroup $\SO(3)_n$ of elements which stabilize the normal $n$. The standard Barrett-Crane projector \cite{DPFKR} is obtained for $n=1$, i.e. for the diagonal $\SO(3)$ subgroup.  

By combining the simplicity projector $\widehat{S}:= \widehat{S}_1$ with the standard closure projector, used also in the definition of the Ooguri topological model, one gets the field $\Psi(x)\equiv(\widehat{S} \star \widehat{C}\star\vphihat)(x)$ which can be shown to reproduce the GFT field used in the standard GFT formulation of the Barrett-Crane model. More precisely,  combining the interaction term: 
\[
\frac{\lambda}{5!}\int \Psi_{1234}\Psi_{4567}\Psi_{7389}\Psi_{962\,10}\,\Psi_{10\,851}
\]
with the possible kinetic terms:
\[
\frac{1}{2}\int \Psi_{1234}\Psi_{4321} \; , \hspace{0.1cm} \frac{1}{2} \int (\widehat{C}\star\vphihat)_{1234} (\widehat{C}\star\vphihat)_{4321} \, \hspace{0.1cm} \text{or}\hspace{0.2cm} \frac{1}{2} \int \vphihat_{1234}\vphihat_{4321} \]
one obtains, respectively, the versions of the BC model derived in \cite{DPFKR}, in \cite{PR} and in \cite{valentin}. Notice that in all cases one fixes the normal vector to the time gauge $n=(1,0,0,0)$ in all (the simplicity constraints acting on) GFT fields.
The origin of these different versions can be understood geometrically, in the Lie algebra representation of the GFT.
For $h\in \SO(4)$, we have:
\[
(e_h \star \widehat{S}_n)(x) = (\delta_{\minus n h_\minus^\inv \cdot_{\minus} h_\minus n^\inv }(h_\plus^\inv \cdot_\plus h_\plus) \star e_h)(x) 
= (\widehat{S}_{h \rhd n} \star e_h)(x)
\]
where $h \rhd n := h_\plus n h_\minus^{\minus 1}$. This expresses the fact that, after rotation by $h$,  simple bivectors with respect to the normal $n$ become simple with respect to the rotated normal $h \rhd $. So, closure and simplicity constraints do not commute. Geometrically it means that the normal variables $n$ are not transformed alongside the bivectors under $SO(4)$ frame rotations, as they should.
As a result the BC model couples the bivector variables $x$ across 4-simplices; it does not, however, correlate the normal vectors $n$ associated to the same tetrahedron in different 4-simplices. This implies a missing geometric condition on connection variables $h_{\tau\sigma}$. This point can be corrected using an extended formulation of the model, in which normal vectors to tetrahedra are added to the bivectors as arguments of the GFT field, the closure condition (which translates into gauge invariance of the tetrahedra under $\SO(4)$ rotations) is generalized to a simultaneous rotation of both bivectors and normals, and the normals themselves are identified (by the GFT kinetic term) across 4-simplices. The result is again a specific version of the Barrett-Crane amplitudes, with uniquely specified measure, but defined using projected spin networks \cite{projectedSN}. All this, including an extended discussion of remaining problematic features of the resulting model, from a geometric perspective, and their possible interpretation, can be found in \cite{ioaristide2}.
However, in order to keep the presentation simple and avoid adding further technical ingredients, we stick to the (less satisfactory) formulation outlined above.

A simplicial path integral formulation of the Barrett-Crane model, for, say, the BC version \cite{valentin}, is obtained by using the propagator and vertex:
\[
\prod_{i=1}^4 (\delta_{\minus x^{\pm}_i})(y^\pm_i), 
\quad 
\int \prod_t \extd h_{\tau \sigma} \, \prod_{i=1}^{10} ( \delta_{\minus x^\pm_i} \star \widehat{S} \star \, \e_{h_{\tau\tau'}})(y^\pm_i)
\]
leading to an integral over connection variables $h_{\tau \sigma}$ and Lie algebra elements $x_f$ of the product  of face amplitudes\footnote{The symbol \lq\lq $\bullet$\rq\rq indicates that the result of the $\star$-product of delta functions with plane waves has to be evaluated on the Lie algebra element at the end of the calculation.}
\begin{align*}
A_f &= 
\int \extd g \, (e_{g^\inv} \star \{\delta_{\minus \bullet^{\minus}}(\bullet^\plus) \star \e_{h_{01}}  \star \cdots \delta_{\minus \bullet^{\minus}}(\bullet^\plus) \star \e_{h_{N0}}\} \star \e_g)(x_f) \\
&= \int \extd g \, (\cO_f \star e_{H})(gx_fg^\inv)\,=\,A_f(x_f, h_{\tau\sigma})
\end{align*}
where the observable $\cO$ is a $\star$-product of $N+1$ simplicity constraints written in the frame of the $N+1$ tetrahedra, and suitably parallel-transported:
\[
\cO_f(x) = \bigstar_{j=0}^N \, \delta_{\minus h_{0j}^{\minus \inv} x^{\minus} h_{0j}^{\minus}}(h_{0j}^{\plus \inv} x^\plus h_{0j}^{\plus}) 
\]
The integers $j = 0, .., N$ label a ordered sequence of tetrahedra around the face $f$; $x_f$ is the metric variable attached to $f$ in the frame of the reference tetrahedron $0$. 

Finally the whole Feynman amplitude  takes the form of a non-commutative observable insertion in BF theory
\[
Z_{BC} := \int \prod_{\tau \sigma} \extd h_{\tau \sigma} \int \prod_f \extd^3 x_f \, (\cO_f \star e_{H_f)} (x_f)
\]

Therefore we see that the Lie algebra representation of the GFT model allows: 1) a geometrically transparent imposition of the constraints on the $x$ bivector variables, which moreover takes into account their non-commutativity; 2) to identify some dubious aspects in the standard GFT derivations of the BC model \cite{ioaristide} and to a natural solution of them \cite{ioaristide2}; 3) to obtain automatically a simplicial path integral re-writing of the BC spin foam amplitudes.

Incidentally, we notice again that the mentioned modification of the formalism that leads to a solution of the covariance issue pointed out above implies immediately that the boundary states of the resulting spin foam model  are projected spin networks \cite{projectedSN}.

The same construction can be generalized to obtain models for the topological sector and for the case of finite Immirzi parameter \cite{ioaristide3}. Again, the GFT formalism gives immediately the simplicial path integral formulation of them, which can then be re-written, using Peter-Weyl decomposition, in spin foam form.
The result in both cases is a model that differs, when written in spin foam form, from both the EPRL and the FK models mentioned in the previous section. Clearly, lots remains to be understood about the quantum geometry of all these models, their respective merits and problematic issues, and, most important, the physics they encode.

\section{A selection of research directions and recent results}
We now present a selection of recent results, obviously reflecting (and limited by) our own taste and knowledge.

\subsection{Making sense of the quantum field theory: combinatorics of Feynman diagrams, divergences, scaling and GFT renormalization}
A first area of recent develpments \cite{ren,borel,JosephVincent, razvan2,linearized} has been the application of quantum field theory techniques to GFTs, to gain a better understanding and control over its perturbative expansion, using tools from renormalization theory, which is and will be relevant both for a mathematical definition of the theory and for the study of its continuum approximation. The aim is thus to address the remaining main shortcomings of tensor models, within the GFT approach. 

As we said, GFTs define a sum over simplicial complexes 1) of arbitrary topology and 2) that includes pseudo-manifolds, i.e. complexes contain topological singularities at the vertices (equivalently, the dual d-cells have boundaries which are not (d-1)-spheres). As we have seen, the issue of controlling the sum over topologies, and of identifying an approximation in which simple topologies dominate, has an analogue in the context of matrix models \cite{mm}, and it was solved by the definition of the large-N limit, in which diagrams of trivial topology ($S^2$ in the compact case) dominate the Feynman amplitudes of the theory. The ultimate goal, in this respect, in the GFT context, would be to define some generalized version of this scaling limit. In turn, this requires understanding in detail the degree of divergence of arbitrary GFT diagrams, and possibly characterizing it in terms of topological quantities.
Notice that the issue of controlling the relative weight of manifolds and pseudo-manifolds in the perturbative sum, and possibly identifying a regime in which the second are subdominant arises instead only in dimensions $d>2$. 
Notice also that these issues, which are basically the issues of divergence and renormalization of spin foam amplitudes \cite{matteovalentin}, can now be tackled using standard field theory language and techniques, in the GFT framework; more precisely they become a problem in GFT renormalization. This point of view is complementary to the one focused on spin foam amplitudes per se, as lattice systems or as covariant formulation of the dynamics of Loop Quantum Gravity.

The work of \cite{ren} started making the first steps toward solving these three issues, starting a systematic study of GFT renormalization, in the context of the Boulatov model for 3d (Riemannian) quantum gravity (but the results apply to a wider class of models). These first steps aimed at establishing power counting results for the divergences of the corresponding Feynman amplitudes.  Many more developments followed, and a great deal of results. For an up to date review of them, we refer to \cite{tensorReview,matteovalentin} and to the rest of the literature. Here we limit to a very sketchy, and not so up to date summary. 

We have seen that the Feynman diagrams of the theory are, by construction, 3d triangulations, while the corresponding Feynman amplitudes are given by the Ponzano-Regge spin foam model \cite{PR1,barrettPR,thesis}, which provides a quantization of 3d BF theory or 1st order gravity discretized on the simplicial complex dual to the given Feynman diagram.
The divergences of this (still rather simple) model, or equivalently, the scaling of the amplitudes with respect to a cut-off introduced in the spectral decomposition of the propagator, are very complicated to characterize, mainly due to the very complicated topological structure of $3d$ simplicial complexes. They can immediately be seen to be related to the topology of the bubbles (3-dimensional cells), dual to vertices of the simplicial complex, and this where the attention of the initial investigations was focused. 
What was achieved in \cite{ren} is the following: 
1) a detailed algorithm is given for identifying bubbles (3-cells) in the Feynman diagrams of the model, together with their boundary triangulations, which in turn can be used to identify the topology (genus) of the same boundary.
2) the identification of a subclass of Feynman diagrams which allow for a complete contraction procedure, and thus the ones that allow for an almost standard renormalization, and represent a natural generalization of the 2d planar graphs of matrix models.
3) for this class of diagrams, an exact power counting of divergences, according to which their divergence is of the order;

$$ A_{\Gamma}\,=\,\left(\delta^{\Lambda}(I)\right)^{|{\cal B}_{\Gamma}|-1}$$
where $| {\cal B}_\Gamma|$ is the number of bubbles in the diagram $\Gamma$, and $\delta^\Lambda(I)$ is the delta function on the group, with cut-off $\Lambda$, evaluated at the identity $I$.

These diagrams were then shown \cite{linearized} to be manifolds of spherical topology.

A different perspective on divergences in GFT amplitudes was taken in \cite{borel}, which also tackles the difficult issue of the summability of the entire perturbative sum (thus including the sum over topologies). The authors consider both the Boulatov model and a modification of the same proposed in \cite{freidellouapre}. The modification amounts to
adding a second interaction term in the action, given by:

\begin{equation}
+\frac{\lambda\,\delta}{4!}\prod_{i=1}^{6}\int dg_i\left[
\, \phi(g_1,g_2,g_3)\phi(g_3,g_4,g_5)\phi(g_4,g_2,g_6)\phi(g_6,g_5,g_1)\right].
\end{equation}

The new term corresponds simply to a slightly different
recoupling of the group/representation variables, geometrically corresponding to the only other possible way of gluing 4
triangles to form a closed surface. Even if it has no clear physical interpretation
yet from the quantum gravity point of view, it is indeed a very mild modification,
and most importantly one likely to be forced upon us by
renormalisation group arguments, that usually require us to
include in the action of our field theory all possible terms that are
compatible with the symmetries. This modification gives rise to a Borel summable partition function \cite{freidellouapre}, which shows that a control over the sum over topologies and a non-perturbative definition of the corresponding GFT is feasible.

For both the Boulatov model and the modified one, the authors of \cite{borel} manage to establish very general perturbative bounds on amplitudes using powerful and elegant constructive techniques, rather than focusing on explicit power counting or Feynman evaluations, or on the combinatorial structures of the diagrams. They find that, using the same regularization used in \cite{ren}, the amplitudes of the Boulatov model for a diagram with $n$ vertices, are bounded, with cut-off $\Lambda$, by $K^n \Lambda^{6+3 n/2}$, for some arbitrary positive constant $K$, while the modifed model has amplitudes bounded by $K^n \Lambda^{6+3 n}$. Both bounds can be saturated. in fact, those that saturate the bound in the Boulatov case are type 1 graphs in the definition of \cite{ren}. This result shows that the Freidel-Louapre modification (BFL), even though Borel summable, is more divergent that the original Boulatov model from the perturbtive point of view.

The second main result of \cite{borel} relies again on constructive field theory techniques. A cactus expansion of the BFL model is obtained, and used to prove the Borel summability of the free energy of the model and to define its Borel sum. We can expect more aplication of these techniques to other GFT models, also in higher dimensions.  

\

An important step forward in the study of the combinatorial and then statistical properties of the GFT perturbative expansion, came with the definition of a modified version of the  above group field theory, in any dimension \cite{razvan}, called \lq colored\rq.  This has already been proven to be very useful in the topological analysis of the GFT Feynman diagrams, in GFT renormalization, and in the study of GFT symmetries.

In particular, concerning this last issue, an important development is represented by the identification of the GFT counterpart of the simplicial diffeomorphism symmetry of discrete gravity, for topological models \cite{ioaristideflorian}. This local symmetry of the simplicial gravity path integral corresponds to a global translation-like symmetry of the GFT action, in accordance with the 3rd quantization interpretation of the same \cite{3rd}.

The colored version of the 3d gravity GFT model we have presented. This is defined in terms of four complex fields over $\SU(2)^3$: $\varphi_{f}(g_1,g_2,g_3)$, $f = 1,2,3,4$, with the same diagonal invariance (or closure condition), by the action:

\bes 
S_{3d}[\phi]\,&=&\, \frac{1}{2}\sum_t\int[dg]^3
\varphi_t^*(g_1,g_2,g_3)\varphi_t(g_3,g_2,g_1) \;+ \nonumber \\ &+&\;\frac{\lambda}{4!}\int [dg]^6
\varphi_1(g_1,g_2,g_3)\varphi_2(g_3,g_4,g_5)\varphi_3(g_5,g_2,g_6)\varphi_4(g_6,g_4,g_1) \label{boulatovcolored}\;+\;\text{c.c.}\; ,
\ees
where $*$ denotes complex conjugation.

This model corresponds to having \lq colored\rq the four triangles $t$ of each tetrahedron, and having imposed that only triangles of the same color can be identified when gluing tetrahedra together in the Feynman expansion.
Because of this restriction, and making use of the complex structure to define an orientation in the Feynman diagrams, it turns out that only orientable complexes are generated by the perturbative expansion, regardless of the ordering of arguments chosen in each field.

The colored models can be seen as the GFT equivalent of the multi-matrix models \cite{mm}.

These colored GFT models (in any dimension) have several other nice features. 

\begin{itemize}

\item First of all, the possibility of giving a clear definition of bubbles, i.e. of the 3-cells of the Feynman diagrams of any dimension, which
means that the colored Feynman diagrams identify a full d-dimensional cellular complex, and not just
a 2-complex, i.e. a spin foam. 

\item Second, the possibility of defining a computable cellular homology for each Feynman diagram. This can be used to show that the dual simplicial complexes are manifolds or pseudo-manifolds, as usual, but
with only point-like singularities allowed in the non-manifold case. 

\item Moreover, one can prove the absence of generalized \lq\lq tadpoles\rq\rq and of \lq\lq tadfaces\rq\rq \cite{JosephVincent}. All this simplifies greatly the analysis of divergences in these models. 

\item Indeed, the authors of \cite{JosephVincent} can improve considerably, for colored models, the scaling bounds obtained in \cite{borel} for non-colored models.

\item If the field $\varphi$ is assumed to be fermionic, the model enjoys an SU(4) global symmetry in the field indices, which is however absent in the bosonic case.

 
\item One can also define a related homotopy transformation and show the link between the GFT amplitudes to the fundamental group of the cellular complex.

\end{itemize}
 
Moreover, in \cite{razvan2} the boundary graphs of the same colored models are identified, and the topological (Bollobas-Riordan) Tutte polynomials associated to (ribbon) graphs are generalized to topological polynomials adapted to colored group field theory diagrams in arbitrary dimension.

Also, more specific scaling bounds have been established \cite{iosylvain}, that show how pseudo-manifold configurations are generically suppressed with respect to manifolds, highlighting a  mechanism that, if at play also in 4d gravity models, can lead to a dynamical explanation of the emergence of the nice topological features of our macroscopic spacetime. 

Concerning power counting and scaling of the Feynman amplitudes, we mention the beautiful series of works \cite{matteovalentin} which led, in both colored and non-colored topological GFT models in any dimension, to a precise and complete power counting theorem for the related divergences, completing in a sense the path initiated in \cite{ren}, and generalizing most of the above results. This result shows clearly how the divergences of such topological models have both a purely topological nature, in that they depend on the fundamental group of the whole simplicial complex representing the Feynman diagram and on the structure of its bubbles, and a depends on the specific presentation of this topology, i.e. on the combinatorial details of the simplicial complex chosen, for given topology.
  
\

Finally, a very strong result has been proven very recently \cite{largeN} for the colored topological GFT models, in any dimension, which represents the GFT generalization of the large N limit of multi-matrix models. This was based on many of the results mentioned above and uses heavily the coloring of Feynman diagrams. The result is that the limit of large cut-off of the amplitudes of these GFT models is indeed dominated by diagrams of trivial topology and corresponding to manifolds. Although a better control over the perturbative GFT expansion is still needed, in particular over the sum over topologies it encodes, we now know that GFTs can solve also the other mentioned shortcomings of tensor models. This has been the basis for a variety of new developments, and rigorous results, even though mainly confined to simplified versions of the GFT models we have discussed, called \lq\lq independent identically distributed\rq\rq models, and basically obtained from topological GFT models describing BF theory by removing the closure/gauge invariance constraint that characterizes them. The amplitudes become then rather trivial, and take a form very closely related to that of the dynamical triangulations approach. Among the recent results on these models, we mention only: 

1) the precise identification of the type of simplicial complexes (of spherical topology) that dominate in the perturbative expansion in the scaling limit, and are very likely to dominate also in similar, more involved models, called \lq\lq melons\rq\rq \cite{critical};

2) their exact re-summation leading to the identification of the critical behaviour of the partition function and free energy of these models, corresponding to a large volume limit, for a specific value of the coupling constant, and to the computation of the critical exponent;

3) the identification of a tensor model generalization, again for i.i.d. models, of the Virasoro algebra that characterize the large N limit of matrix models \cite{virasoro}.

Despite these many developments, which confirm that the field is growing and progressing fast, it remains clear that much more needs to be done, especially in the context of the physically more interesting GFT models and for what concerns the continuum limit of the same. In particular, the study of divergences of the new 4d GFT models for gravity, with similar techniques, and of their critical behaviour has recently started \cite{VincentRenorm}, but it remains in its infancy.

\subsection{Extracting physics via mean field approximation: deriving non-commutative effective matter field theories from GFT?}
The other set of results we want to mention are interesting steps in the direction of bridging the gap between the microscopic description of quantum space, as provided by the GFT perturbative expansion (and the language of spin networks, simplices, spin foams, etc) and macroscopic continuum physics, described by General Relativity and quantum field
theories for matter. This is the outstanding problem faced by {\it all} current discrete approaches
to quantum gravity, as well as by Loop Quantum Gravity, despite its continuum nature \cite{libro}.

One would expect \cite{gftfluid} a generic continuum spacetime to be formed by
zillions of Planck size building blocks, rather than few
macroscopic ones, and thus to be, from the GFT point of view, a
many-particle system whose microscopic theory is given by some
fundamental GFT action. This also suggests us to look for
ideas and techniques from statistical field theory and condensed
matter theory, and to try to apply/reformulate/re-interpret them
in a GFT context. A key approximation scheme is given by mean field theory, and it is this idea that has started to be explored in recent years \cite{eterawinston, noi, iolorenzo, ioeterajimmy}, from a variety of perspectives. The results in this direction are still quite scarce and still rather tentative, but they represent a starting point in a promising, in our opinion, research direction. 
Here we outline only one such line of exploration.

\

Condensed matter theory also provides specific examples of systems in which the
collective behaviour of the microscopic constituents in some
hydrodynamic approximation gives rise to effective emergent
geometries \cite{analog}. The collective variables of the fluid, in
these background configurations, can be recast as the component
functions of an {\it effective metric}.
Moreover, the effective dynamics of perturbations (quasi-particles, themselves
collective excitations of the fundamental constituents of the
fluid) around the same background configurations take the form of matter field theories in curved spacetimes, with the effective metrics
obtained from the collective background parameters of the fluid.
Inspired by these results, we ask: assuming that a given GFT model describes
the microscopic dynamics of a
{\it discrete quantum space}, and that some solutions of the
corresponding fundamental equations can be interpreted as
identifying a given quantum spacetime configuration (this is justified also by the interpretation of the GFT equations of motion, as we have discussed),  can we
obtain an effective macroscopic {\it continuum} field theory for
matter fields from it, using a similar strategy? If so, what is the effective
spacetime and geometry that these emergent matter fields see?

The answer is affirmative \cite{eterawinston, noi}. The effective matter
field theories that we obtain most easily from GFTs are quantum field
theories on non-commutative spaces of Lie algebra type.

The basic point is the use of the natural duality between Lie algebra and corresponding Lie group, the same duality at the basis of the Lie algebra representation of GFTs, which is the non-commutative version of the usual duality between coordinate and momentum space. We have already mentioned that this duality has its basis in the classical phase space of simplicial gravity and loop quantum gravity, strictly related to the cotangent  bundle $\mathcal{T}^*\SU(2)$. This would lead to a natural interpretation of the group manifold as configuration space and the Lie algebra as momentum space. This setting can also be turned upside-down. If we have a non-commutative spacetime of Lie algebra type $[X_\mu, X_\nu] = C_{\mu\nu}^\lambda X_\lambda$, the momentum space is naturally identified with the corresponding Lie group, in such a way that the non-commutative coordinates $X_\mu$ act on it as (Lie) derivatives. From this perspective, we understand the origin of the spacetime non-commutativity to be the curvature of the corresponding momentum space, a sort of Planck scale \lq\lq co-gravity\rq\rq \cite{majid}. The link with GFTs is then obvious: in momentum space the field theory on such non-commutative spacetime will be given, by definition, by some sort of group field theory, stripped of its combinatorially non-local aspects. This is a general fact. The task is to derive interesting matter field theories from interesting GFT models of quantum spacetime. 

\subsubsection{3d case}
In 3 spacetime dimensions the results obtained concern a euclidean non-commutative spacetime given by the $\su(2)$ Lie algebra, i.e. whose spacetime coordinates are identified with the $\su(2)$ generators with $[X_i,X_j]= i \f{1}{\kappa} \epsilon_{ijk} X_k$. Momenta are instead identified with group elements $\SU(2)$ \cite{majid}, acquiring a non-commutative addition property following the group composition law. The duality is realised by the same non-commutative Fourier transform we introduced earlier. A scalar field theory  in momentum space is then given by a group field theory of the type: 
\be
S[\psi]\,=\,
\f{1}{2}\int_{\SU(2)} dg  \psi(g)\kk(g)\psi(g^{-1})-\f{\lambda}{3!}\int[dg]^3\,
\psi(g_1)\psi(g_2)\psi(g_3)\delta(g_1g_2g_3), 
\ee
in the 3-valent case, where the integration measure is the Haar measure on the group, and $\kk$ is some local kinetic term.  

The Feynman amplitudes of the above scalar field action (with simple kinetic terms) can be derived from the Ponzano-Regge spin foam model coupled to point particles \cite{PR3}, in turn obtainable from an extended GFT formalism \cite{iojimmy}. We will see that the GFT construction to be presented allows to bypass completely the spin foam formulation of the coupled theory.

\medskip

Take now the Boulatov model we have introduced earlier, for a real field $\phi: \SU(2)^3\rightarrow \R$  invariant under the diagonal right action (we use right instead of left action to stay closer to the results as presented in \cite{eterawinston}).

Now \cite{eterawinston} we look at two-dimensional variations of
the $\phi$-field around classical solutions of the corresponding equations of motion:     

\be \phi(g_3,g_2,g_1)\,=\,\f{\lambda}{3!}\int
dg_4dg_5dg_6 \phi(g_3,g_4,g_5)\phi(g_5,g_2,g_6)\phi(g_6,g_4,g_1).
\ee 
That is, calling $\phi^{(0)}$ a generic solution to this equation, we
look at field perturbations
$\delta\phi(g_1,g_2,g_3)\equiv\psi(g_1g_3^{-1})$ which do not
depend on the group element $g_2$.
We consider a specific class of classical solutions, named ``flat"
solutions (they can be interpreted, from a variety of (non-conclusive) arguments, as {\it quantum flat space} on some a priori non-trivial topology): 
\be
\phi^{(0)}(g_1,g_2,g_3)\,=\,\sqrt{\f{3!}{\lambda}}\,\int dg\;
\delta(g_1g)F(g_2g)\delta(g_3g), \quad F:G\rightarrow\R\;\;\;, \ee 
with $F(g)$ assumed to be invariant under conjugation $F(g)=F(hgh^{-1})$.
As shown in \cite{eterawinston}, this ansatz gives solutions to the field
equations as soon as $\int F^2=1$.

This leads to an effective action for the 2d variations $\psi$: \be
S_{eff}[\psi]\,=\,
\f12\int\psi(g)\kk(g)\psi(g^{-1})-\f\mu{3!}\int[dg]^3\,
\psi(g_1)\psi(g_2)\psi(g_3)\delta(g_1g_2g_3)
-\f{\lambda}{4!}\int[dg]^4\, \psi(g_1)..\psi(g_4)\delta(g_1..g_4),
\ee with the kinetic term and the 3-valent coupling given in term
of $F$ (and $\lambda$):
$$
\kk(g)\,=\,1-2\left(\int F\right)^2-\int dh F(h)F(hg), \qquad
\f\mu{3!}\,=\,\sqrt{\f{\lambda}{3!}}\,\int F\;\;\; ,
$$
while the 4th order interaction is independent of the solution chosen.
Such an action defines a non-commutative quantum field theory invariant
under the quantum double of $\SU(2)$ (a quantum
deformation of the Poincar\'e group) \cite{PR3,majid,eterawinston,NCFourier}.
Being an invariant function, $F$ can be expanded in group characters: \be F(g)=\sum_{j\in\N/2}
F_j\chi_j(g) \qquad F_j=\int dg\,
F(g)\chi_j(g), \ee where the $F_j$'s are the Fourier coefficients
of the Peter-Weyl decomposition in irreducible representations of $\SU(2)$, labelled by $j\in\N/2$.
The kinetic term reads then: \be \kk(g)=1- F_0^2-\sum_{j\ge 0}\f{F_j^2
}{d_j} \chi_j(g)=\sum_{j\ge 0}
F_j^2\left(1-\f{\chi_j(g)}{d_j}\right)-2F_0^2\,\equiv\,
Q^2(g)-M^2. \ee It is easy to check that $Q^2(g)\ge 0$ with $Q(\id)=0$. We interpret this term as the generalized ``Laplacian" of the theory
while the 0-mode $F_0$ defines the mass $M^2\equiv 2 F_0^2$. The same operator can be written in terms of momentum coordinates $\vec{p}\,=\,\Tr( g \vec{\sigma})$.

If we choose the simple solution (other choices will give more complicated kinetic terms) \be
F(g)=a+b\chi_1(g), \qquad \int F= a^2+b^2=1, \ee we obtain \be
\kk(g)=\f43(1-a^2)\,\vec{p} \,^2 -2a^2, \ee
i.e. a standard Klein-Gordon kinetic operator, with bounded momentum.

In general, however, we get a relativistic field theory with a deformed Poincar\'e symmetry and thus with both  deformed dispersion relations and modified scattering thresholds.

\subsubsection{4d case}
The above procedure is not restricted to $\SU(2)$ (nor to groups, as it can be applied to homogeneous spaces as well). If one wants to apply it to more physical situations, we need to identify the relevant group and non-commutative space for 4d effective field theories, and, later, interesting GFT models from which these can be derived. It is fair to say that we have no strong or conclusive result to exhibit in this domain, but work in this direction has started \cite{noi}. We now illustrate briefly this initial step.

\

A 4-dimensional non-commutative spacetime that is of direct relevance for Quantum Gravity phenomenology is so-called $\ka$-Minkowski \cite{majid,QGPhen}. We recall here some of its features, and refer to \cite{noi} for further details and references. The non-commutative coordinates have the Lie algebra structure
\be
\label{an3} [X_0,X_k]\,=-\f{i}{\ka}X_k,\qquad [X_k,X_l]\,=0, \quad
k,l=1,...,3. \ee
$\ka$-Minkowski space-time can also be identified with the Lie algebra $\an_{3}$, which is
a subalgebra of $\so(4,1)$. Indeed, if $J_{\mn}$ are the generators of $\so(4,1)$,  the $\an_{3}$ generators are: \be
X_0=\f{1}{\ka}J_{40},\quad X_k=\f{1}{\ka}(J_{4k}+J_{0k}), \quad
k=1,...,3 \,\, , \ee 
Using this, we can then define non-commutative plane waves with the $\AN_3$ group elements as $h(k_\mu)=h(k_0,k_i)\,\equiv\,e^{ik_0X_0}e^{ik_iX_i}$, thus identifying the coordinates
on the group $k_\mu$ as the wave-vector (in turn related to
the momentum). From here, a non-commutative addition of wave-vectors follows from the group multiplication of the corresponding plane waves.
The construction of measure and action of the related non-commutative field theory rests on this embedding of $\AN_3$ into $\SO(4,1)$, about which we need then to say a bit more. 
The Iwasawa decomposition \cite{klymik} relates $\SO(4,1)$ and $\AN_{3}$ as: \be \SO(4,1)\,=\,
\AN_{3}\,\SO(3,1)\,\cup\,\AN_{3}\cM\,\SO(3,1), \ee where
the two sets are disjoint and $\cM$ is the diagonal
matrix with entries $(-1,1,1,1,-1)$
in the fundamental 5d representation of $\SO(4,1)$. 
An arbitrary point $v$ on the coset $\SO(4,1)/\SO(3,1)$, 
can be uniquely obtained as: \be v\,=\,(-)^\epsilon h(k_\mu).v^{(0)}=
h(k_\mu)\cM^\epsilon.v^{(0)},\qquad \epsilon=0\textrm{ or }1,\quad
h\in\AN_{3}, \ee where we have taken a reference space-like vector
$v^{(0)}\equiv(0,0,0,1)\in \R^4$, such that its little group is the Lorentz group $\SO(3,1)$ and the action of
$\SO(4,1)$ is transitive, and defined the
vector $v\,\equiv\, h(k_\mu).v^{(0)}$ with coordinates:
\be
v_0\,=\, -\sinh\f{k_0}{\ka}+\f{{\bf k}^2}{2\ka^2}e^{k_0/\ka} \;\;\;\;
v_i\,=\, -\f{k_i}{\ka} \;\;\;\;\;
v_4 \,=\, \cosh\f{k_0}\ka -\f{{\bf k}^2}{2\ka^2}e^{k_0/\ka}. \nn
 \ee The sign $(-)^\epsilon$ corresponds to the two
components of the Iwasawa decomposition. 
We then introduce the set $\AN^c_{3}\equiv\AN_{3}
\cup\AN_{3}\cM$, such that the Iwasawa decomposition reads
$\SO(4,1)=\AN^c_3\,\SO(3,1)$. One can check that $AN^c_{3}$ is itself a group.
A crucial point is that the component $v_4$ of the above vector is left invariant by the action of the Lorentz group $\SO(3,1)$. This suggests to use this function of the \lq\lq momentum\rq\rq $k_\mu$ as a new (deformed) invariant energy-momentum  (dispersion) relation, in the construction of a deformed version of particle dynamics and field theory on $\ka$-Minkowski spacetime.  This is the basis for much current QG phenomenology \cite{QGPhen}. 
Finally, we need an integration measure on $\AN_3$ in order
to define a Fourier transform \cite{noi}. 
For the free real scalar field $\phi: G\rightarrow \R$, we
define the action \be \label{action DSR 1} \ss(\phi)= \int
dh \, \phi(h)\, \kk(h)\, \phi(h), \quad \forall h\in G, \ee
where $dh$ is a left invariant
measure. We then interpret $G=\AN_3,\AN_3^c$ as the momentum space. We
demand $\kk(h)$ to be a function on $G$ invariant  under
Lorentz transformations, which suggests to use some 
function $\kk(h)=f(v_4(h))$.  Two common choices are
 \be \kk_1(h)= (\kappa
^2-\pi_4(h)) -m^2, \quad \kk_2(h)= \kappa ^2-(\pi_4(h))^2-m^2,
\quad \pi_4=\kappa v_4. \ee
The above action is then Lorentz invariant if we choose a Lorentz invariant measure $dh_L$.

One can define a non-commutative Fourier transform on $\AN_c^c$, using the above plane-wave $h(k_\mu)$ \cite{majid,FKG}. The
group field theory action on $G$ can now be rewritten as a
non-commutative field theory on $\ka$-Minkowski (in the $\AN_3$ case) \be \label{action DSR 2}
\ss(\phi)= \int dh_L \, \phi(h)\, \kk(h)\, \phi(h) = \int \,
d^4X\, \left(
\partial_\mu\hat\phi(X)
\partial^\mu\hat\phi(X) + m^2 \hat\phi^2(X)\right). \ee The
Poincar\'e symmetries are naturally deformed in order to be
consistent with the non-trivial commutation relations of the
$\ka$-Minkowski coordinates \cite{FKG}.

\

Now we should derive this class of theories from GFT models describing quantum space.

We start then from the group field theory describing $\SO(4,1)$
BF-theory.

From the quantum gravity perspective, there are several reasons of interest in this model: 1) the McDowell-Mansouri formulation (as well as related ones
\cite{artem}) 
defines 4d gravity with cosmological constant as a BF-theory for $\SO(4,1)$ plus a potential
term which breaks the gauge symmetry from $\SO(4,1)$ down to the
Lorentz group $\SO(3,1)$; this suggests to try to define Quantum
Gravity in the spin foam context as a perturbation of a
topological spin foam model for $SO(4,1)$ BF theory. These ideas
could also be implemented directly at the GFT level, and the starting point would necessarily be a GFT for $SO(4,1)$ of
the type we use here. 2) we expect \cite{iogft,laurentgft} any
classical solution of this GFT model to represent quantum De
Sitter space on some given topology, and such configurations
would be physically relevant also in the
non-topological case. 3) the spinfoam/GFT model for $\SO(4,1)$ BF-theory
seems the correct arena to build a spin foam model for 4d
quantum gravity plus particles on De Sitter space \cite{artemkg},
treating them as topological curvature defects
for an $SO(4,1)$ connection, similarly to the 3d case \cite{PR3}.

For a general 4d GFT related to topological BF quantum
field theories with gauge group ${\cal G}$, the action is: \bes S_{4d}&=&\f12\int
[dg]^4\,\phi(g_1,g_2,g_3,g_4)\phi(g_4,g_3,g_2,g_1) \\&&
-\f{\lambda}{5!} \int [dg]^{10}
\phi(g_1,g_2,g_3,g_4)\phi(g_4,g_5,g_6,g_7)\phi(g_7,g_3,g_8,g_9)\phi(g_9,g_6,g_2,g_{10})\phi(g_{10},g_8,g_5,g_1),
\nn \ees where the field is again required to be gauge-invariant,
$\phi(g_1,g_2,g_3,g_4)=\phi(g_1g,g_2g,g_3g,g_4g)$ for any $g\in{\cal G}$. The relevant group for our construction will be $\SO(4,1)$ (which requires some
regularization to avoid divergencies).

\

We generalize to 4d the ``flat solution" ansatz of the 3d group field
theory as \cite{eterawinston}: \be
\phi^{(0)}(g_i)\,\equiv\, {}^3\sqrt{\f{4!}{\lambda}}\int
dg\,\delta(g_1g)F(g_2g)\tlF(g_3g)\delta(g_4g), \ee with $(\int F\tlF)^3=1$. 
A simpler special case of the classical solution above is obtained choosing
$\tF(g)=\delta(g)$ while keeping $F$ arbitrary but with $F(\id)=1$. Calling $c\equiv\int F$, the
effective action for 2d perturbations $\psi(g)$, around such background, becomes \cite{noi}: \bes
S_{eff}[\psi]&=& \f12\int
\psi(g)\psi(g^{-1})\left[1-2c^2-2cF(g)F(g^{-1})\right]
+\; \text{interactions}. \label{compacteffaction}
\ees

In order to make contact with deformed special relativity, we now specialize this construction to one that gives an effective field
theory based on the momentum group manifold $AN_3$.

Following the above procedure we naturally
obtain an effective field theory living on $\SO(4,1)$. We want then to obtain from it an effective theory
on $AN_3^{(c)}$. 
We choose: \be F(g)\,=\, \alpha(v_4(g)+a)\vartheta(g),
\qquad \tF(g)\,=\,\delta(g). \ee The function $v_4$ is defined as
matrix element of $g$ in the fundamental (non-unitary)
five-dimensional representation of $SO(4,1)$, $v_4(g)\,=\,\la
v^{(0)}|g|v^{(0)}\ra$, where $v^{(0)}=(0,0,0,0,1)$ is, as
previously, the vector invariant under the $SO(3,1)$ Lorentz
subgroup. $\vartheta(g)$ is a cut-off function providing a
regularization of $F$, so that it becomes an integrable function. Assuming that
$\vartheta(\id)=1$, we require $\alpha=(a+1)^{-1}$ in order for the normalization condition
to be satisfied.

Then we can derive the effective action around such classical
solutions for 2d field variations: \bes S_{eff}[\psi]&=&
\f12\int
\psi(g)\psi(g^{-1})\left[1-2c^2-\f{2c\vartheta^2(g)(a+v_4(g))^2}{(a+1)^2}\right]
+\; \text{interactions},
\label{noncompacteffaction}\; . \ees 
We recognize the
correct kinetic term for a DSR field theory. However, the
effective matter field is still defined on a
$SO(4,1)$ momentum manifold. The only remaining issue is therefore
to understand the ``localization" process of the field $\psi$ to
$\AN^c_3$.
This can be obtained in a variety of more or less satisfactory ways for a description of which we refer to \cite{noi}.
In any case, restricting to group elements $h_i\in\AN^c_3$,
we arrive at a proper DSR field theory on $\kappa$-Minkowski, with a
$\ka$-deformed Poincar\'e symmetry, obtained from the GFT for $\SO(4,1)$ topological
BF-theory.

Notice that if we had started from a GFT model for BF theory with gauge group $\AN(3)$, we would have obtained the above result without any effort, like in the 3d case. However,  we would have lost any direct connection with quantum gravity models, as it is unclear how such BF theory is related to gravity.

\

Work in this direction, including these preliminary results, is a step towards bridging the gap between our
fundamental discrete theory of spacetime and the
continuum description of spacetime we are accustomed to, thus hopefully bringing this class of models a closer to quantum gravity phenomenology and
experimental falsifiability. 

For example, as mentioned earlier, in a similar spirit with the above construction, the application of mean field theory ideas for the extraction of effective geometrodynamics equations from GFTs (and effective dynamics for generic perturbations), has recently started along two different directions:

1) using GFT classical solutions as background configurations and looking for a geometrodynamics interpretation (effective Hamiltonian constraint for spin networks) in the effective theory for perturbations around it \cite{ioeterajimmy};

2) using loop quantum gravity coherent states as background configurations, and following the paradigm of the derivation of hydrodynamics (Gross-Pitaevski) equations in Bose condensates \cite{iolorenzo}.

These are just first steps, of course. Much more work to be done lies ahead.

\

We point out that contrary to the situation in analog gravity models in condensed matter \cite{analog}, our GFT models are non-geometric and far from usual
geometrodynamics in their formalism, but at the same time
are expected to encode quantum geometric information and to
determine, in particular in their classical solutions, a (quantum
and therefore classical) geometry for spacetime. We
are, in other words, far beyond a pure analogy, here, at least as a matter of principle.

\section{A selection of open issues}
\noindent We list now, with no presumption of completeness, several open issues in the GFT approach that we deem important (and some ideas for tackling them). We hope the brave and talented reader will pick them up and join our ongoing efforts to address them.

\

\noindent Obviously, the first issue is the {\it construction of a convincing GFT model for 4d quantum gravity}.
By this we do not mean that such convincing model should be derived by \lq quantizing\rq GR, or that its dynamics should be given by some operator version of the Einstein's equation in the continuum. We do not believe that GFTs should be understood in this way. GFTs are, in our view, candidates for the microscopic dynamics of the building blocks, the atoms of quantum space, while GR is an effective macroscopic dynamics for large collections of them.
From the point of view presented in this article, this means obtaining a GFT model whose Feynman amplitudes have a compelling expression as a simplicial path integral for 4d gravity, with a clear geometric meaning of the various contributions to the amplitude. Also, its dual spin foam expression should be derivable and possibly manageable, showing the translation of the geometric content of the amplitude in the algebraic language of group representation theory, at the quantum level.
Finally, its space of boundary states should be clearly identified, and related to the space of states of canonical Loop Quantum Gravity.

\

\noindent A second issue is then to achieve {\it a rigorous and physically transparent link between the canonical LQG framework (or, more generally, a canonical theory based on spin networks) and the GFT/spin foam one}.

We would expect, in fact, that a proper covariant reformulation of the canonical dynamics of LQG quantum states, for any graph-changing Hamiltonian (or Master) constraint operator \cite{thomas} would lead to a GFT formalism, in the same way as the dynamics of any system of particles, including the possibility of creation and annihilation of them, is best captured in a QFT formalism. To realize this correspondence, one would need for example a rigorous definition and a detailed understanding of the space of GFT states as a Fock space, and to compare this with kinematical state space of LQG, showing in what exact sense the former is a second quantization of the latter. Having this, one could attempt a derivation of the GFT path integral from LQG using coherent states, as often done in the usual QFT case. For this, the non-commutative GFT representation  seems particularly suited. Any such derivation should lead to a better understanding of how any realistic LQG Hamiltonian/Master constraint is encoded in GFT action, and thus in its equations of motion.

\

\noindent This brings us to the next issue, that of  {\it solutions of the GFT equations of motion}.
 
Obviously, first of all it is necessary to identify more of them and understand their physical/geometric meaning (including the ones already found \cite{eterawinston, noi}). Also, this should clarify in what sense they correspond to solutions of Hamiltonian/Master constraint. In particular, simplified field configurations, possibly corresponding with more symmetric configurations of quantum space, would play a special role, and lead to simplified effective dynamics. A better grasp over the space of solutions of the classical GFT equations can be obtained, possibly, by learning how to control and use the perturbative GFT tree level expansion. Finally, this could be an avenue to investigate the physical (gravitational) meaning of GFT coupling constant.

\

\noindent The problem of gaining control over the tree level expansion of the GFT is of course part of the more general issue, that we discussed above, of {\it gaining control and understanding of the GFT Feynman expansion}.

As we have shown, this is a complex matter, even for simple GFT models. It includes identifying the contribution of non-manifold configurations to the sum, and possibly suppressing it in some way. It involves addressing in full the problem of divergences: of individual Feynman amplitudes (simplicial path integrals/spin foam models), i.e. the perturbative GFT renormalization, and of the total perturbative sum, i.e. its (Borel) summability and non-perturbative definition.
It means understanding the role of topology change and of its physical consequences, and controlling somehow the sum over topologies. Once more, work on these aspects of the GFT perturbative expansion could also shed light on the meaning of the GFT coupling constant, which has been suggested to govern exactly the topology changing processes \cite{laurentgft,iogft}. These are the task of {\it GFT renormalization}, if one studies GFTs as bona fide quantum field theories, or of statistical methods, if one sees them (in the spirit of matrix models) as statistical systems of random complexes. Of course, this type of study will be crucial also for the issue of critical behaviour and phase transitions, and for the continuum limit of the theory, that we will discuss again below.

\

\noindent  
As we have mentioned, it is only recently \cite{ioaristideflorian} that some symmetries of discrete gravity, already identified at the level of the GFT amplitudes, i.e. at the level of simplicial gravity path integrals have been identified also at the level of the corresponding GFT action. These are the local rotation and the translation symmetry that characterize BF theory, in any dimension. In particular, translation symmetry is crucial for the topological invariance of the theory, and it is also closely related to diffeomorphism symmetry, to the point that it could almost be identified with it \cite{laurentdiffeo}. Therefore, this result is a starting clue for a more extensive {\it analysis of the GFT symmetries and their relation with the symmetries of simplicial gravity; in particular, the identification of diffeomorphism in GFTs}. In particular, having identified diffeomorphism symmetry in 3d gravity, and translation symmetry in 4d BF, we have now to understand whether and how exactly they are broken \cite{bianca} when passing to 4d gravity models. In fact, it can already be seen that the imposition of the gravity constraints on 4d BF theory, in general, breaks the BF translation symmetry. It remains however to study if there is any remnant of such symmetry, and its relation to simplicial diffeomorphisms, the details of its breaking, and how it can be recovered is some (geometric) regime. 
The possibility of working in a simplicial path integral representation for the GFT amplitudes is going to be crucial. Having identified the symmetries of various GFT models, one should develop a systematic analysis of these symmetries (and others) at the quantum GFT level, using QFT tools (e.g. Ward identities on GFT n-point functions, which we know correspond to states of quantum space, and thus may encode some version of the Hamiltonian constraint of GR, and thus the quantum dynamics of geometry).
This analysis should also provide clues for the study of the continuum limit of the theory, which, if indeed gives back some (modified) version of General Relativity, is almost fully characterized by the presence of diffeomorphism symmetry.

\

\noindent The problem of {\it the continuum approximation and of the link with General Relativity (and matter field theories} is the real outstanding open issue in the GFT framework, like in most other approaches to non-perturbative and background independent quantum gravity. 

Here there are two possibilities. As we have seen, the GFT perturbative expansion around the vacuum is an expansion in simplicial complexes of higher and higher complexity. The question is then: are a few simplices or some simple spin networks enough, in order to compute approximate, but physically meaningful {\it continuum} geometric quantities? 

The first possibility is that the answer is positive, i.e. we can give a physical, continuum meaning to working with a simplicial complex, in terms of some precise truncation of the degrees of freedom of the full continuum theory, or in terms of some large scale approximation. This is, for example, the point of view behind the applications of spin foam models to graviton propagation and scattering and to cosmology in \cite{graviton, SFcosmology}. If this is the case, then it could be enough, at least for a subset of physically interesting questions, to work at a given (low?) order in the GFT perturbation theory, and to compute physical continuum geometric quantities using a fixed simplicial complex, or a finite number of them. Then we can use many results in simplicial gravity to do so, and the coherent state techniques for fixed spin network graphs or encoding semi-classical simplicial geometry will be crucial also to study the continuum semi-classical approximation of the theory, and to extract physics from GFT models. 

A different possibility is that small numbers of simplices or simple graphs do not capture in an adequate way continuum information, especially concerning the dynamics of the theory (and thus continuum macroscopic physics). In order to do so, then, one would have to use highly refined simplicial complexes made out of very high numbers of simplices \cite{biancacoarse}. From the GFT point of view, this means that the physics of continuum spacetimes has to be looked for in the regime of many GFT particles, thus far from the perturbative (Fock) vacuum around which the GFT Feynman expansion is defined, which correspond to the $\varphi =0$ \lq\lq no-space\rq\rq state. 

\

\noindent In this case, one should study {\it statistical GFT, i.e. the physics of many GFT quanta}. 

In particular, the question of the continuum approximation becomes the question of identifying the correct {\it GFT phase(s), the corresponding phase transition(s), and the relevant regime(s) of dynamical variables} around which a continuum approximation becomes valid and {\it the effective dynamics} of the GFT system is described by (maybe modified) General Relativity. Again, if the physics of the continuum spacetime is the physics of large ensembles of GFT quanta, then the correct conceptual strategy is to treat quantum space as a sort of weird condensed matter system with microscopic, atomic description given by some (class of) GFT models. Notice that this is exactly what happens in matrix models. The issue is then first of all to develop the appropriate mathematical tools to study the thermodynamic limit of GFTs, identify the relevant phases and extract the effective GFT dynamics around them. Second, one need to devise methods (and probably an appropriate conceptual framework) to re-express this effective dynamics in spacetime and geometric terms, i.e. from the GFT \lq\lq pre-geometric\rq\rq language to the language of continuum General Relativity. We are already in a rather speculative setting, here. But we could speculate further \cite{gftfluid,FQXiEssay} that GFTs could provide the right framework to realize the idea \cite{volovik, hu} of continuum spacetime as a sort of condensate, in the precise form of a condensed or fluid phase of (very many) GFT quanta, simplices or spin network vertices, and of GR as a sort of hydrodynamics (or thermodynamics \cite{ted}) for these fundamental quanta of space in such regime.

\

\noindent Whether these speculations are correct or not, and whether we can realize them fully and rigorously or not, the really important point is to be able to {\it extract physical predictions from GFTs, possibly together with a better understanding of the fundamental nature of space and time}, and say something new and interesting about our world.

\section{Conclusions}
To conclude, we have introduced the key ideas behind the group field theory approach to quantum gravity, and the tentative microscopics of quantum space it suggests. We have introduced the basic elements of the GFT formalism, focusing on the 3-dimensional case for simplicity. We have reported briefly on the current status of the work devoted to the construction of interesting 4-dimensional GFT models. Finally, we have also briefly reported on some recent results obtained in this approach, concerning both the mathematical definition of these models as bona fide field theories, and  possible avenues towards extracting interesting physics from them. We hope that, our outline shows clearly that, while much more work is certainly needed in this area of research, the new direction toward quantum gravity that group field theories provide is exciting and full of potential.

\section*{Acknowledgements}
We thank A. Baratin, V. Bonzom, G. Calcagni, S. Carrozza, B. Dittrich, L. Freidel, F. Girelli, S. Gielen, R. Gurau, E. Livine, R. Pereira, M. Raasakka, V. Rivasseau, C. Rovelli, J. Ryan, L. Sindoni, M. Smerlak, J. Tambornino, and many other colleagues for useful discussions and collaboration on the topics dealt with in this contribution. A very special thank goes to C. Restuccia for a careful reading of the draft and a very detailed and useful list of comments and corrections. We also thank the editors of this volume for their patience with a difficult contributor. Support from the A. von Humboldt Stiftung through a Sofja Kovalevskaja Prize is gratefully acknowledged.

\end{document}